\documentclass[iop]{emulateapj}

\usepackage{lineno}
\usepackage{graphicx}
\usepackage{subfigure}
\usepackage{amssymb}
\usepackage{amsmath}
\usepackage{mathrsfs}
\usepackage{color}
\usepackage{hyperref}
\hypersetup{
    bookmarks=true,
    bookmarksopen=false,
    colorlinks=true,        % false: boxed links; true: colored links
    linkcolor=blue,          % color of internal links
    citecolor=blue,        % color of links to bibliography
    filecolor=magenta,      % color of file links
    urlcolor=cyan           % color of external links
}
\allowdisplaybreaks[2]
\usepackage{morefloats}

\newcommand{\newtext}[1]{{#1}}

\shorttitle{Kinetic study of radiation-reaction-limited particle acceleration}
\shortauthors{Yuan et al.}

\begin{document}

\title{Kinetic study of radiation-reaction-limited particle acceleration during the relaxation of unstable force-free equilibria}
\author{Yajie Yuan\altaffilmark{1}, Krzysztof Nalewajko\altaffilmark{1,2,3}, Jonathan Zrake\altaffilmark{1}, William E. East\altaffilmark{1}, Roger D. Blandford\altaffilmark{1}}

\altaffiltext{1}{Kavli Institute for Particle Astrophysics and Cosmology, SLAC National Accelerator Laboratory, Stanford University, 2575 Sand Hill Road M/S 29, Menlo Park, CA 94025, USA}
\altaffiltext{2}{Nicolaus Copernicus Astronomical Center, Bartycka 18, 00-716 Warsaw, Poland}
\altaffiltext{3}{NASA Einstein Postdoctoral Fellow}

\begin{abstract}
Many powerful and variable gamma-ray sources, including pulsar wind nebulae, active galactic nuclei and gamma-ray bursts, seem capable of accelerating particles to gamma-ray emitting energies efficiently over very short time scales. These are likely due to rapid dissipation of electromagnetic energy in a highly magnetized, relativistic plasma. In order to understand the generic features of such processes, we have investigated simple models based on relaxation of unstable force-free magnetostatic equilibria. In this work, we make the connection between the corresponding plasma dynamics and the expected radiation signal, using 2D particle-in-cell simulations that self-consistently include synchrotron radiation reaction. We focus on the lowest order unstable force-free equilibrium in a 2D periodic box. We find that rapid variability, with modest apparent radiation efficiency as perceived by a fixed observer, can be produced during the evolution of the instability. The ``flares'' are accompanied by an increased polarization degree in the high energy band, with rapid variation in the polarization angle. Furthermore, the separation between the acceleration sites and the synchrotron radiation sites for the highest energy particles facilitates acceleration beyond the synchrotron radiation reaction limit. We also discuss the dynamical consequences of radiation reaction, and some astrophysical applications of this model. Our current simulations with numerically tractable parameters are not yet able to reproduce the most dramatic gamma-ray flares, e.g., from Crab Nebula. Higher magnetization studies are promising and will be carried out in the future.
\end{abstract}


\section{Introduction}
Many powerful gamma-ray sources are found to produce dramatic flares that suggest rather high radiation efficiency. One example is the Crab Nebula \citep[e.g.,][]{Abdo:2011aa-Crab,Tavani:2011aa,Buhler:2014aa}, which flares in the energy range $\sim$100 MeV$-$1 GeV roughly once per year. The biggest flare reached an isotropic luminosity $L_{\gamma}=1\%$ of the pulsar's spin down power, and the flux doubling time scale $t_v$ can be as short as a few hours \citep{Buehler:2012aa}. This suggests an isotropic fluence $\mathcal{E}_{\rm rad}=L_{\gamma}t_v=3.6\times10^{40}L_{36}t_{10 hr}$ erg, taking $L_{\gamma}=10^{36}L_{36}$ erg s$^{-1}$ and $t_v=10 t_{10 hr}$ hr. If we define the radiation efficiency $\epsilon$ as the ratio between $\mathcal{E}_{\rm rad}$ and the magnetic energy $\tilde{\mathcal{E}}_B$ contained in the volume $(ct_v)^3$, we would get $\epsilon=\mathcal{E}_{rad}/\tilde{\mathcal{E}}_B\sim10^3 L_{36}B_{-3}^{-2}t_{10 hr}^{-2}$, assuming an average magnetic field of $B=10^{-3}B_{-3}$ G. The radiation mechanism is generally believed to be synchrotron, but the spectral peak goes beyond the classical synchrotron radiation reaction limit $\hbar\omega_{\rm syn,lim}=9mc^2/(4\alpha_F)\approx160$ MeV \citep[e.g., ][]{landau_classical_1975}. This, together with the fast variability, large total energy and non-detection in other wavebands \citep[e.g.,][]{Weisskopf:2013aa,Rudy:2015aa}, presents great challenges to existing particle acceleration theories. Rapid variability is also notably detected in many Active Galactic Nuclei (AGN). For example, the flat spectrum radio quasars (FSRQ) 3C 273 \citep[e.g.,][]{Rani:2013aa} and 3C 279 \citep{Hayashida:2015aa} produce GeV flares on hourly time scales; the radio galaxy IC310 \cite{Aleksic:2014IC310}, the FSRQ PKS 1222+21 \citep{Aleksic:2011aa} and several BL Lac objects, e.g. PKS 2155-304 \citep[e.g.,][]{Aharonian:2007aa} and Markarian 501 \citep{Albert:2007aa}, are observed to flare in TeV band with variability time scales $\sim$ a few minutes. More strikingly, quite a few flares from blazars are found to be accompanied by large polarization angle swings in the optical band \citep{Abdo:2010-3C279polarization,Marscher:2008aa,Blinov:2015aa, kiehlmann_polarization_2016}, suggesting that the dissipation region has a coherent, strong magnetic field.

In many of these cases, the rapid gamma-ray variability is likely to be a consequence of efficient electromagnetic dissipation in a highly magnetized, relativistic outflow from the central engine (neutron star or black hole). Our group has proposed a general idea, called \emph{magnetoluminescence}, to account for the most dramatic gamma-ray flaring events \citep{Blandford:2014aa,Blandford:2015aa}. Magnetoluminescence refers to large scale, catastrophic conversion of electromagnetic energy into high-energy, non-thermal radiation. It could be triggered by some macroscopic, ideal instabilities in the outflow, which leads to formation of regions with non-ideal electric field that accelerate particles efficiently and process through a large volume rapidly. In the radiation-reaction-limited regime, the energy is quickly removed by radiation and, so, this might conclude with an implosion.

Currently in the literature, the process of electromagnetic dissipation in a relativistic plasma has been primarily studied in terms of magnetic reconnection at a planar current layer \citep[e.g.,][]{Kagan:2015aa}. In such a configuration, fast reconnection and energy release proceeds through tearing instability of the current layer. It has been found that in relativistic regime, a fast reconnection rate, $\sim0.2-0.3$, can be reached, especially when the magnetization is high; efficient particle acceleration produces a power law distribution of non-thermal particles $dN/d\gamma\propto\gamma^{-p}$, and the power law index $p$ hardens toward 1 as magnetization increases to very high values \citep[e.g.,][]{Guo:2014aa,Sironi:2014aa,Liu:2015aa,Werner:2016aa}. Magnetic reconnection has been proposed to be the underlying mechanisms for the Crab flares \citep[e.g.,][]{Uzdensky:2011aa,Arons:2012aa,Cerutti:2013aa,Baty:2013aa}, and mini-jet models based on magnetic reconnection are invoked to explain the rapid variability from AGN jets \citep{Giannios:2009aa,Nalewajko:2011aa}. However, most of the existing kinetic simulations of magnetic reconnection start from a Harris-type current sheet that is already on kinetic scales; this is quite artificial and does not capture the more typically dynamic nature of current sheet formation and evolution.

Our group has been investigating new configurations that could be more generic in revealing the basic properties of electromagnetic dissipation in a highly magnetized, relativistic plasma. One class of examples is the force-free equilibria in a Cartesian periodic box. Previous analytical analysis and force-free/MHD simulations \citep{East:2015aa,Zrake:2016aa} have found that the higher order states are generally unstable to ideal modes; they release their free magnetic energy within a single dynamic time scale while keeping the helicity constant. Recently, \citet{Nalewajko:2016aa} carried out 2D particle-in-cell (PIC) simulations of the lowest order unstable configuration, in a mildly relativistic pair plasma. Most interestingly, they show the self-consistent formation of current layers, which evolve on Alfven time scales as determined by the global dynamics. Particles are accelerated by the parallel electric field in the current layers, as well as second order processes during the late stage of relaxation. The non-thermal particle fraction increases with the magnetization. Similar studies have also been carried out independently by \citet{Lyutikov:2016aa} lately.

In this work, we extend our 2D kinetic simulations to include synchrotron radiation reaction self-consistently using the PIC code Zeltron, in a regime where individual particles are ultrarelativistic. We extract the detailed radiation signatures and study the dynamical consequences of radiation reaction systematically. Some aspects of the radiation effect have been investigated before in PIC simulations of Harris-type reconnection \citep{Cerutti:2012aa,Cerutti:2014aa,kagan_beaming_2016}; here we see new features as a consequence of the dynamic evolution of the current layers. Also, for the first time, we calculate the polarization signals from the PIC simulations.

The rest of the paper is organized as follows. We introduce our setup in \S\ref{sec:setup}, and present the results in \S\ref{sec:results}, including the evolution of the instability (\S\ref{subsec:evolution}), mechanisms of particle acceleration and electromagnetic dissipation (\S\ref{subsec:mechanism}), the radiation signatures (\S\ref{subsec:radiation}) and the dynamical consequences of radiation reaction (\S\ref{subsec:dynamical}). We discuss the astrophysical applications in \S\ref{sec:discussion} and present our conclusions in \S\ref{sec:conclusion}.

\section{Setup of the problem}\label{sec:setup}
\subsection{Kinetic description, scalability and numerical method}
For a relativistic pair plasma, the evolution of electromagnetic field is governed by the Maxwell equations with the charge and current densities given self-consistently by the distribution function $F_s$ for each species $s$ as
\begin{gather}
\rho=\sum_s q_s\int F_s d^3u,\label{eq:rho}\\
\mathbf{J}=\sum_s q_s\int F_s\frac{\mathbf{u}}{\gamma}d^3u,\label{eq:J}
\end{gather}
where $\mathbf{u}$ and $\gamma$ are the proper velocity and Lorentz factor of individual particles, respectively. The distribution functions satisfy the conservation of particles in phase space
\begin{equation}
\frac{\partial F_s}{\partial t}+\nabla_{\mathbf{x}}\cdot(\mathbf{v}F_s)+\nabla_{\mathbf{u}}\cdot(\frac{d\mathbf{u}}{dt}F_s)=0,
\end{equation}
and the acceleration of individual particles is determined by
\begin{equation}
m_s\frac{d\mathbf{u}}{dt}=q_s(\mathbf{E}+\frac{\mathbf{v}}{c}\times\mathbf{B})+\mathbf{F}_{\rm rad}.
\end{equation}
We use the Abraham-Lorentz-Dirac formalism for the radiation reaction force \citep{Jackson:1999aa}
\begin{equation}
F_{\mu}^{rad}=\frac{2e^2}{3c^2}\left[\frac{d^2u_{\mu}}{d\tau^2}+u_{\mu}\left(\frac{du_{\nu}}{d\tau}\frac{du^{\nu}}{d\tau}\right)\right],
\end{equation}
where $\tau$ is the proper time. In the case of synchrotron radiation or Inverse Compton (IC) scattering in the Thomson regime, in the ultrarelativistic limit, the first term (jerk) can be neglected and the radiation reaction force becomes essentially a drag opposing the particle motion. For synchrotron radiation,
\begin{equation}
\mathbf{F}_{\rm syn}=-\frac{2e^2}{3c^5}\gamma^2a_{L\perp}^2\mathbf{v},
\end{equation}
where $\mathbf{a}_{L\perp}=\frac{q}{m}(\mathbf{E}+\mathbf{v}\times\mathbf{B}/c)_{\perp}$ is the component of acceleration that is perpendicular to the particle velocity. For IC in the Thomson regime with isotropic background photon field,
\begin{equation}
\mathbf{F}_{\rm IC}=-\frac{4}{3}\sigma_{T}U_{ph}\gamma^2\mathbf{v}/c,
\end{equation}
where $\sigma_{T}$ is Thomson cross section and $U_{ph}$ is the soft photon energy density.
% Synchrotron power for a single electron is $2\sigma_{T}U_{B}\gamma^2c\beta_{\perp}^2$, Only after averaging over electron velocities do we get $\frac{4}{3}\sigma_{T}U_{B}\gamma^2c$.
In the ultra-relativistic limit ($\gamma\gg1$), the distribution function $F_s$ is scalable with particle energy $\gamma$, therefore the equations can be made dimensionless if we measure lengths in terms of the system size $L$, time in units of $L/c$, and let $\mathbf{E}=B_0\mathbf{\tilde{E}}$, $\mathbf{B}=B_0\mathbf{\tilde{B}}$, $\gamma=\gamma_0\tilde{\gamma}$, $F_s=n_0\tilde{F}_s$:
\begin{gather}
\nabla\cdot\mathbf{\tilde{E}}=\frac{L}{r_L\sigma}\tilde{\rho},\\
\nabla\times\mathbf{\tilde{E}}=-\frac{\partial\mathbf{\tilde{B}}}{\partial t},\\
\nabla\cdot\mathbf{\tilde{B}}=0,\\
\nabla\times\mathbf{\tilde{B}}=\frac{\partial\mathbf{\tilde{E}}}{\partial t}+\frac{L}{r_L\sigma}\tilde{\mathbf{J}},\\
\frac{\partial \tilde{F}_s}{\partial t}+\nabla_{\mathbf{x}}\cdot(\mathbf{v}\tilde{F}_s)+\nabla_{\mathbf{\tilde{u}}}\cdot(\frac{d\mathbf{\tilde{u}}}{dt}\tilde{F}_s)=0,\\
\frac{r_L}{L}\frac{d\mathbf{\tilde{u}}}{dt}=(\mathbf{\tilde{E}}+\frac{\mathbf{v}}{c}\times\mathbf{\tilde{B}})-\eta\tilde{\gamma}^2\frac{\mathbf{v}}{c}\left[(\mathbf{\tilde{E}}+\frac{\mathbf{v}}{c}\times\mathbf{\tilde{B}})_{\perp}^2+\frac{2U_{ph}}{3U_{B}}\right],
\end{gather}
where
\begin{gather}
\tilde{\rho}=\frac{\rho}{n_0e}=\sum_s\rm{sgn}(q_s)\tilde{F}_sd^3\tilde{u},\\
\tilde{\mathbf{J}}=\frac{\mathbf{J}}{n_0ec}=\sum_s\rm{sgn}(q_s)\int\tilde{F}_s\frac{\tilde{\mathbf{u}}}{\tilde{\gamma}}d^3\tilde{u},
\end{gather}
\\and $r_L\equiv\gamma_0mc^2/eB_0$ is the nominal Larmor radius, $\sigma\equiv B_0^2/(4\pi n_0\gamma_0mc^2)$ is the nominal magnetization parameter, $U_B=B_0^2/8\pi$ is the nominal magnetic energy density, $U_{ph}$ is the background photon energy density, $\eta$ characterizes the strength of synchrotron cooling
\begin{equation}
\eta\equiv\frac{\omega_{\rm syn}}{\omega_{\rm syn,lim}}=\frac{4\alpha_F \hbar\omega_{\rm syn}}{9mc^2}=\frac{1}{\omega_gt_{\rm cool}}=\frac{2e^3\gamma_0^2B_0}{3m^2c^4}.
\end{equation}
Here the characteristic frequency of synchrotron radiation is $\omega_{\rm syn}=3\gamma_0^3c/(2r_L)$, particle gyro frequency is $\omega_g=eB_0/\gamma_0mc$, and the synchrotron radiation reaction limit occurs at $\hbar\omega_{\rm syn,lim}=9mc^2/(4\alpha_F)\approx160$ MeV. Thus, the evolution of the system is determined by the dimensionless parameters $L/r_L$, $\sigma$, $\eta$ and $U_{ph}/U_B$, making the problem fully scalable.

We use the electromagnetic, relativistic Particle-In-Cell code Zeltron\footnote{http://benoit.cerutti.free.fr/Zeltron/index.html} \citep{Cerutti:2013aa} to evolve the plasma system in time. Zeltron uses the Yee mesh to solve the Maxwell equations, which conserves $\nabla\cdot\mathbf{B}$ to machine precision \citep{Yee:1966aa}. However, Gauss' Law is not automatically satisfied so an explicit Poisson correction is applied every time step. Linear weighting is used in both the charge/current density and force calculations. The equations of motion for the particles are integrated using a modified Boris push to include the radiation reaction force. 

\subsection{Initial conditions}
\begin{figure}
\centering
\includegraphics[width=0.4\textwidth]{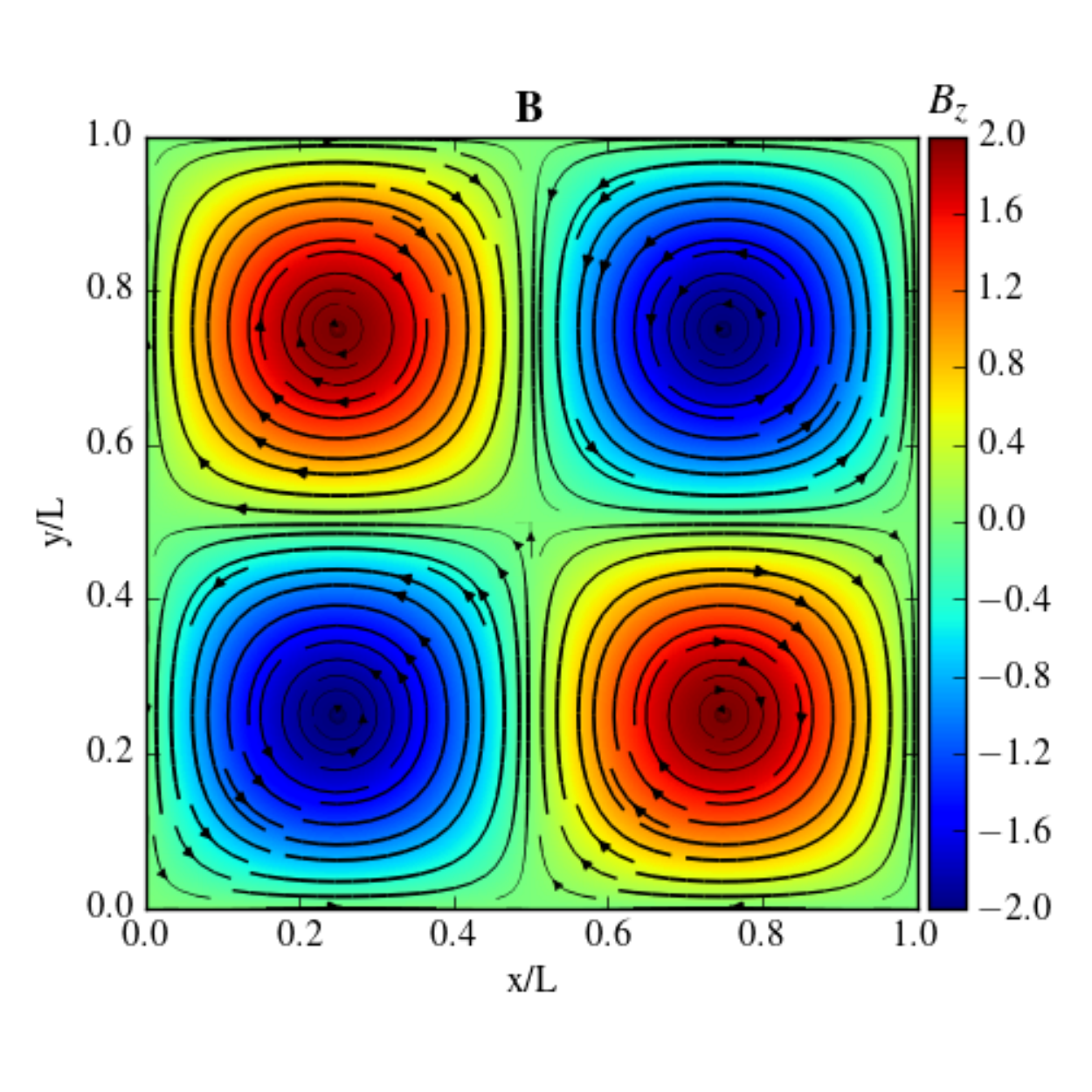}
\caption{Initial magnetic field configuration. The direction and thickness of the streamlines indicate the direction and magnitude of the in plane magnetic field, while the color represents $B_z$. The field strength is in units of $B_0$ (same below).}\label{fig:B0}
\end{figure}

\begin{deluxetable*}{lccccccccc}[hbt]
\tablewidth{0pt}
\tablecaption{Parameter specification for each run, results of instability growth rate and energy partition\label{table:energy partition}}
\tablehead{
\colhead{Run ID} & $r_L/\Delta x$ & \colhead{$\eta$} & \colhead{$t_{LC}/t_{\rm cool}$} & \colhead{$\omega_i$} &
\colhead{$\Delta\mathcal{E}_B/\mathcal{E}_B(0)$} & \colhead{$\mathcal{E}_E/\mathcal{E}_B(0)$} & 
\colhead{$\Delta\mathcal{E}_k/\mathcal{E}_B(0)$} & \colhead{$\mathcal{E}_{\rm rad}/\mathcal{E}_B(0)$} & 
\colhead{$\mathcal{E}_{\rm rad}(\omega>10\omega_0)/\mathcal{E}_B(0)$}
}
\startdata
1	& 2.56		& $1.10\times10^{-8}$	&$8.79\times10^{-6}$	& 2.173		&0.224	&0.014	&0.210	&$1.80\times10^{-4}$	& $7.63\times10^{-5}$\\
1-hr	& 5.12		& $1.10\times10^{-8}$	&$8.79\times10^{-6}$	& 2.179		&0.230	&0.017	&0.212	&$1.61\times10^{-4}$	& $7.57\times10^{-5}$\\
1-lr	& 1.28		& $1.10\times10^{-8}$	&$8.79\times10^{-6}$	& 2.144		&0.223	&0.011	&0.213	&$1.58\times10^{-4}$	& $7.12\times10^{-5}$\\
2	& 2.56		& $1.10\times10^{-6}$	&$8.79\times10^{-4}$	& 2.175		&0.220	&0.012	&0.193	&0.0150	& 0.0066\\
3	& 2.56		& $4.40\times10^{-6}$	&$3.52\times10^{-3}$	& 2.178		&0.230	&0.018	&0.165	&0.0468	& 0.015\\
4	& 2.56		& $2.75\times10^{-5}$	&$2.20\times10^{-2}$	& 2.195		&0.228	&0.025	&0.0295	&0.173	& 0.033\\
5	& 2.56		& $1.10\times10^{-4}$	&$8.79\times10^{-2}$	& 2.235		&0.232	&0.0355	&-0.115	&0.312	& 0.024
\enddata
\tablecomments{The columns from left to right are: run id, the grid resolution $r_L/\Delta x$, the parameter $\eta=1/\omega_gt_{\rm cool}$, the ratio between the light crossing time scale $t_{LC}$ and the cooling time scale $t_{\rm cool}$, the growth rate of the electric field during the linear evolution stage where $E\propto e^{\omega_itc/L}$, energy partition measured at $t=6.21L/c$---including the change of magnetic energy $\Delta\mathcal{E}_B/\mathcal{E}_B(0)$, electric energy $\mathcal{E}_E/\mathcal{E}_B(0)$, the change of particle kinetic energy $\Delta\mathcal{E}_k/\mathcal{E}_B(0)$, total radiated energy $\mathcal{E}_{\rm rad}/\mathcal{E}_B(0)$, and the efficiency of high energy radiation $\mathcal{E}_{\rm rad}(\omega>10\omega_0)/\mathcal{E}_B(0)$, where $\mathcal{E}_B(0)$ is the initial magnetic energy contained in the simulation domain, and $\omega_0=\eta\omega_{syn,lim} (\gamma_{RMS}/\gamma_0)^2 (B_{RMS}/B_0) \approx 22\eta\omega_{syn,lim}$ is the peak synchrotron frequency of the initial distribution.}
\end{deluxetable*}

We consider the lowest-order unstable modes of linear force-free equilibria in a 2D periodic box with size $L\times L$. One particular example is described by
\begin{align}
B_x(x,y)&=\sqrt{2}B_0\sin(k x)\cos(k y),\\
B_y(x,y)&=-\sqrt{2}B_0\cos(k x)\sin(k y),\\
B_z(x,y)&=-2B_0\sin(k x)\sin(k y),
\end{align} 
where $k=2\pi/L$ and the field satisfies
\begin{equation}
\nabla\times\mathbf{B}=-\sqrt{2}k\mathbf{B}.
\end{equation}
The topology of the equilibrium can be readily seen as two pairs of flux tubes on which the magnetic field lines wind helically with opposite orientation but the same sense of helicity (Fig. \ref{fig:B0}). The magnetic field magnitude is not uniform in the box: its peak is $2B_0$ and the root-mean-square value is $\langle B\rangle_{\rm RMS}=\sqrt{2}B_0$.

In the equilibrium $\mathbf{E}=\mathbf{0}$ and the distribution function $F_s$ for each species $s$ should satisfy steady state Maxwell-Vlasov equation (neglecting radiation reaction for the moment):
\begin{gather}
\frac{\mathbf{u}}{\gamma}\cdot\nabla_{\mathbf{x}} F_s+\frac{q_s\mathbf{u}\times\mathbf{B}}{m_sc\gamma}\cdot\nabla_{\mathbf{u}}F_s=0,\label{eq:Vlasov1}\\
\rho=\sum_s q_s\int F_s d^3u=0,\label{eq:Gauss1}\\
\mathbf{J}=\sum_s q_s\int F_s\frac{\mathbf{u}}{\gamma}d^3u=\frac{c}{4\pi}\nabla\times\mathbf{B}.\label{eq:Ampere1}
\end{gather}
In practice, we follow the approach of \citet{Nalewajko:2016aa} and approximate the particle distribution function as
\begin{equation}
F_s(\mathbf{u},\mathbf{x})d^3ud^3x=n_0\frac{f(\gamma)g(\mu)}{4\pi}d\gamma d\mu d\phi d^3x
\end{equation}
where $\mu$ is the cosine of particle pitch angle with respect to local magnetic field, such that equations (\ref{eq:Gauss1})(\ref{eq:Ampere1}) and the first two moments of equation (\ref{eq:Vlasov1}) are satisfied. We assume that the particle distribution is uniform in space; in energy space $f(\gamma)$ satisfies the Maxwell-J\"{u}ttner distribution
\begin{equation}
f(\gamma)=\frac{\gamma u}{\gamma_0 K_2(\gamma_0)}e^{-\gamma/\gamma_0}.
\end{equation}
where $\gamma_0$ is related to the temperature as $\gamma_0= k_B T/m c^2$. Since we are working in the ultrarelativistic regime, $\gamma_0\gg1$ so the distribution function becomes simply
\begin{equation}
f(\gamma)=\frac{\gamma^2}{2\gamma_0^3}e^{-\gamma/\gamma_0}
\end{equation}
which is free from the rest mass scale. The angular distribution is assumed to be
\begin{equation}
g(\mu)=1+a(\mathbf{x})\mu
\end{equation}
which guarantees that the pressure is isotropic and uniform, while the average drift velocity is
\begin{equation}
\langle v_d/c\rangle=\frac{a}{3}\int\frac{u f(\gamma)}{\gamma}d\gamma=\frac{a}{3}
\end{equation}
Assuming that electrons and positrons have equal number density and drift in opposite directions, we get from Equation (\ref{eq:Ampere1})
\begin{equation}\label{eq:ne}
n_e=n_p=\frac{\sqrt{2}kBc}{8\pi e\langle v_d\rangle}=\frac{3\sqrt{2}B}{4Lea}
\end{equation}
We set the initial number density using the nominal values $B_0$ and $a_0$ ($|a_0|\le0.5$ to make sure $F_s>0$):
\begin{equation}
 n_0=\frac{3\sqrt{2}B_0}{4Lea_0}.
\end{equation}
With this assumption, the nominal magnetization parameter as defined before becomes
\begin{equation}\label{eq:nominal-sigma}
 \sigma\equiv\frac{B_0^2}{4\pi n_0\gamma_0 m c^2}=\frac{a_0L}{6\pi \sqrt{2}r_L}
\end{equation}
where $r_L=\gamma_0 m c^2/eB_0$ is the characteristic gyro radius of a thermal particle. We note that the ratio between the skin depth $d_e=c/\omega_p=c/\sqrt{4\pi ne^2/\gamma_0 m}$ and $r_L$ is
\begin{equation}
 \frac{d_e}{r_L}=\sqrt{\frac{B_0^2}{4\pi\gamma_0 n_0 m c^2}}=\sqrt{\sigma}.
\end{equation}
Therefore, the computational grid should resolve $r_L$ or electron skin depth, depending on the value of $\sigma$.

Given the functional form of $F_s$, the evolution of the system is now fully determined by the following dimensionless parameters: $\eta$, $U_{ph}/U_B$ and two of the other three parameters: $L/r_L$, $\sigma$ and $a_0$. Among the latter we choose $\sigma$ and $a_0$: $\sigma$ indicates the initial partition between magnetic energy and particle kinetic energy, while $a_0$ is a parameter characterizing the charge multiplicity, or equivalently, level of opposite drifting between electrons and positrons: small $a_0$ implies large multiplicity/small relative drift. We have performed a series of simulations with different values for these dimensionless parameters, to explore the underlying scaling relations. In this paper, we focus on one particular choice of $\sigma$ and $a_0$, consider only synchrotron radiation, and explore the dynamical effects of different $\eta$ as well as the radiation signatures. Specifically, we set $L/r_L=800$, $\sigma=7.52$, $a_0=0.25$---the corresponding box-averaged, warm magnetization is $\bar{\sigma}_w=\langle B^2/(4\pi w)\rangle=\sigma/2=3.76$, and $\bar{\beta}_{\rm plasma}\equiv8\pi \langle P\rangle/\langle B^2\rangle=0.13$, where $w=4P$ is the enthalpy density of the plasma. Specification of $\eta$ for each run is listed in Table \ref{table:energy partition}. We use a grid resolution $\Delta x=r_L/2.56$, correspondingly $d_e/\Delta x\approx7$. We've tested the convergence using both lower and higher resolution grids with $\Delta x=r_L/1.28, r_L/5.12$ and we find that our main results do not depend on the resolution. The number of particles per cell for each species is 144 \citep[slightly larger than][]{Cerutti:2013aa,Nalewajko:2016aa}. The time step is set to $\Delta t=0.9\Delta x/\sqrt{2}c$ to satisfy the CFL stability condition \citep{Courant:1967aa}. During the evolution, the instability as found by \citet{East:2015aa} does grow spontaneously from the initial conditions we use, but to reduce the computational cost in most of the simulations we introduce a small amplitude, long wavelength perturbation of the form $\mathbf{\delta B}=0.01B_0(\sin y, \sin x, -\cos x+\cos y)$ when $t=L/c$, and turn on radiation reaction at the same time.

 \begin{figure*}
  \centering
  \subfigure[]
    {
        \includegraphics[width=0.45\textwidth]{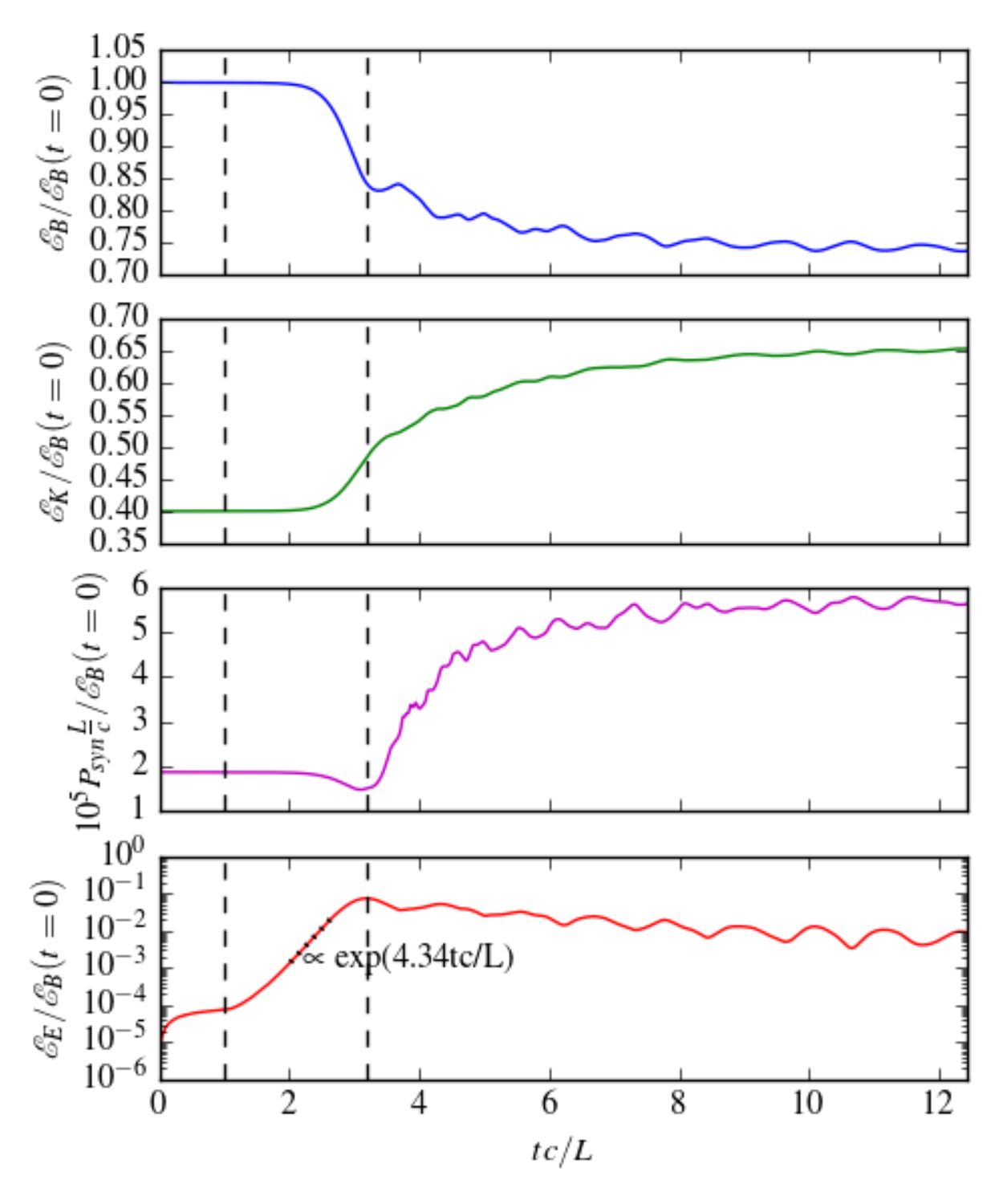}
    }
    \subfigure[]
    {
        \includegraphics[width=0.45\textwidth]{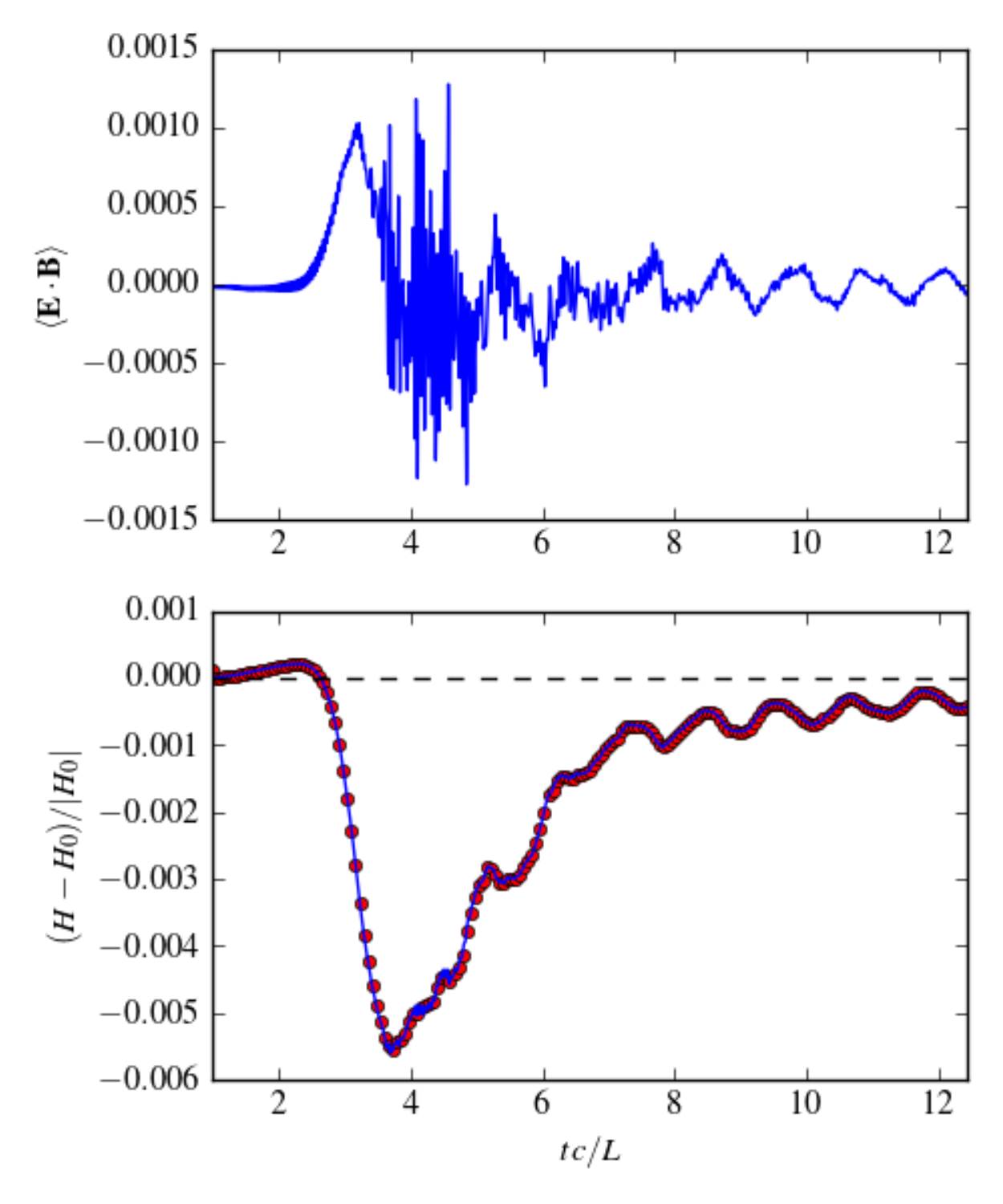}
    }
  \caption{From run 1: (a) Evolution of magnetic energy, particle kinetic energy, emitted synchrotron power and electric energy as a function of time. The first vertical dashed line corresponds to the time point $t=L/c$ when the perturbation sets in; the second vertical dashed line is the saturation point $t=3.19L/c$. In the lowest panel, the black dotted line indicates the measured linear growth rate. (b) The upper panel shows the evolution of $\mathbf{E}\cdot\mathbf{B}$, and the lower panel compares the measured helicity change (red dots) with the theoretical value $-2\int\!\int \mathbf{E}\cdot\mathbf{B}\, dV dt$ (blue line).}\label{fig:energy_helicity}
\end{figure*}

 \begin{figure*}
  \centering
        \includegraphics[width=\textwidth]{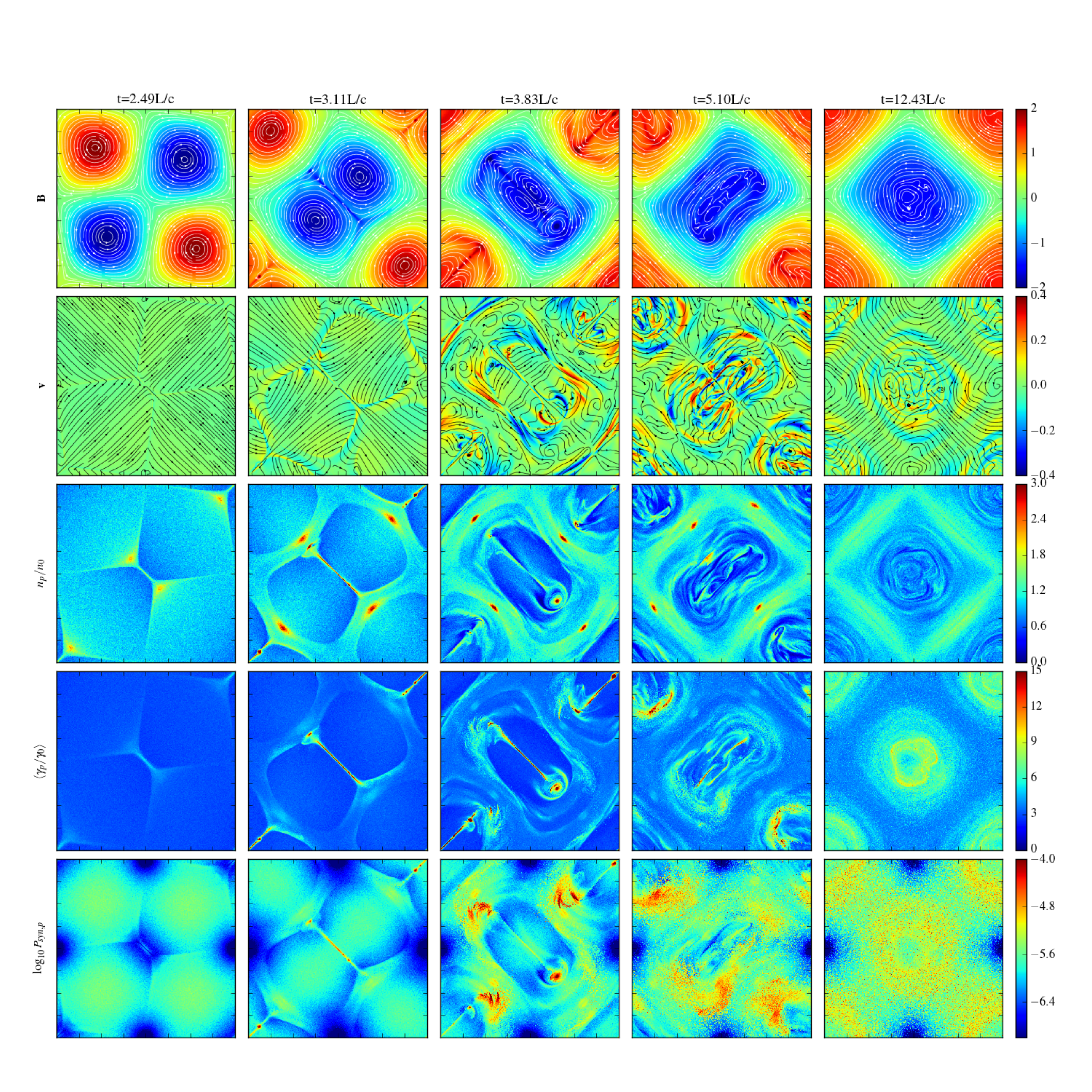}
     \caption{Snapshots from run 1. From top to bottom: magnetic field, center of rest mass velocity, positron number density, positron average Lorentz factor and positron synchrotron power map. For both the magnetic field and the velocity, streamlines indicate the in-plane field and color represents the z component.}\label{fig:Bvng}
\end{figure*}

\begin{figure*}
  \centering
        \includegraphics[width=\textwidth]{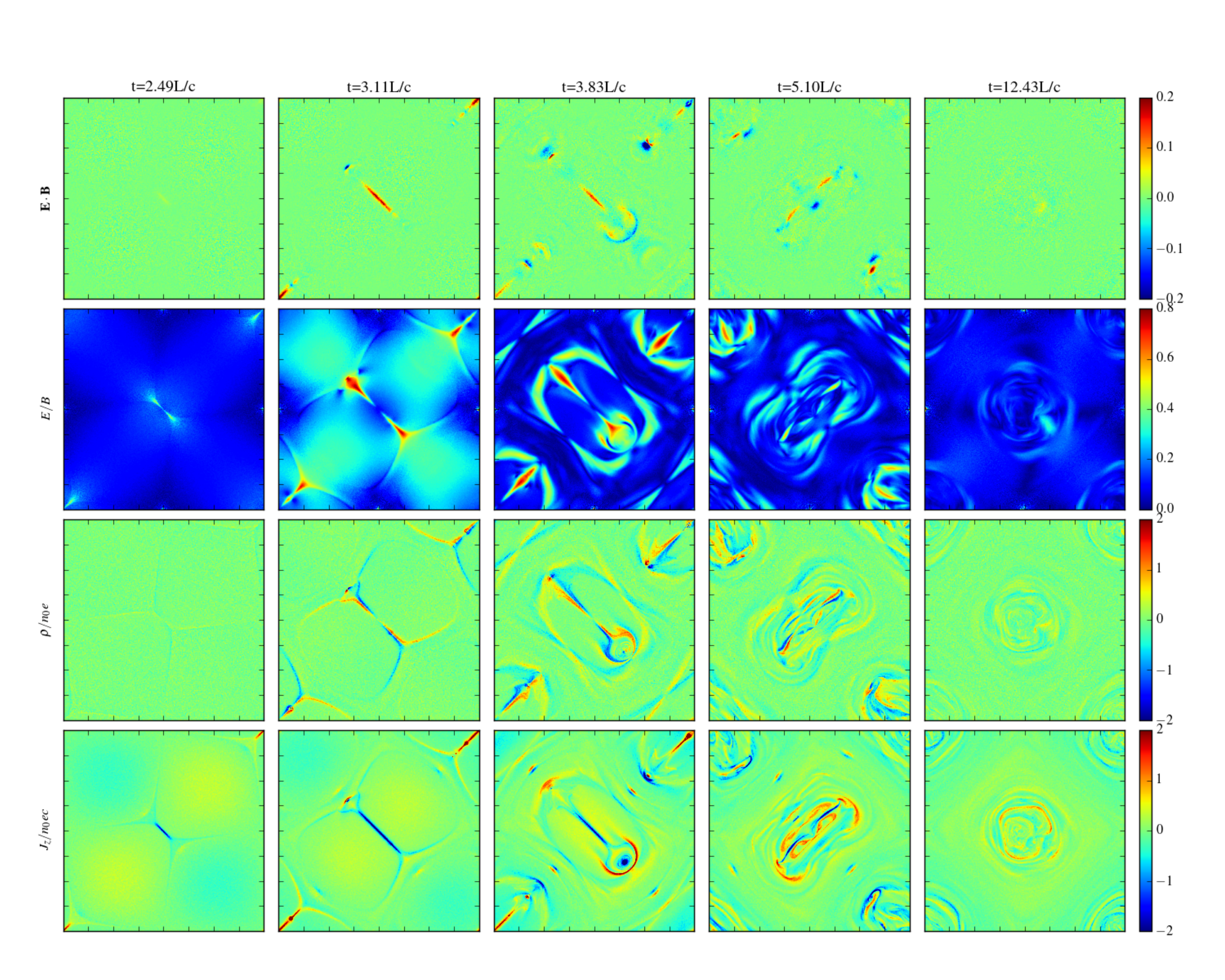}
     \caption{Snapshots from run 1 (continued). From top to bottom: $\mathbf{E}\cdot\mathbf{B}$, $E/B$, charge density and current density.}\label{fig:EBrJ}
\end{figure*}

\section{Results of the simulations}\label{sec:results}
In this section, we first consider the case where the radiation reaction is not yet dynamically important (run 1). We summarize briefly the evolution of the instability in \S\ref{subsec:evolution}, then discuss in \S\ref{subsec:radiation} the radiative signatures of the instability, including variability, spectra and angular distribution, as well as polarization. We analyze the dynamical consequences of radiation reaction systematically in \S\ref{subsec:dynamical}.

\subsection{Evolution of the instability--mildly radiative cases}\label{subsec:evolution}
The overall evolution of the field configuration is qualitatively similar for the range of parameters we explored, and also resembles the mildly relativistic cases studied by \citet{Nalewajko:2016aa}. Take Run 1 as an example. Figure \ref{fig:energy_helicity}(a) shows the evolution of different energy components. At the start of the simulation, the random noise in charge density due to finite number of particles cause the electric energy to grow to a plateau $\sim10^{-4}$ times the initial total magnetic energy $\mathcal{E}_B$. The initial small fluctuations roughly settle down after one light crossing time scale, and the instability grows from the perturbation. 

During the linear growth of the instability, the two pairs of flux tubes with the same sign of $B_z$ start to merge; each pair produces a layer in between with enhanced current and number density (Figure \ref{fig:Bvng}, first column). The electric field grows exponentially as $E\propto e^{\omega_ict/L}$, with $\omega_i=2.17$ in this particular case. Non-ideal regions with $\mathbf{E}\cdot\mathbf{B}\ne0$ start to be produced in the current layer, while $E$ remains less than $B$ due to the advection and compression of $z$ magnetic flux into the current layers. The exhausts of the current layers are the locations where $E/B$ becomes maximal (Figure \ref{fig:EBrJ}). Some charge density appears surrounding the flux tubes, because the in-plane motion of the flux tubes gives rise to a Lorentz force that moves positive and negative charges in opposite directions along the $x-y$ field lines, accumulating different signs of charges at different sections of the flux tube boundaries. The boundaries separating the flux tubes where relative motion occurs turn out to be tangential discontinuities.

The current layer continues to get thinner and longer, as plasmas are pushed into the layer from both sides and ejected along the exhausts. Eventually the thickness of the current layer is determined by the skin depth of the plasma in the sheet: $\lambda=\sqrt{(P_{xx}+P_{yy})/(8\pi n^2e^2)}$. When the aspect ratio of the current layer becomes large enough, plasmoids---namely, small islands with local concentration of current in this 2D case---start to form and grow in the current layer (Figures \ref{fig:Bvng} and \ref{fig:EBrJ}, second and third columns). They, too, get ejected from the ends of the current layer, and when they collide with the ambient magnetic field, secondary current layers are produced, as well as fast waves that propagate into the neighboring magnetic domain.

The initial current sheet only lasts for about one dynamical time scale. It gets destroyed as the overall field structure continues on a large amplitude damped oscillation, with part of the energy going back and forth between magnetic form and electric plus kinetic form. During the oscillation, transient current layers with $\mathbf{E}\cdot\mathbf{B}\ne0$ are still produced. In the end, the system roughly settles into the longest wavelength equilibrium, with a relatively dense, cool plasma near the magnetic separatrices and a hot, dilute plasma near the center of the magnetic domain (Figure \ref{fig:Bvng}, last column).

During the whole evolution, deviation from helicity conservation is less than 0.6\% (Figure \ref{fig:energy_helicity}b).  At the end, $\sim25\%$ of the magnetic energy has been dissipated. As a comparison, theoretically the initial state has an energy $\sqrt{2}$ times that of the true ground state (the longest wavelength solution allowed in the periodic box), so the maximum amount of magnetic free energy is $1-1/\sqrt{2}=0.293$. The actual released magnetic energy is smaller because in 2D there are additional topological constraints making the final state different from the true ground state \citep{Zrake:2016aa}.

\begin{figure*}
  \centering
  \subfigure[]
  {
        \includegraphics[width=0.27\textwidth]{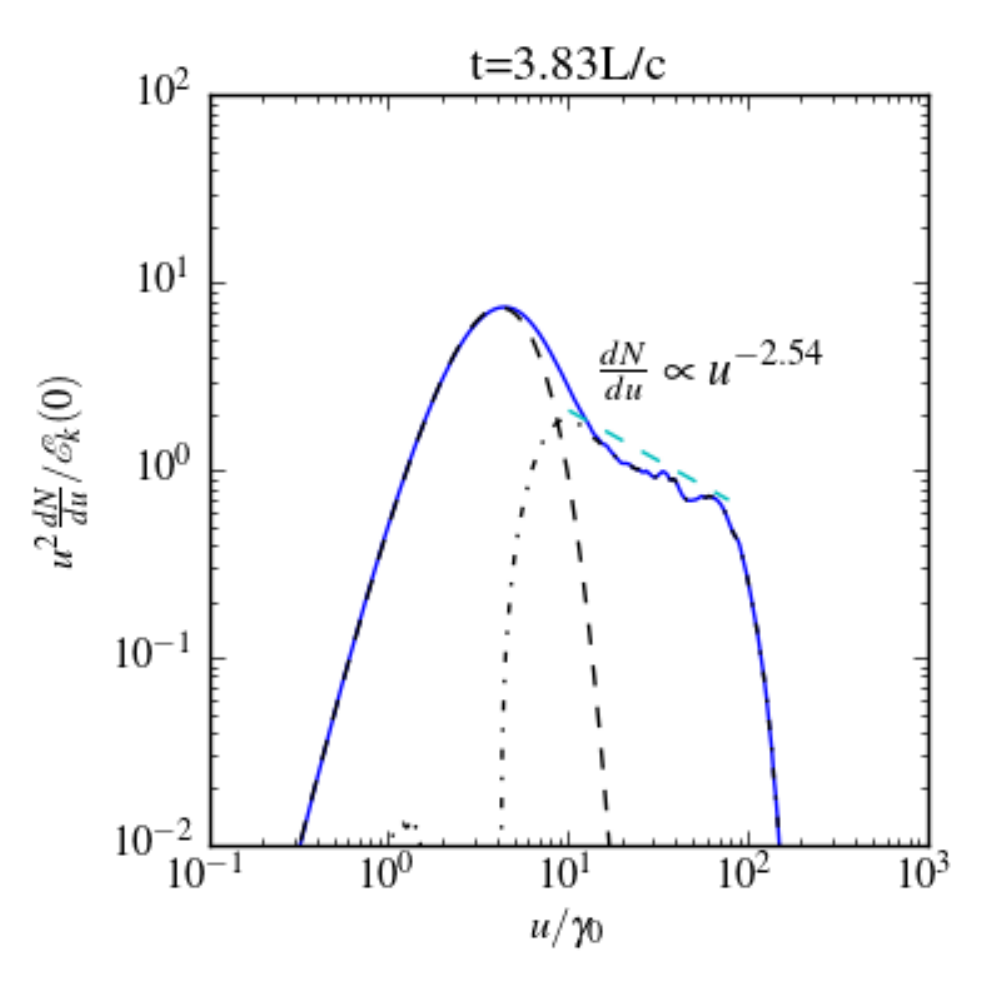}
  }
   \subfigure[]
  {
        \includegraphics[width=0.33\textwidth]{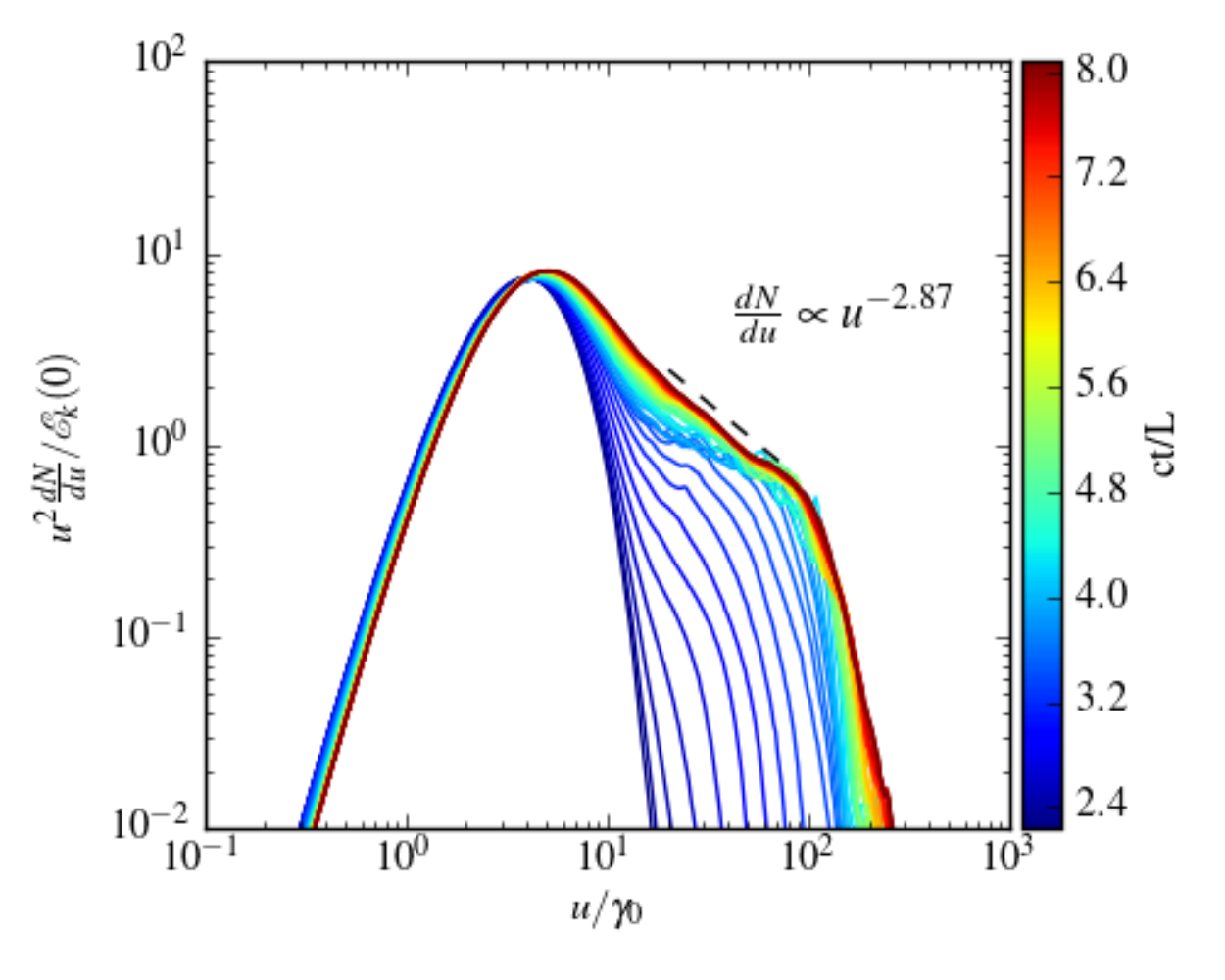}
  }
 \subfigure[]
  {
        \includegraphics[width=0.33\textwidth]{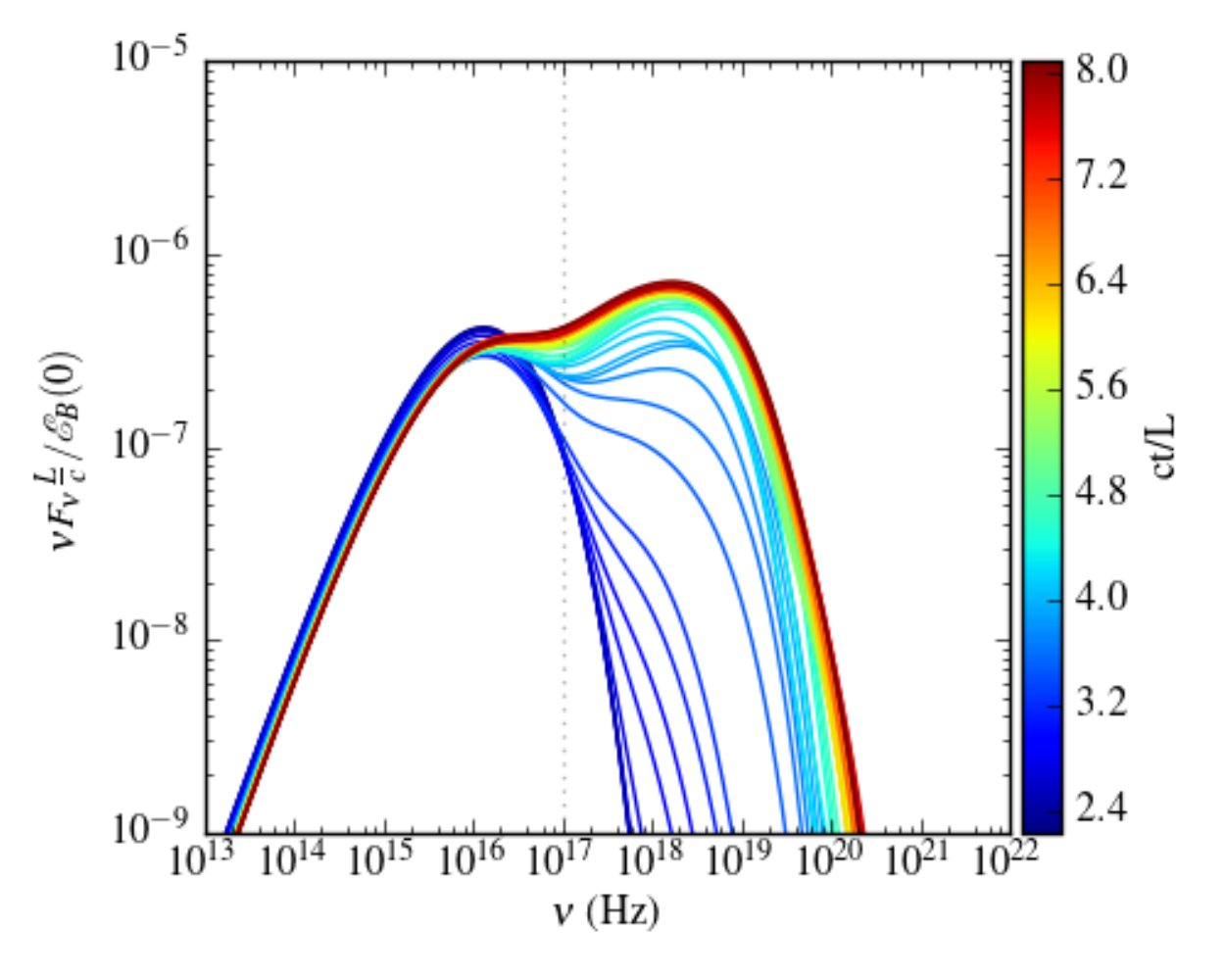}
  }
     \caption{From run 1: \newtext{(a) Particle spectrum, integrated over the whole box and over all solid angles, at the time point $t=3.83L/c$ when the main current layer is in the plasmoid dominated reconnection stage. The blue line is the total spectrum; the black dashed line is the thermal Maxwellian; the black dash-dotted line is the difference between the two. The nonthermal high energy tail has a roughly power-law like distribution spanning about one decade in energy, with a spectral index 2.54.} (b) Isotropic particle spectra (similar to (a)) at different times during the evolution. Colors from blue to red correspond to equally spaced time points from $t=2.24L/c$ to $t=8.08L/c$. At the end of the simulation non-thermal particles comprise 12\% in number and 36\% in energy. (c) Instantaneous radiated spectrum, integrated over all particles and all solid angles. Note that this is different from the \emph{observed} spectrum. The vertical dotted line indicates the separation between the low energy band and high energy band we refer to when calculating light curves and power spectra.}\label{fig:particle spectrum}
\end{figure*}

\begin{figure*}
  \centering
        \includegraphics[width=\textwidth]{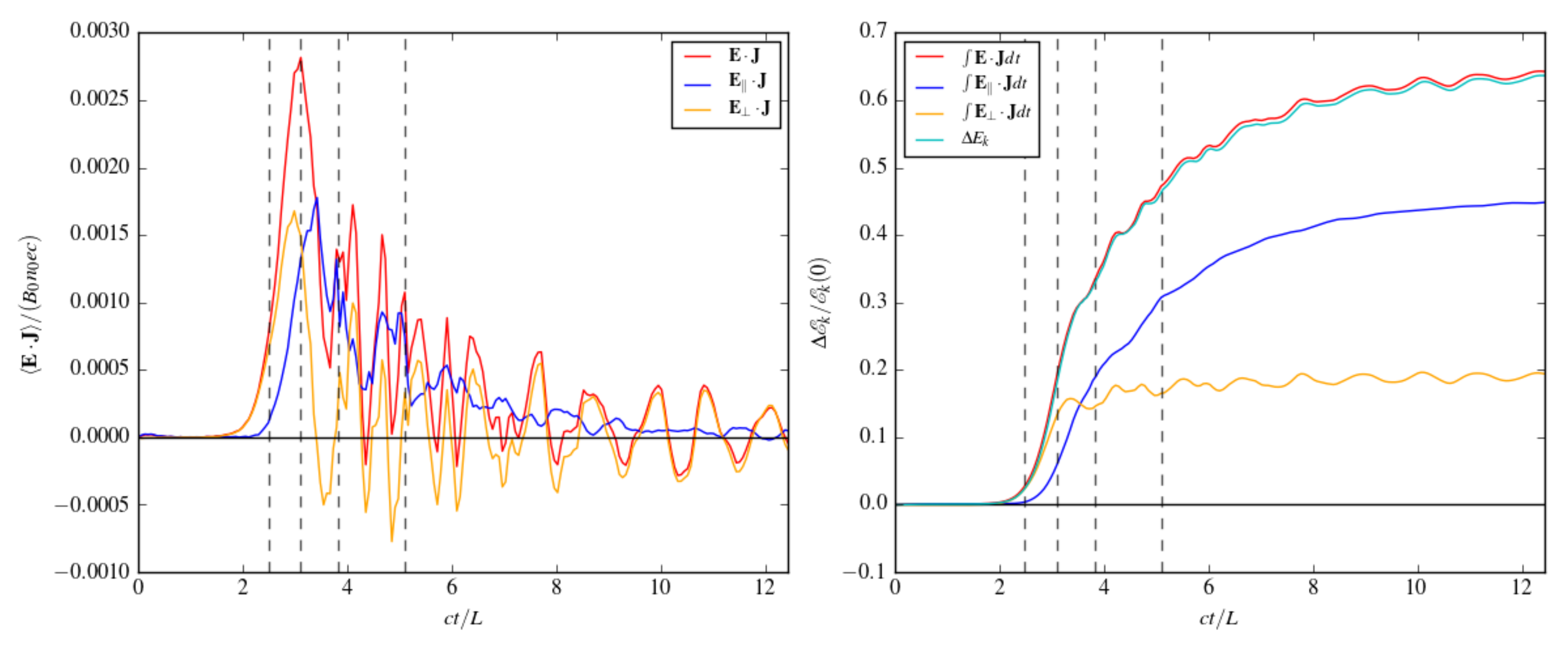}
     \caption{Left panel: the rate of energy transfer from electromagnetic field to particles as a function of time, for run 1. The red line is the total power $\mathbf{E}\cdot\mathbf{J}$, the blue/magenta lines correspond to that contributed by the component of the electric field parallel/perpendicular to the magnetic field. Right panel: corresponding work done by the electromagnetic field on the particles as a function of time, compared with the particle kinetic energy change. All are scaled to the initial total kinetic energy. It can be seen that parallel electric field acceleration is the dominant dissipation mechanism here.}\label{fig:EJ}
\end{figure*}

\begin{figure*}
  \centering
  \subfigure[]
    {
        \includegraphics[width=\textwidth]{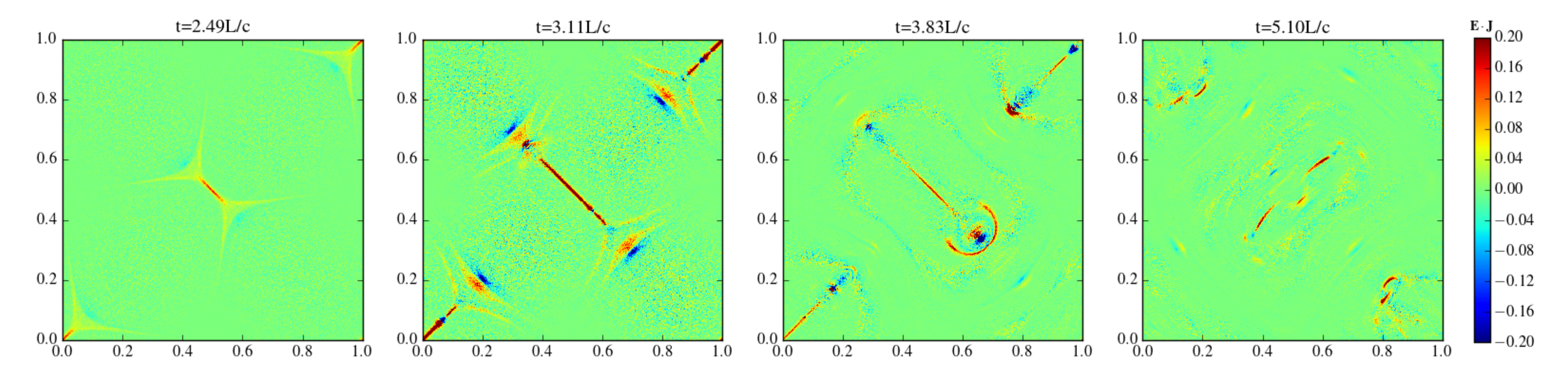}
    }\\
    \subfigure[]
    {
        \includegraphics[width=\textwidth]{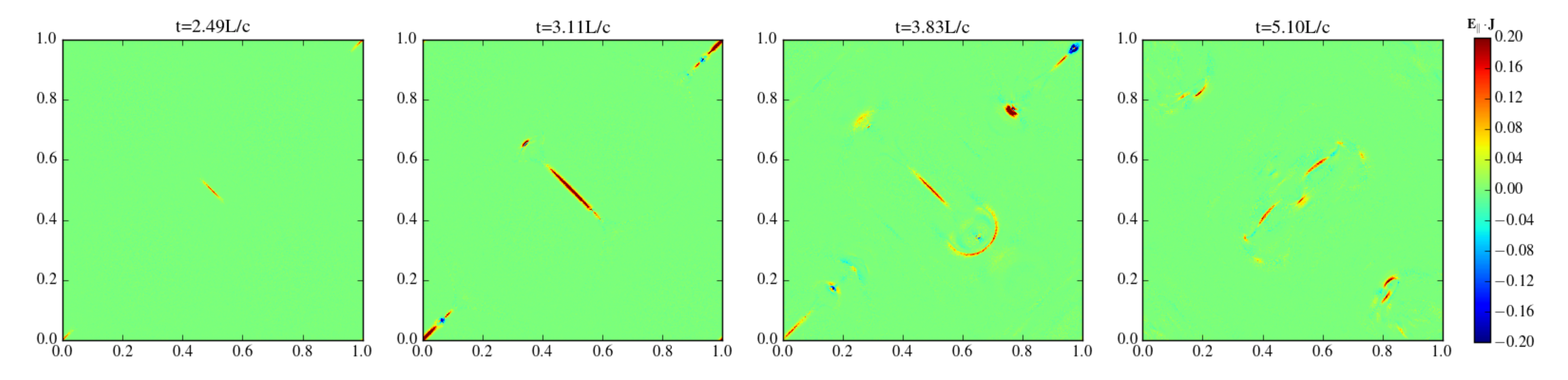}
    }
  \caption{(a) Maps of $\mathbf{E}\cdot\mathbf{J}$ for run 1, at four time points indicated by dashed lines in Figure \ref{fig:EJ}. (b) Maps of $\mathbf{E_{\parallel}}\cdot\mathbf{J}$ at the same time points.}\label{fig:EJmap}
\end{figure*}

\begin{figure*}
  \centering
         \includegraphics[width=\textwidth]{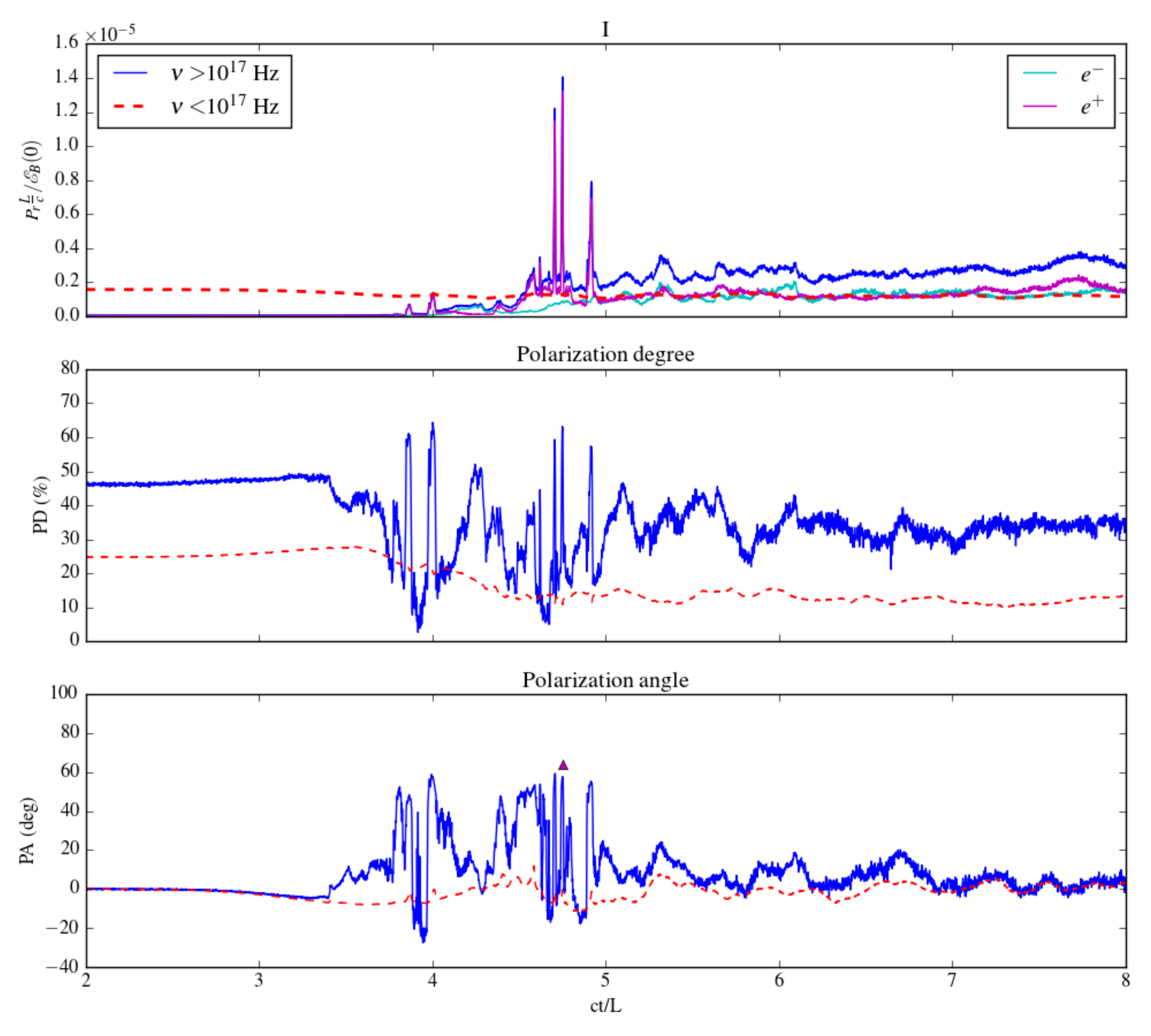}
  \caption{From run 1: light curve, polarization degree and polarization angle as a function of time seen by an observer located at $+x$. The blue line is for the high energy band ($\nu>10^{17}$ Hz) and the red line is for the low energy band ($\nu<10^{17}$ Hz), both summed over electrons and positrons. In the top panel, we also show the high energy contribution from electrons and positrons separately. The polarization angle is measured from the $x-y$ plane (same blow). \newtext{The magenta triangle in the bottom panel indicates the polarization angle of the emission from the bunch of high energy particles tracked in Figure \ref{fig:particle history}.}}\label{fig:polarization_x}
\end{figure*}

\begin{figure*}
  \centering
         \includegraphics[width=\textwidth]{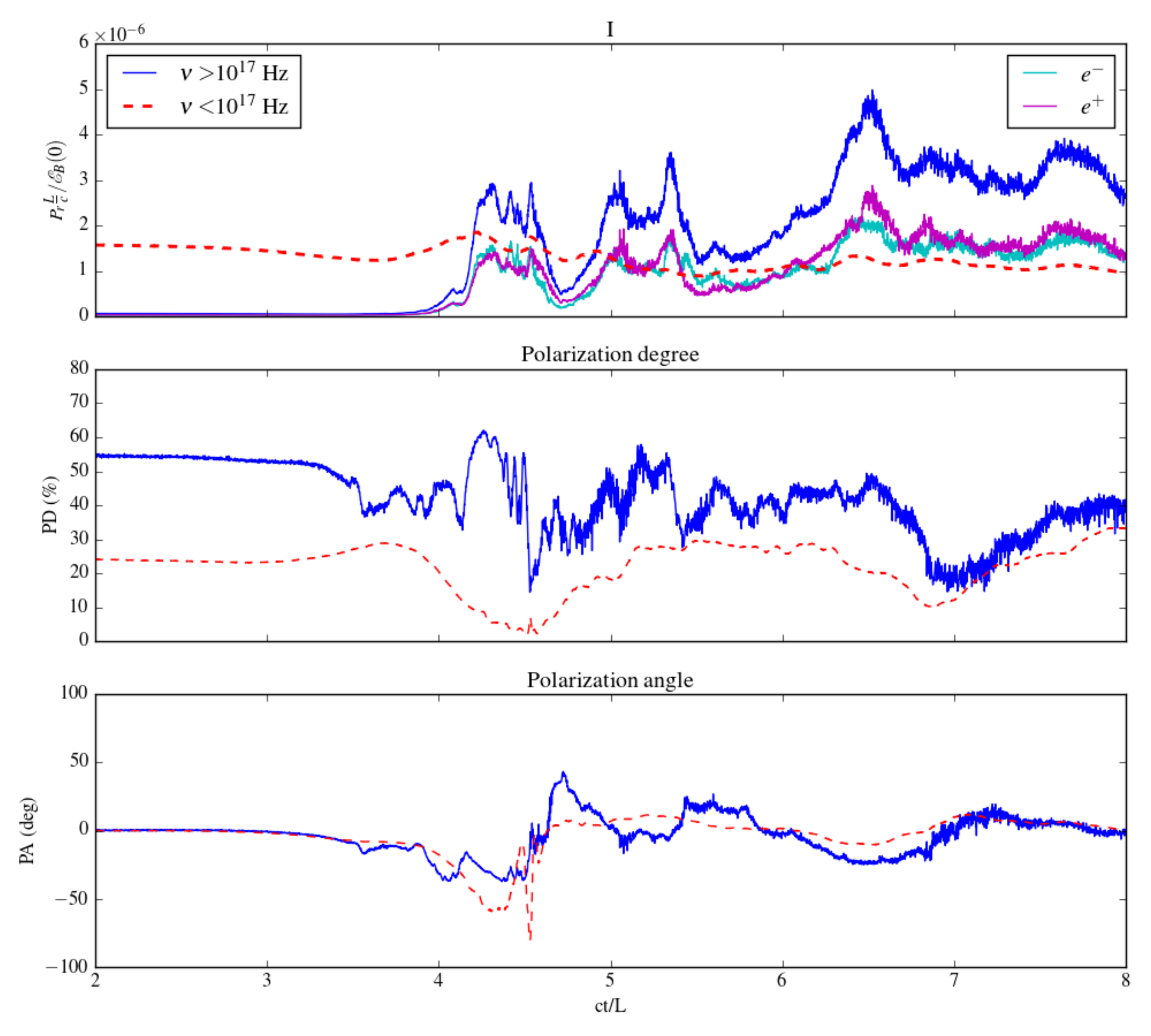}
  \caption{From run 1: light curve, polarization degree and polarization angle as a function of time seen by an observer located at $45^{\circ}$ from $+x$. (The sharp change in polarization signals at $t\approx4.6L/c$ has something to do with emitting features crossing the box boundary and should be ignored.)}\label{fig:polarization_L45}
\end{figure*}
% Features crossing over the box boundary do leave imprint on the polarization degree and polarization angle...

\begin{figure*}
  \centering
         \includegraphics[width=\textwidth]{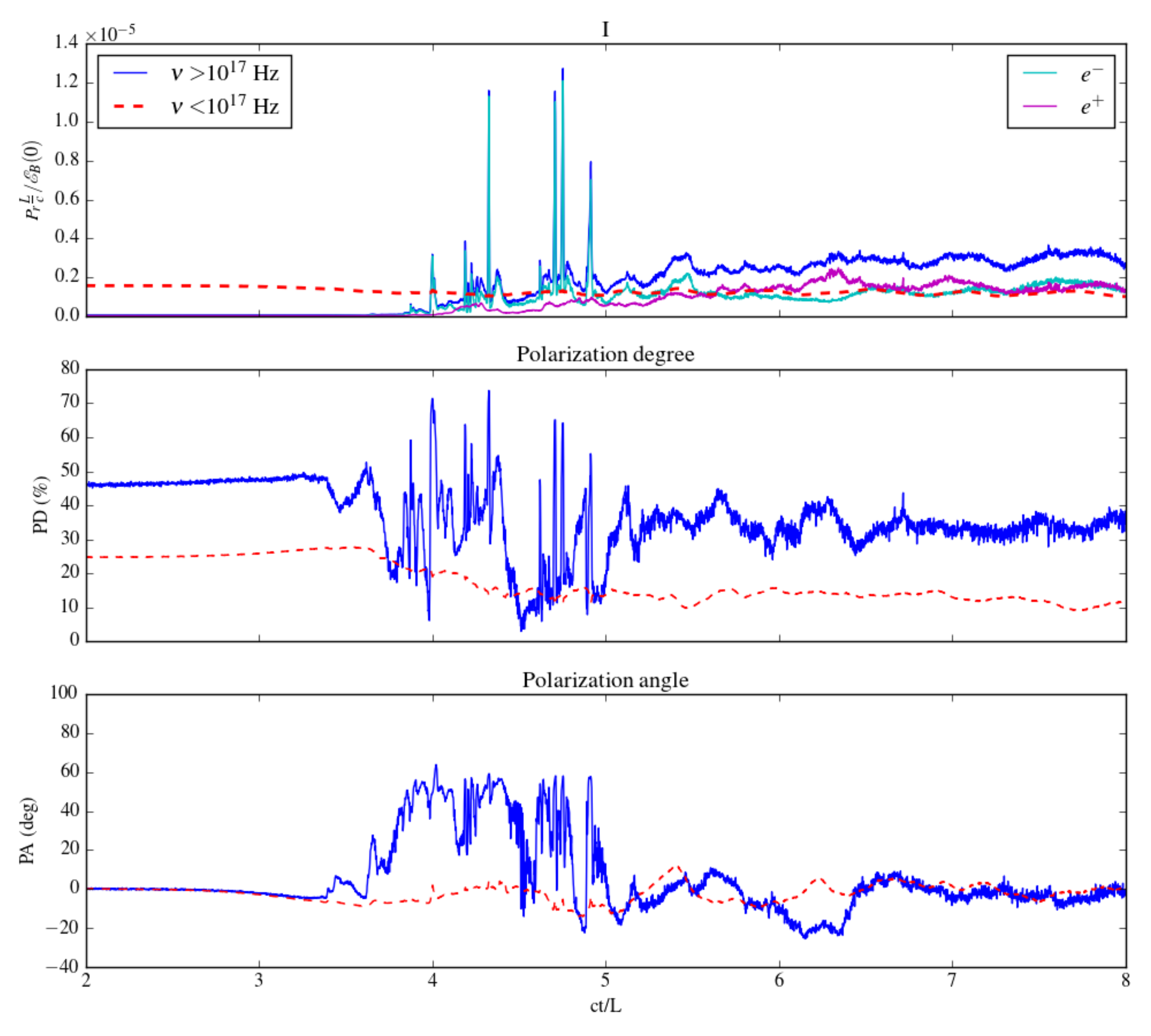}
  \caption{From run 1: light curve, polarization degree and polarization angle as a function of time seen by an observer located at $+y$.}\label{fig:polarization_y}
\end{figure*}

\subsection{Mechanisms of particle acceleration and dissipation of electromagnetic energy}\label{subsec:mechanism}
The isotropic particle spectra at a series of simulation times are shown in Figure \ref{fig:particle spectrum}. Consistent with \citet{Nalewajko:2016aa}, we find that high energy particles are first accelerated in the initial current layers by the nonideal electric field $\mathbf{E}_{\parallel}$, forming a bump on the tail of the distribution. \newtext{This appears more clearly after we subtract the thermal Maxwellian component (Figure \ref{fig:particle spectrum}a). The high energy non-thermal tail expands in energy range and the spectrum gets harder as the current sheet continues to stretch. Eventually the tail reaches an extent of about one decade in energy, within which a roughly power-law-like distribution is established. The hardest power-law index $\sim2.5$ is achieved when the main current sheets are in the plasmoid dominated reconnection stage. After the main current layers dissolve, the system enters a turbulent relaxation process, particles diffuse in momentum space, gradually forming a steeper power law connecting the high energy bump with the thermal Maxwellian. There's also an overall heating of the background plasma \citep[see also,][]{Lyutikov:2016aa}. At the end of the simulation, the power law has a spectral index $\sim2.9$, and the nonthermal particles comprise 12\% in number and 36\% in energy. Due to the modest $\sigma$ in our simulations, the particle spectrum is relatively soft. It has been shown by \citet{Nalewajko:2016aa} that the spectrum does get harder as $\sigma$ increases.}

Figure \ref{fig:EJ} shows the rate of energy transfer from the electromagnetic field to the particles, $\mathbf{E}\cdot\mathbf{J}$, as well as that contributed by $\mathbf{E}_{\parallel}$ and $\mathbf{E}_{\perp}$---the components of electric field parallel and perpendicular to the magnetic field, respectively. \newtext{Since the initial magnetic configuration contains null points where $B=0$, at these locations we let $\mathbf{E}_{\parallel}=\mathbf{E}$. However, we find that as the evolution starts, $B_z$ is advected into the current layer, so $B$ does not vanish in the current layer, $E$ is always less than $B$, and $\mathbf{E}_{\parallel}$, $\mathbf{E}_{\perp}$ are well defined.} It can be seen that although $\mathbf{E}_{\perp}\cdot\mathbf{J}$ dominates at earlier times when the ideal instability just starts to develop, most of the dissipation happens upon the saturation of the instability and is dominated by $\mathbf{E}_{\parallel}\cdot\mathbf{J}$. Figure \ref{fig:EJmap} shows maps of $\mathbf{E}\cdot\mathbf{J}$ and $\mathbf{E_{\parallel}}\cdot\mathbf{J}$ at representative time points. We observe that $\mathbf{E_{\parallel}}$ mostly operates at the primary current sheets, and also the secondary current layers formed when the plasmoids collide with neighboring flux tubes.

\subsection{Radiation signatures---mildly radiative case}\label{subsec:radiation}
As a benchmark example, in this section we analyze the radiation signatures of the mildly radiative case, run 1. While the total emitted power as a function of time is quite smooth overall (Figure \ref{fig:energy_helicity}a), the radiation received by a fixed observer can be highly variable, as we now show.

\begin{figure*}
  \centering
  \subfigure[]
    {
        \includegraphics[width=0.45\textwidth]{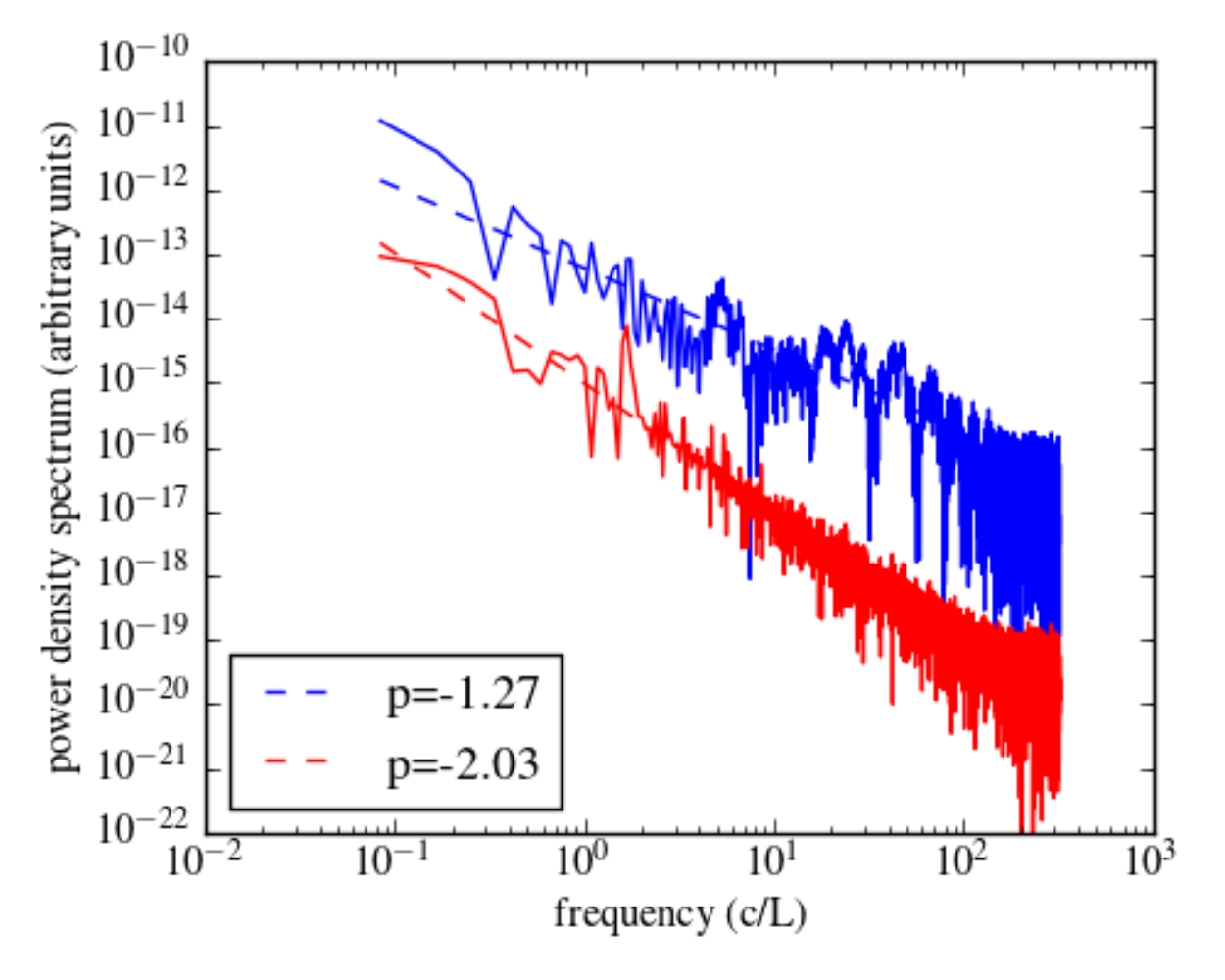}
    }
    \subfigure[]
   {
        \includegraphics[width=0.45\textwidth]{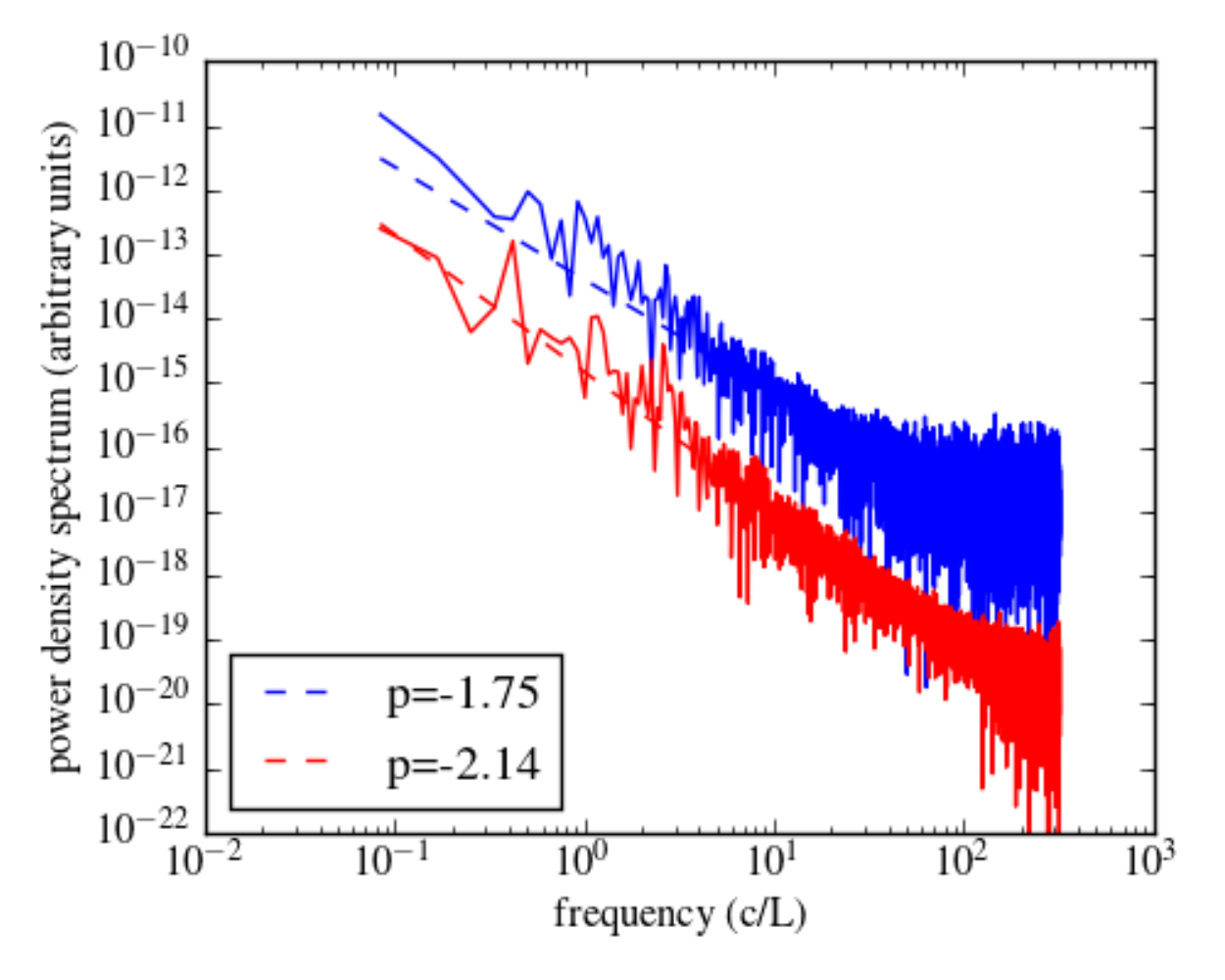}
    }
  \caption{From run 1: power spectra of received radiation in different wavebands for the two different observers corresponding to Figures \ref{fig:polarization_x} and \ref{fig:polarization_L45}, respectively. Blue is the flux in the high energy band above $1.05\times 10^{17}$ Hz; red is the flux in the low energy band below $1.05\times 10^{17}$ Hz. The straight dashed/dotted lines are power laws of the form $f^{p}$, fitted to the power spectrum. The frequency range spans from the inverse of the duration of the simulation, to the resolution we used to calculate the light curves $1/(0.0017L/c)$.}\label{fig:powerspectra}
\end{figure*}

\begin{figure*}
  \centering
         \includegraphics[width=0.6\textwidth]{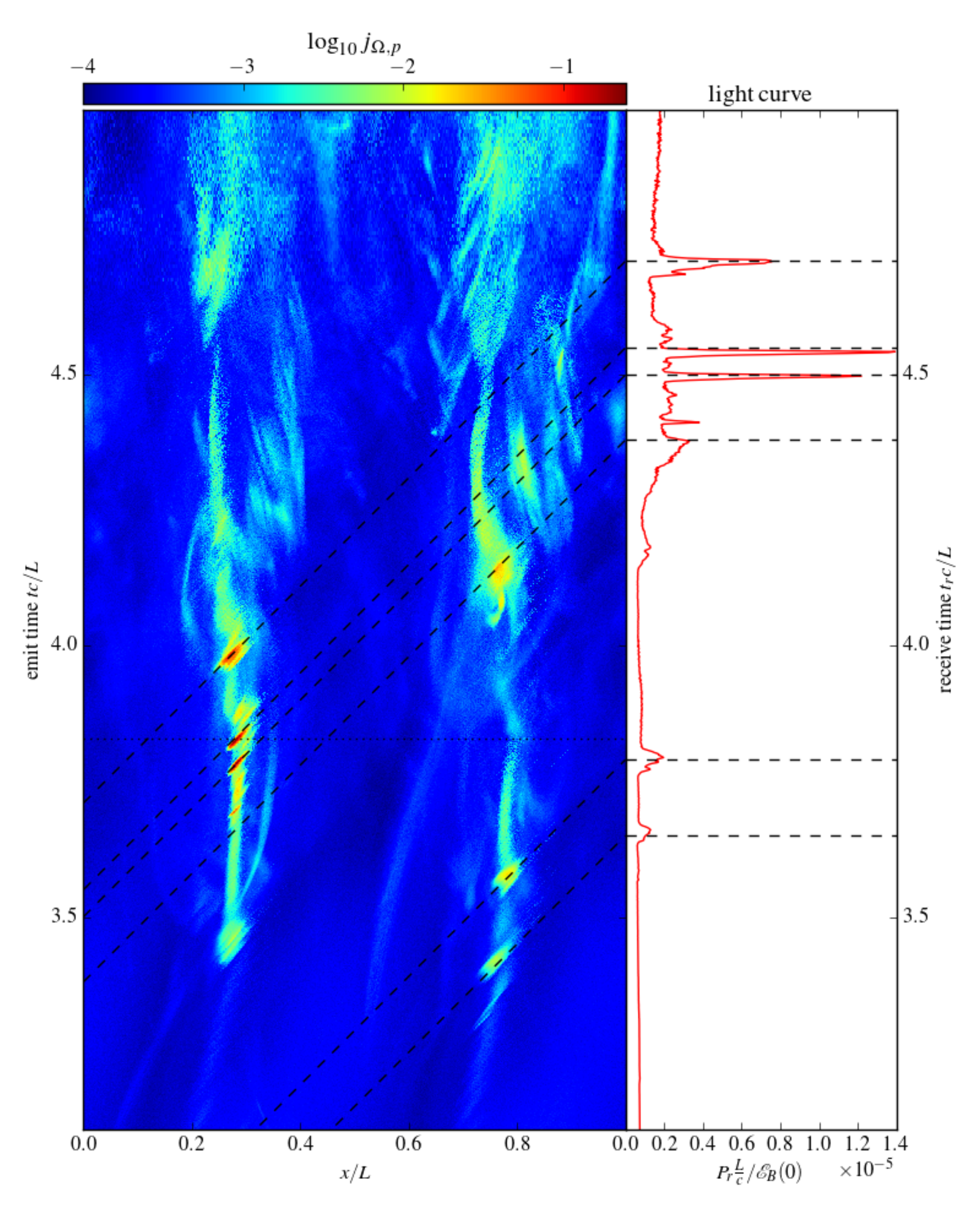}
  \caption{From run 1: the left part is the spacetime diagram of emitted power along $+x$ direction, for positrons only, integrated over y coordinates and frequency (note that the color is on log scale); the right part is the corresponding light curve. The black dashed lines indicate the correspondence between the emissivity and the light curve. The horizontal dotted line in the left panel indicates the emission time $t=3.83L/c$, and we plot in Figure \ref{fig:emissivity_2Dx} the corresponding 2D maps at this time point.}\label{fig:emissivity_xt}
\end{figure*}

\begin{figure}
  \centering
  \subfigure
   {
         \includegraphics[width=0.5\textwidth]{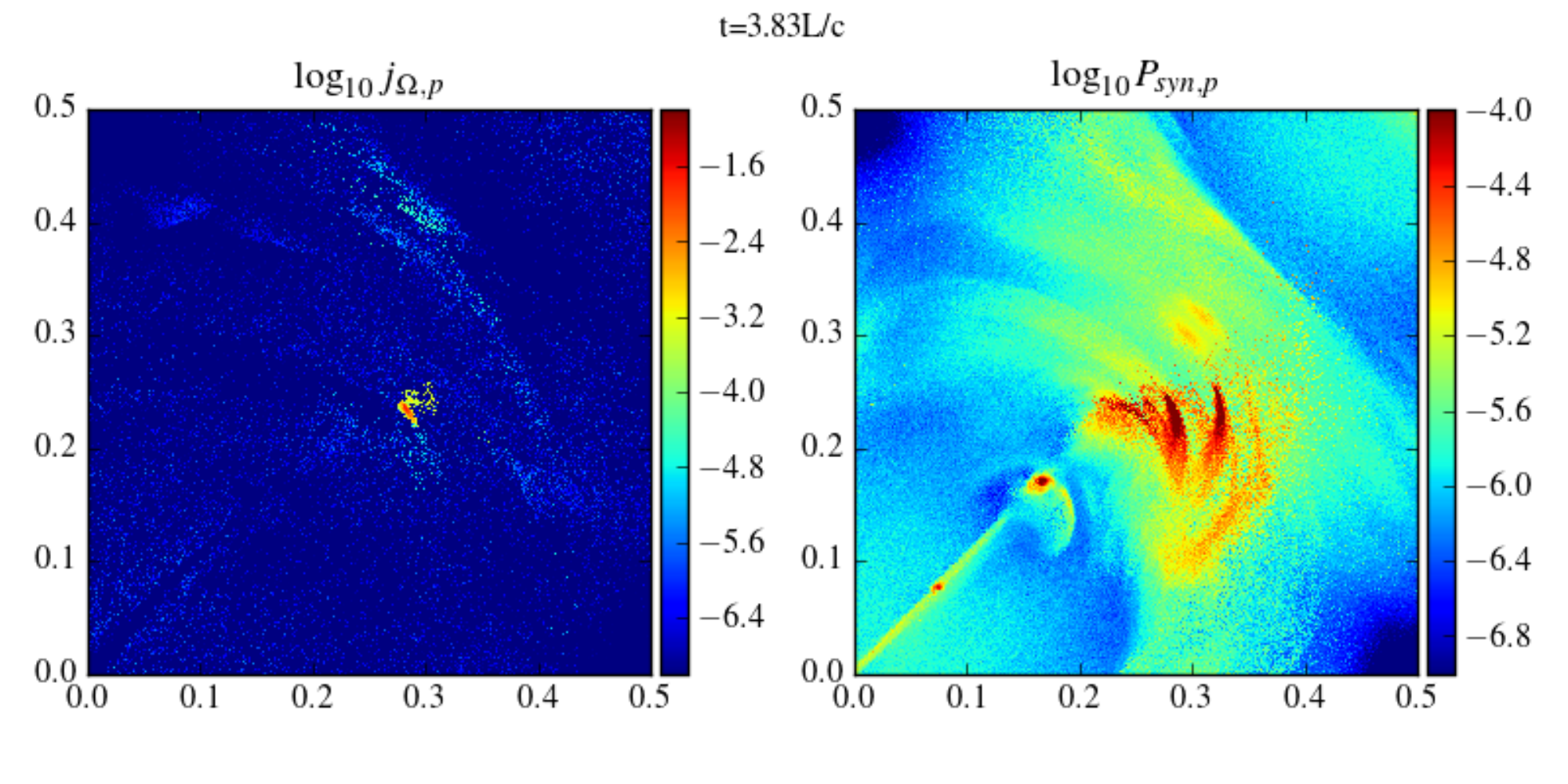}
   }\\
   \vspace{-0.6cm}
    \subfigure
   {
         \includegraphics[width=0.52\textwidth]{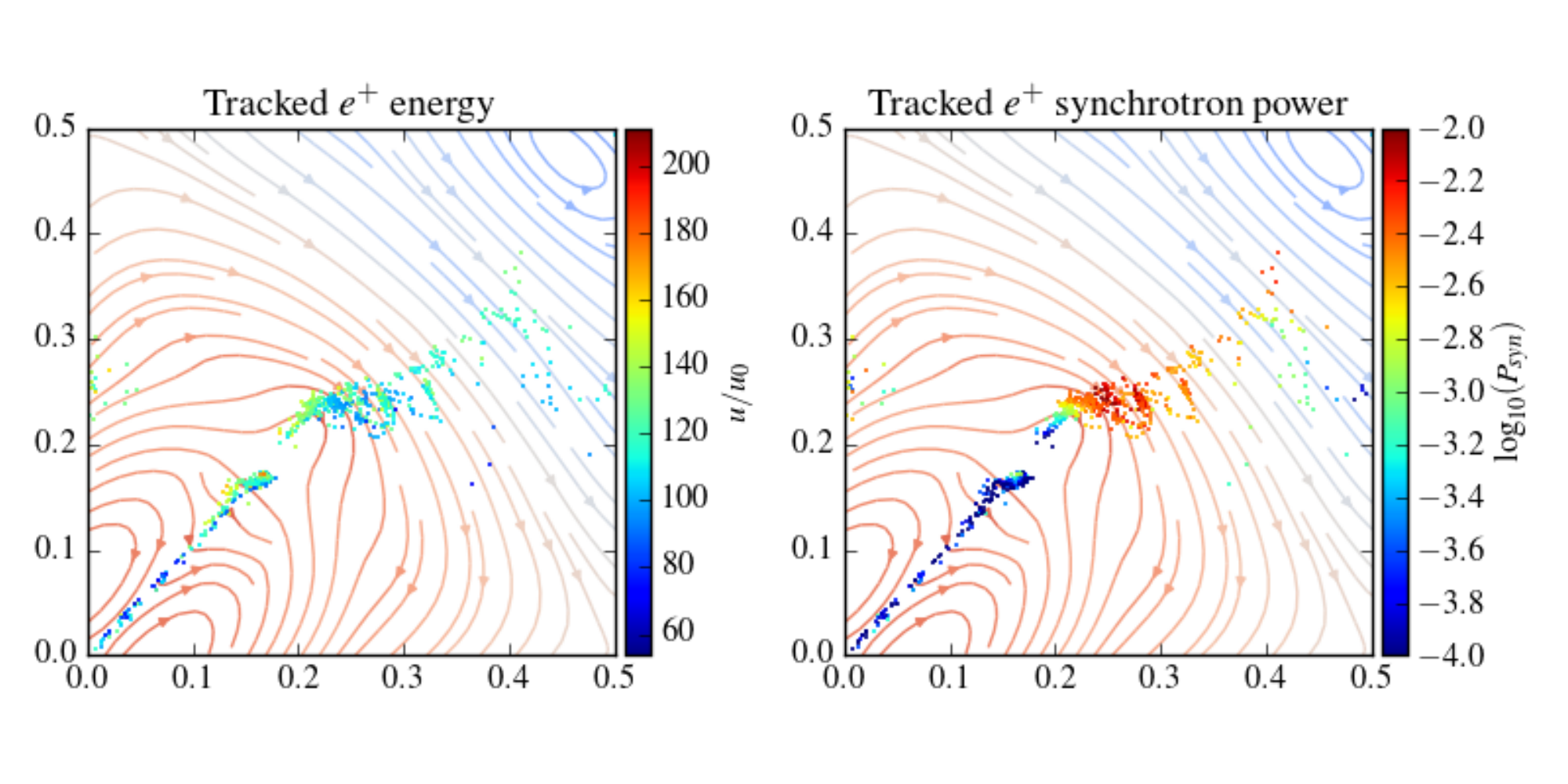}
   }
  \caption{From run 1: we zoom into the lower left corner of the simulation domain where the emitting structure responsible for the highest peak in the $x$ light curve is located, at the time point $t=3.83L/c$ (indicated in Figure \ref{fig:emissivity_xt} by the dotted line). Top: left panel is the 2D emissivity map of positrons; right panel is the total synchrotron power map. Bottom: location of tracked high energy positrons, plotted over the instantaneous field structure, at the same time point. These particles are selected at the end of the simulation who reach $u=250\gamma_0$. In the left panel, the particles are color-coded by their energy while in the right panel they are color-coded by the synchrotron power.}\label{fig:emissivity_2Dx}
\end{figure}

\begin{figure*}
  \centering
  \subfigure
    {
        \includegraphics[width=0.7\textwidth]{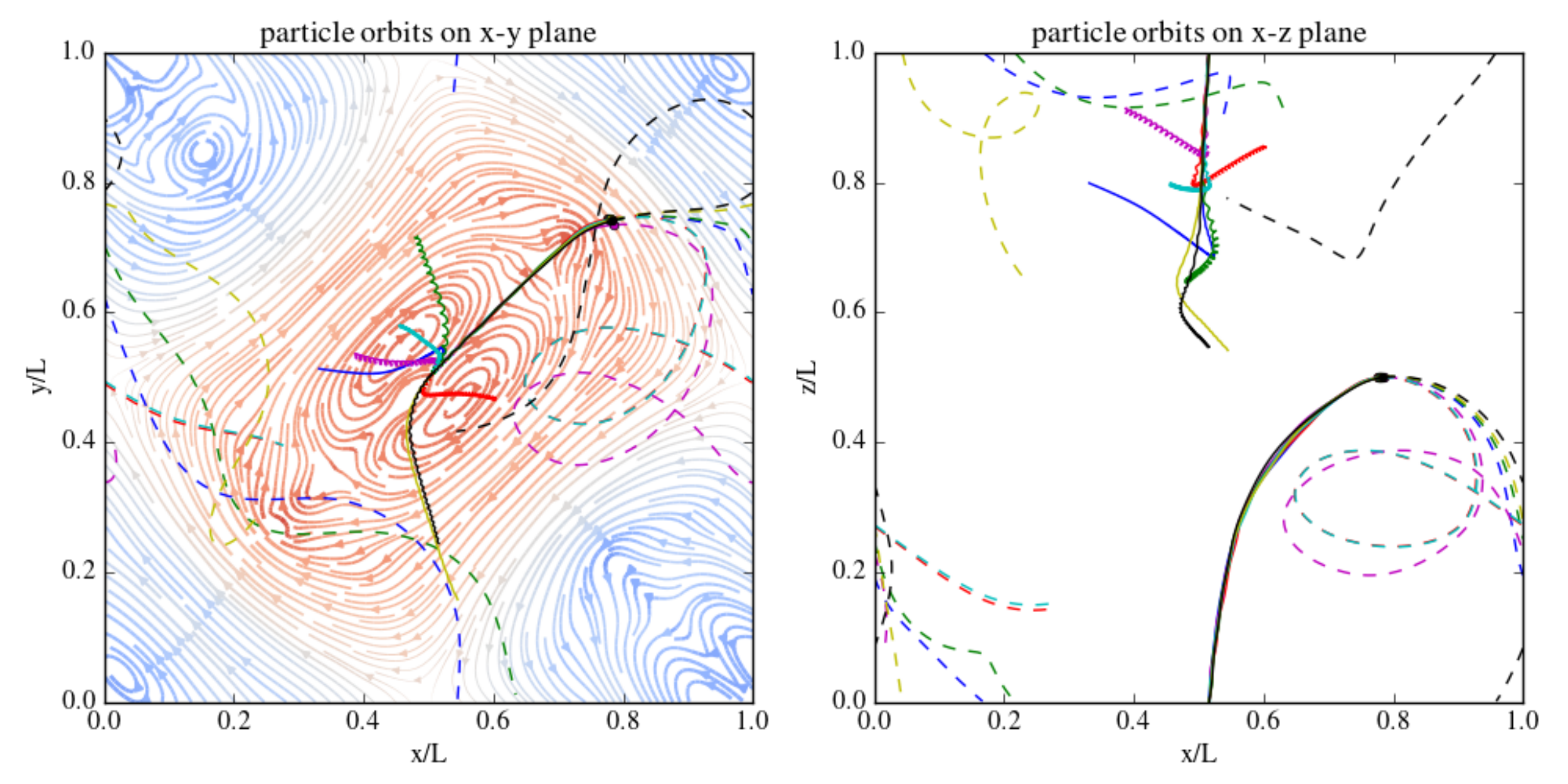}
    }\\
    \vspace{-0.5cm}
  \subfigure
    {
        \includegraphics[width=0.9\textwidth]{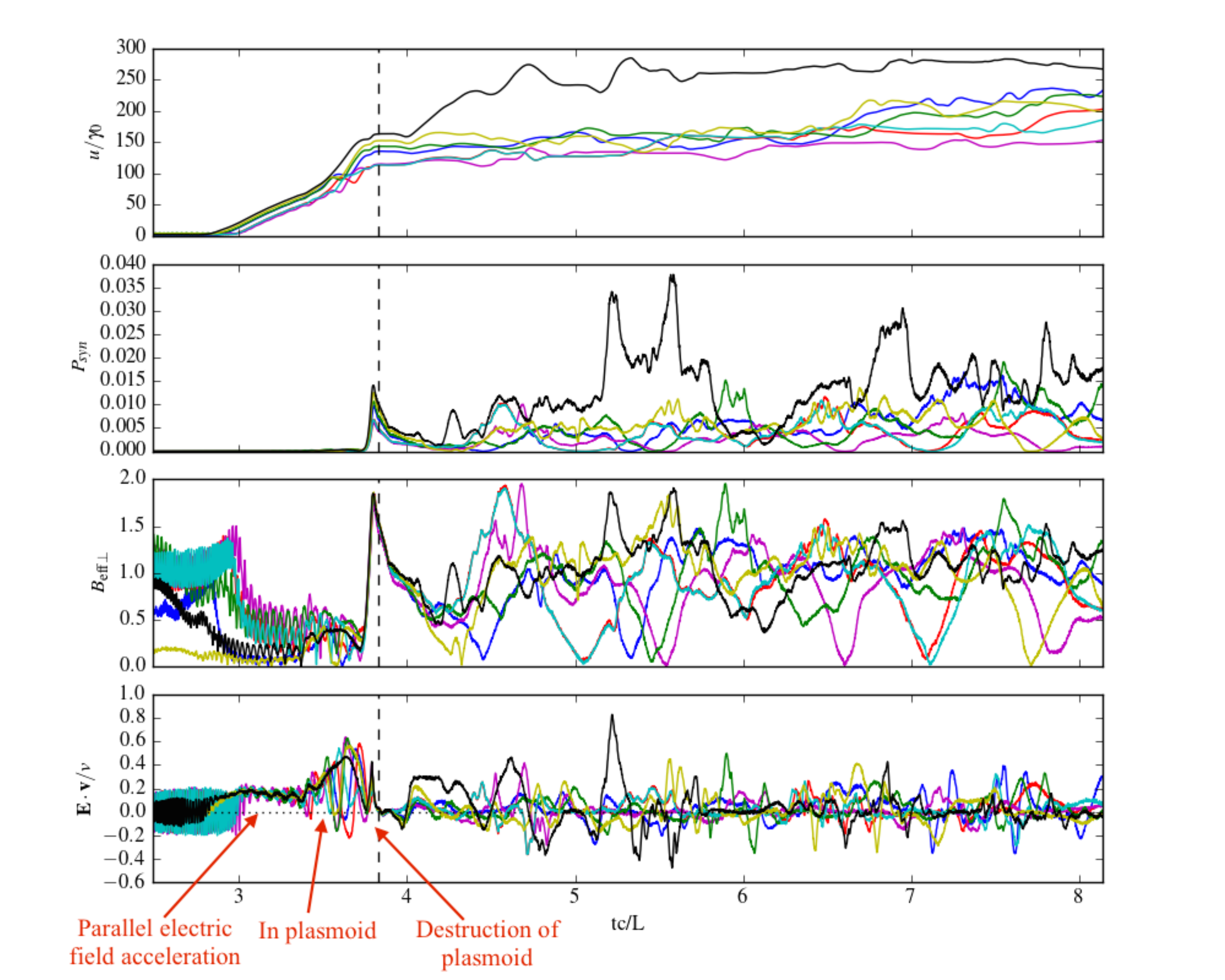}
    }
    \caption{From run 1: history of selected particles within the bunch that is responsible for the highest peak in the $x$ light curve. Top panel shows the particle trajectory on the $x-y$ plane and $x-z$ plane. We have shifted the corner current layer to the center of the box in order to see the particle trajectories better. The dots indicate particle locations at a specific time $t=3.83L/c$ (the field configuration at this time point is also shown on the $x-y$ plane, with streamlines indicating the in-plane field, their color and thickness representing $z$ component and the total magnitude $B$, respectively). Solid lines are past trajectories while dashed lines are future trajectories. The lower panels show particle energy, synchrotron power, effective perpendicular ``magnetic field'' $B_{\rm eff\perp}=ma_{\perp}/e$, and the component of electric field parallel to the particle velocity. The vertical dashed line corresponds to $t=3.83L/c$.}\label{fig:particle history}
\end{figure*}

\subsubsection{shortest variability time scale}
We calculate light curves in different wavebands as received by observers located on the $x-y$ plane. Observers off the $x-y$ plane are not included because the z translational invariance makes it ambiguous to define the emitting/receiving time for the radiation coming into/out of the plane. But fortunately, emission within the $x-y$ plane is already revealing most of the generic behaviors. In Figures \ref{fig:polarization_x}, \ref{fig:polarization_L45} and \ref{fig:polarization_y} top panel, three examples are shown where the observers are looking from $+x$, $45^{\circ}$ counterclockwise of $+x$, and $+y$, respectively. Several remarkable features can be noticed: (1) very sharp, high intensity peaks, with durations $\sim0.01L/c$ (well above the time resolution we use to calculate the light curves), are seen in certain directions. In particular, peaks from positrons are typically observed around $\pm x$ directions, and those of electrons close to $\pm y$ directions; somewhere in between one may not see any of the sharp peaks. (2) These peaks only appear during the early evolution of the system---between the saturation of the linear stage and the destruction of the first current layers. (3) High frequency radiation is dramatically more variable than the low frequency radiation.

The third point is also made clear by the power spectra of the light curves, shown in Figure \ref{fig:powerspectra}. The power spectra in different wavebands can be reasonably fitted by power laws, which get shallower as we go to higher energy band. Such behavior is similar to what has been observed in Harris-type reconnection simulations \citep{Cerutti:2013aa}. In particular, we notice that for radiation emitted along $+x$ and $+y$ directions, there appears to be some power excess on the time scale $0.01-0.1~L/c$.

In the following we investigate carefully the fastest variability seen in the light curves. In Figure \ref{fig:emissivity_xt}, we zoom in around the three biggest peaks in the $+x$ light curve, and compare them directly with the emissivity spacetime diagram (the emissivity has been integrated over y). After identifying the emission event, we pinpoint the responsible features spatially on the 2D emissivity map and total synchrotron power map (Figure \ref{fig:emissivity_2Dx}a). It turns out that these are features $10-20~r_L$ across, moving toward the observer with a speed close to c. These are high energy particle beams ejected from the ends of the current layers, and at this particular time point their trajectories are almost tangent to the line of sight. Here compactness, beam sweeping and light travel time effect are among the factors that cause the sharp emission peaks. In particular, if we look at the three bright tracks on the spacetime diagram that are responsible for the three highest peaks in the light curve (Figure \ref{fig:emissivity_xt}), their span on the emitting time axis can be associated with the time it takes for the opening angle of the particle beam to turn through the observer receiving angle ($\sim10^{\circ}$), but as the tracks lie almost parallel to the light cone, due to the light travel time effect, the beam sweeping time becomes negligible on the receiving time axis; the actual variability time scale is determined by the instantaneous spatial extent of the emitting structure. We have verified that the measured duration of the spikes does not depend on the size of the receiving angle we use for the light curve calculation, as long as it is small enough ($\lesssim10^{\circ}$).

Another point to notice is that, these spatially compact emitting structures form following the merging of a plasmoid into the surrounding magnetic field. From Figure \ref{fig:Bvng} row 3 and 4 we see that the plasmoids contain dense concentration of high energy particles. These plasmoids move out at a speed that reaches $0.6-0.8~c$; the corresponding induced electric field on the back side of the plasmoid tends to accelerate particles along the current while that on the front side tends to decelerate particles. High energy particles are thus bunched by the plasmoids; they experience a further kick at the secondary current layers that form when the plasmoids collide with the ambient field, getting compressed/elongated in a similar manner as the secondary current layers. The spatial bunching and compactness is important in producing the sharp, high magnitude peaks in the light curves, as these are not observed when the ejecta are more diffuse. 

Furthermore, as electrons turn counterclockwise and positrons turn clockwise on the $x-y$ plane when they exit the current layer (for the particular instance we are looking at), observers see emission peaks from electrons mostly near $\pm y$ and positrons near $\pm x$. Figure \ref{fig:emissivity_2Dx}(b) shows some aspects of the particle trajectory bending and beam bunching. We plot on top of the magnetic field lines the position of tracked high energy positrons, and compare their energy and synchrotron power. These particles are selected such that at the end of the simulation they all have $\gamma>250\gamma_0$; they are only the most energetic $6\times10^{-6}$ of all the positrons within the simulation domain. It can be seen that particles that eventually reach high energies are first accelerated in the biggest current layers, where they do not radiate much; synchrotron radiation only becomes important when particles are ejected from the current layers and their trajectories start to bend significantly. Figure \ref{fig:particle history} shows the trajectory and energy history of a few particles that are part of the bunch responsible for the highest peak in the $x$ light curve. The sharp rise in $P_{syn}$ and $\nu_c$ near the dashed line is a result of increased curvature in particle trajectory as they get out of the current layer, and the subsequent quick drop in $P_{syn}$ is not because of cooling (particle energy is almost unchanged) but due to reduced curvature, as indicated by the perpendicular acceleration felt by the particles in Figure \ref{fig:particle history}.  

The energetics of a single spike can be estimated based on its duration $t_v$ and peak flux $P_{\Omega}$. We have seen that $t_v$ is determined by the scale of the ejected plasmoid $r_m$, which varies depending on the history of the plasmoid. Take the sharpest spike as an example: we have $t_v\sim0.01L/c$ and the peak flux is$P_{\Omega}L/c/\mathcal{E}_B(0)\sim10^{-5}$ sr$^{-1}$ (Figure \ref{fig:emissivity_xt}), where $\mathcal{E}_B(0)$ is the initial total magnetic energy. An observer who assumes the radiation to be isotropic would deduce $\mathcal{E}_{spike}=4\pi P_{\Omega}t_v\sim10^{-6}\mathcal{E}_B(0)$. If the observer has a knowledge of the average magnetic field, she could calculate the magnetic energy contained in a volume whose size is determined by the variability time scale, and get $\tilde{\mathcal{E}}_B=(ct_v/L)^2\mathcal{E}_B(0)\sim10^{-4}\mathcal{E}_B(0)$. Thus, the observer would conclude that the apparent radiative efficiency is $\epsilon=\mathcal{E}_{spike}/\tilde{\mathcal{E}}_B\sim0.01$. Of course this depends on $\eta$ as will be shown in \S\ref{subsec:dynamical}. 

\begin{figure*}
  \centering
         \includegraphics[width=\textwidth]{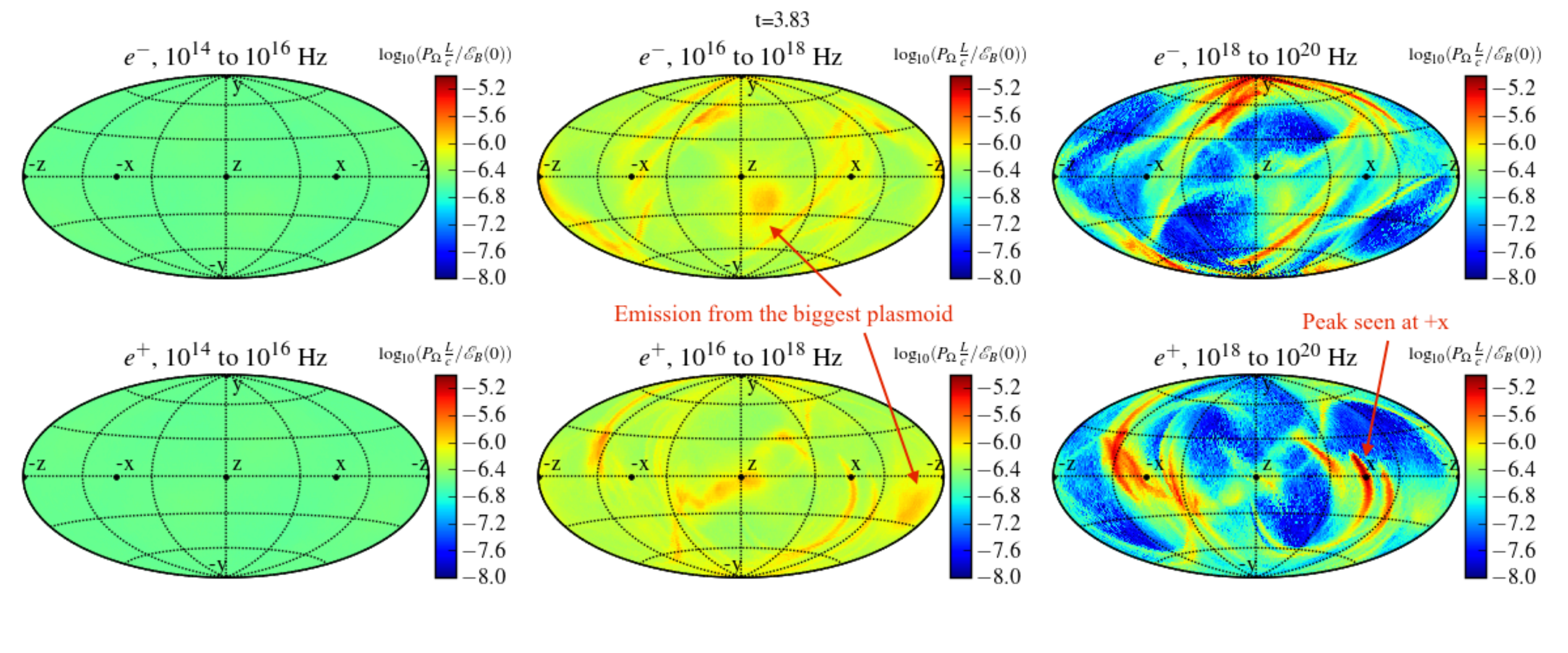}
  \caption{From run 1: angular distribution of emitted synchrotron radiation power in three different wavebands, at the same time point as Figure \ref{fig:emissivity_2Dx}, for electrons and positrons, respectively. We plot the angular distribution using Hammer projection, where $y$ axis is up and $z$ is located at the center of the map (same below).}\label{fig:radangular}
\end{figure*}

The result can be understood from a simple physical argument. Suppose that the average particle number density in a plasmoid is $n_m$ and average Lorentz factor is $\gamma_m$, upon the destruction of the plasmoid particles start to turn in a magnetic field of strength $B_0$, so the observed fluence should be $\mathcal{E}_{spike}=(4\pi/\theta)n_m\pi r_m^2(2e^4\gamma_m^2B_0^2/3c^3m^2)(\gamma_mmc/eB_0)$ where $\theta$ is the beaming angle, determined by the velocity dispersion in the beam. Meanwhile, the inferred magnetic energy content based on the variability time scale is $\tilde{\mathcal{E}}_B=\pi r_m^2(B_0^2/8\pi)$. Assuming the plasmoid gas pressure is $\delta_m$ times the ambient magnetic pressure before its destruction, namely $n_m\gamma_mmc^2\sim\delta_mB_0^2/8\pi$, we have $\mathcal{E}_{spike}/\tilde{\mathcal{E}}_B\sim(4\pi/\theta)\eta\delta_m(\gamma_m/\gamma_0)^2$. In the above example, $\eta=1.1\times 10^{-8}$, $\gamma_m/\gamma_0\sim10^2$, and from Figure \ref{fig:radangular}, the beaming angle is $\theta\sim10^{\circ}\times\sin(30^{\circ})\approx0.03$, so we have $\mathcal{E}_{spike}/\tilde{\mathcal{E}}_B\sim10^{-2}\delta_m$. $\delta_m$ could be of order 1 or larger as the pressure is highly anisotropic and it is the $z$ momentum that dominates. Thus, the simple estimation is roughly consistent with the above measurement.

Despite the attractive high peak intensity and fast variability produced by the small plasmoids, the total energy involved is small. Depending on the viewing angle, an observer may see emission with longer time scales and larger total energetics albeit their small peak flux. The peaks with longer time scales are produced by more diffuse ejecta or smooth field structures lit up by distributed high energy particles, which evolve on dynamic time scales. Another remark to be made is that, in this particular example the current sheet itself is not rotating, but in reality it can be dynamic and can turn around during the evolution. This could introduce further variability.

\begin{figure*}
  \centering
         \includegraphics[width=\textwidth]{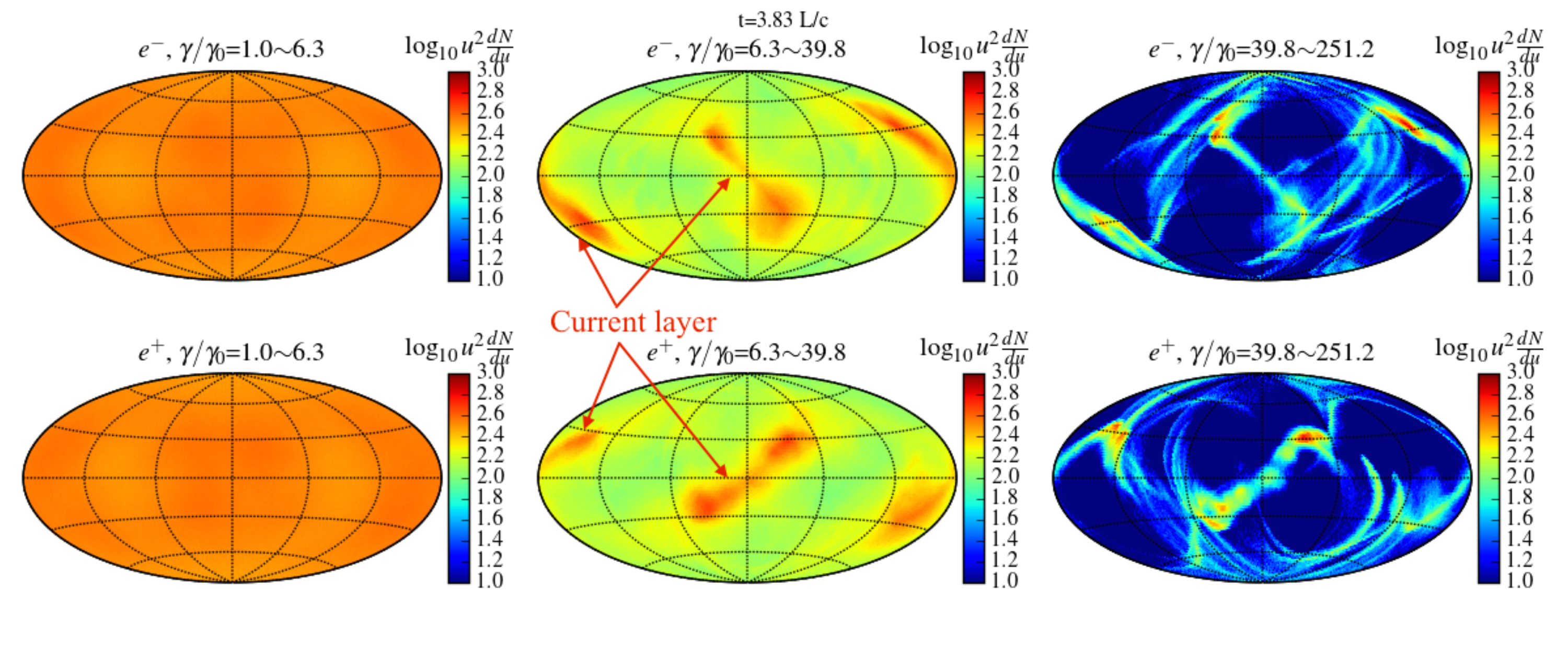}
  \caption{From run 1: angular distribution of particles in three different energy bands, at the same time point as Figure \ref{fig:emissivity_2Dx}, for electrons and positrons, respectively.}\label{fig:particleangular}
\end{figure*}

\begin{figure*}
  \centering
  \subfigure
    {
        \includegraphics[width=0.45\textwidth]{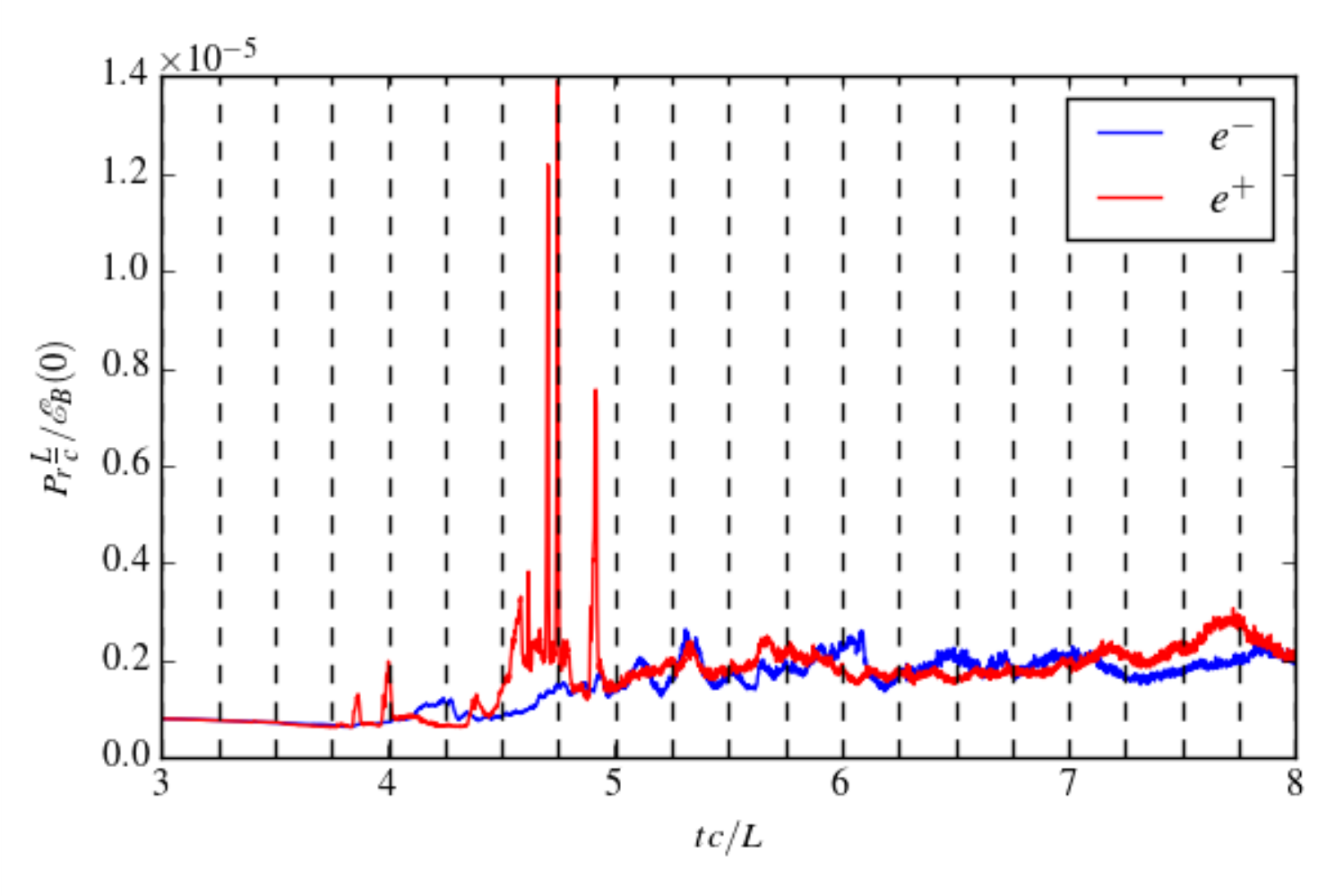}
    }
    \subfigure
   {
        \includegraphics[width=0.45\textwidth]{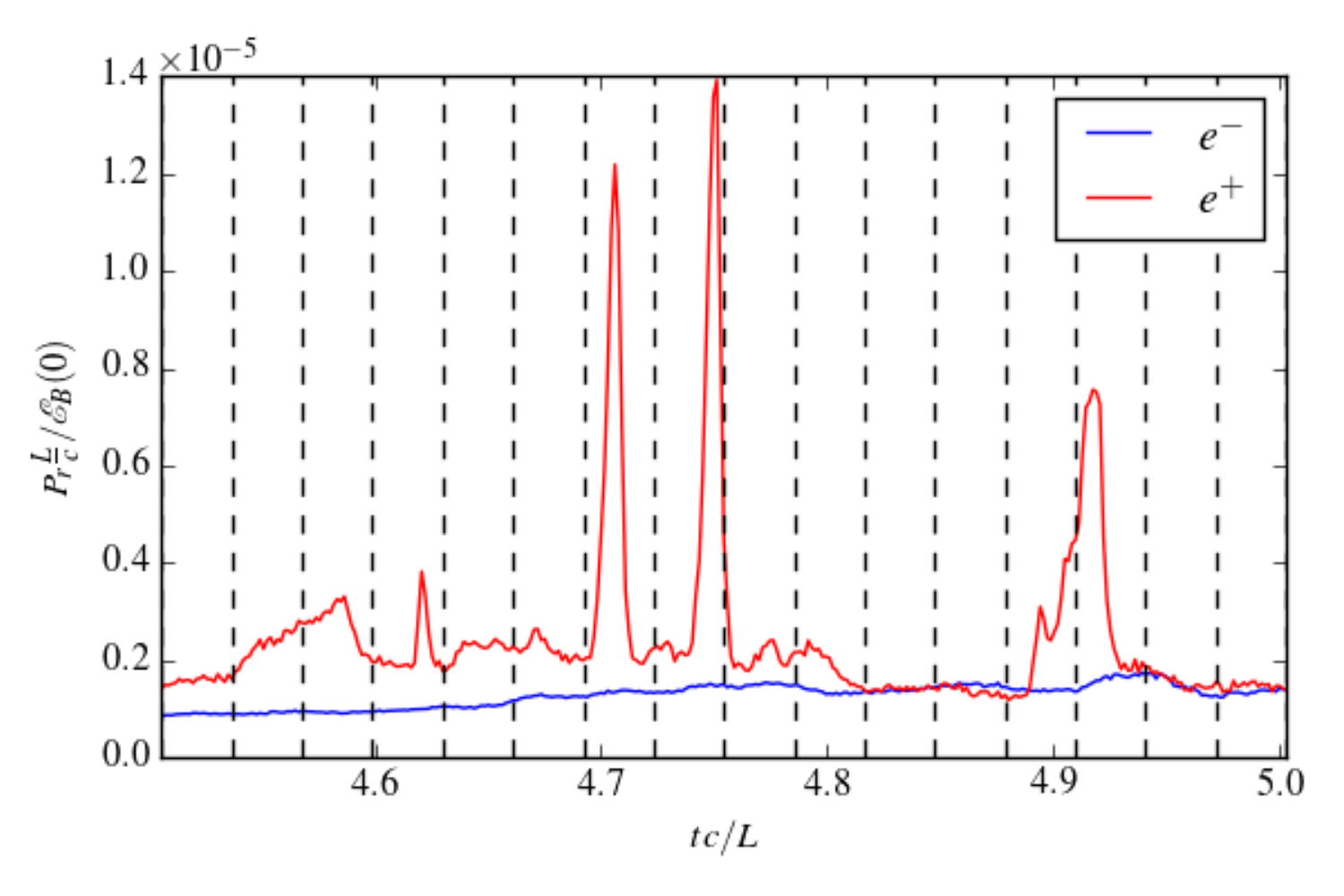}
    }\\
    \subfigure
   {
        \includegraphics[width=0.45\textwidth]{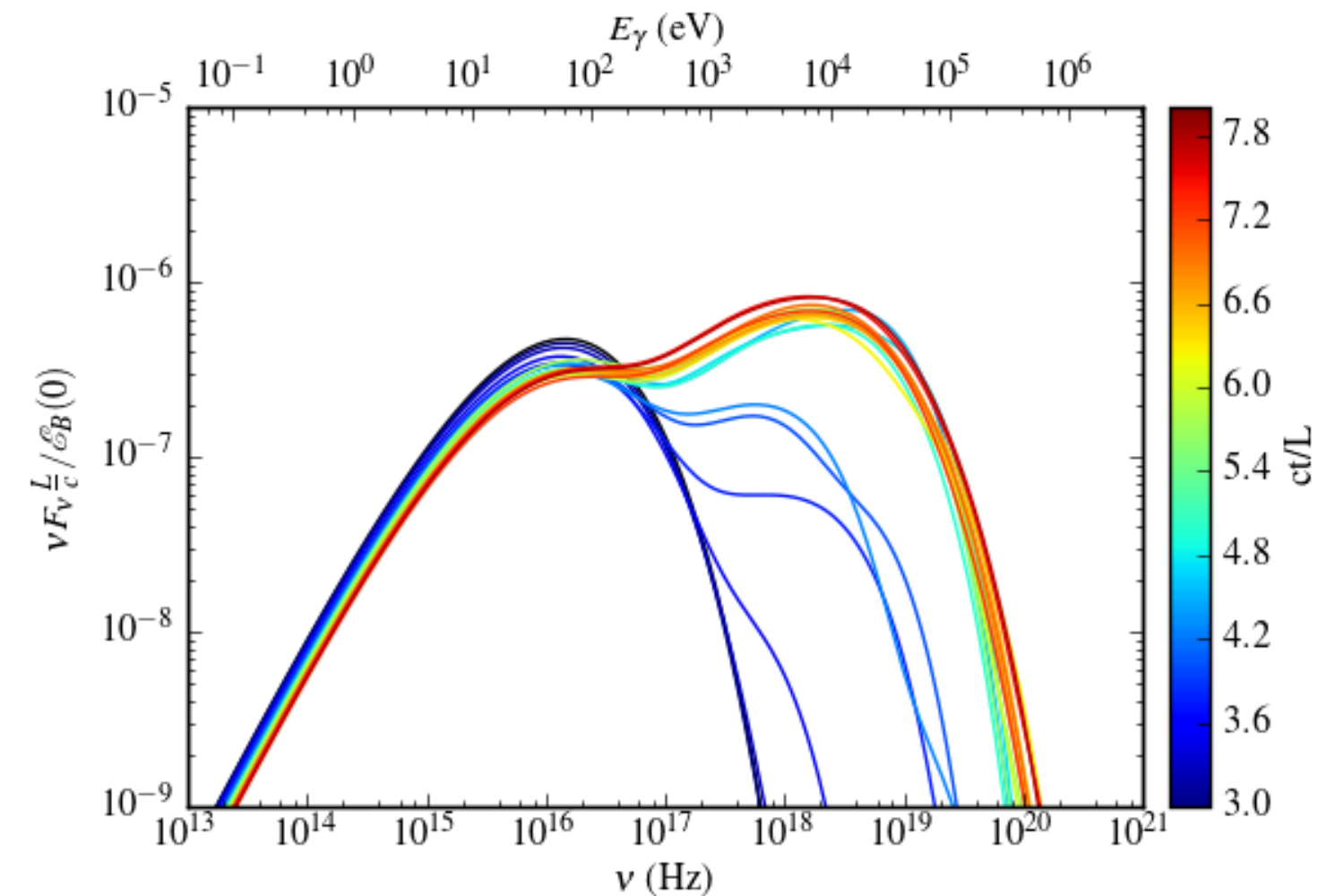}
    }
    \subfigure
   {
        \includegraphics[width=0.45\textwidth]{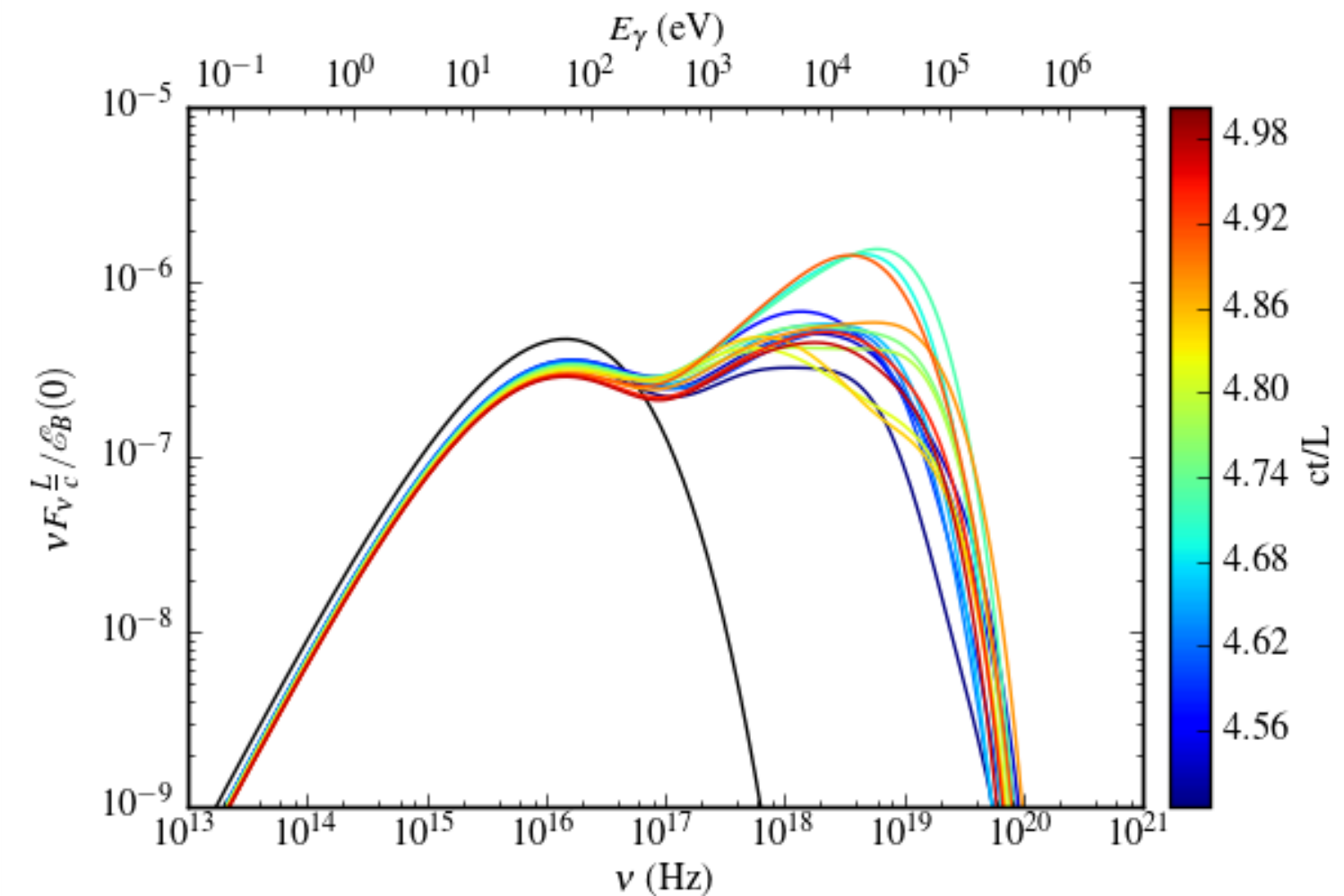}
    }
  \caption{From run 1: we divide the light curve into equally spaced time windows (top) and plot the received radiation spectra in corresponding time windows (bottom). These are for the observer located at $+x$. Left panel spans across the whole simulation duration while the right panel zooms in around the highest peaks. The black line corresponds to the quiescent spectrum.}\label{fig:timespectra}
\end{figure*}

\begin{figure}
  \centering
         \includegraphics[width=0.45\textwidth]{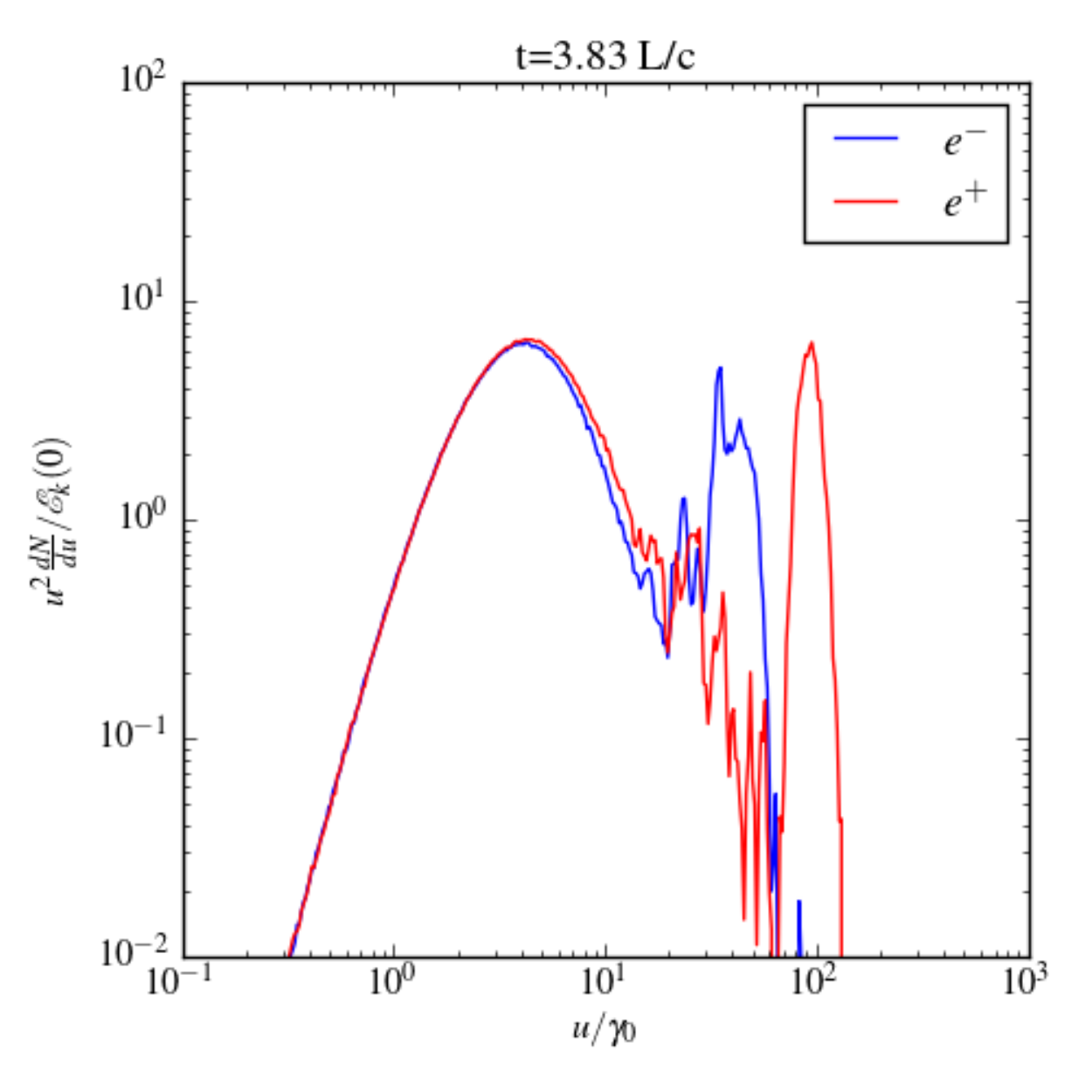}
  \caption{From run 1: energy distribution of particles within $\sim10^{\circ}\times10^{\circ}$ around $+x$, at the same time point as Figure \ref{fig:emissivity_2Dx}, for electrons and positrons, respectively.}\label{fig:particlespectrum_x}
\end{figure}

\subsubsection{Origin and beaming of high energy radiation}
 During the saturation of the linear instability and the early stage of nonlinear evolution, particles are being efficiently accelerated in the current layers by the parallel electric field. Plasmoids are not very good particle accelerators except for those undergoing rapid acceleration themselves, but they could trap particles that are accelerated in the current layer. High energy particles within the current layer have a fan-like angular distribution spanned around $\pm z$; they turn toward the ends of the current layers due to the reconnected magnetic field. These fan-like features are readily seen in Figure \ref{fig:particleangular}, especially in high energy bands. However, particles within the current layers do not produce a large amount of high energy synchrotron radiation---this is demonstrated by the absence of corresponding fan-like structure in the radiation angular distribution (Figure \ref{fig:radangular}). Plasmoids do radiate a significant amount of power, as shown in the synchrotron power map (Figure \ref{fig:Bvng} last row), but this radiation peaks in relatively low frequency---in Figure \ref{fig:radangular} we see emission from plasmoids mainly in the intermediate energy band. The main reason for the lack of high energy radiation from the current layer itself (plasmoids included) is that the curvature of the particle trajectory is small. 
  
 Significant high energy radiation emerges from just downstream of the exhausts of the current layers, where high energy particles are being dumped onto the surrounding magnetic field. The ejecta can be intermittent and compact, as the spatial distribution of high energy particles is modulated by the spontaneously formed plasmoids. In particular, the unrelaxed, small plasmoids formed at the late stage of current sheet evolution produce the most compact, high intensity emission upon their destruction; as the beams turn around in the magnetic field, they give narrow stripes in the angular distribution of radiation (Figure \ref{fig:radangular}) and we see sharp peaks when the beams sweep across the line of sight. On the other hand, the fully relaxed, large plasmoids produce relatively diffuse emission. 
 
 Later on, the initial current layers get destroyed and the system evolves in a more turbulent way. High energy particle beams are gradually dispersed both in configuration space and in momentum space. In this particular example particles do not cool significantly; high energy particles eventually spread over most of the simulation domain. As a result, the high energy radiation becomes more and more diffuse and rapid variability will no longer be observed.

\subsubsection{Time dependent spectrum of observed radiation}\label{subsubsec:time dependent spectrum}
The quiescent radiation spectrum peaks at $\omega_0\approx\eta\omega_{syn,lim}(\gamma_{RMS}/\gamma_0)^2(B_{RMS}/B_0)\approx22\eta\omega_{syn,lim}=10^{16}$ Hz in the ultraviolet; as the instability develops and accelerates a significant fraction of particles to energies $\gg\gamma_0$, a new radiation component also emerges. As an example, we divide the time series of radiation received by the observer located at $+x$ into equally spaced time windows and calculate the synchrotron spectrum within each of these windows. Figure \ref{fig:timespectra} shows the results both for large time windows throughout the simulation duration and for small time windows around the highest peaks. It can be seen that during the evolution, the peak frequency of the high energy radiation can reach $10^3$ times the quiescent value.

Although the overall particle spectrum as shown in Figure \ref{fig:particle spectrum} only exhibits a steep power-law tail, the instantaneous distribution of particles moving along a certain direction can have a dominant component on the tail. Figure \ref{fig:particlespectrum_x} shows the distribution of particles within $10^{\circ}\times 10^{\circ}$ around $+x$ near the time point when the highest peak in the light curve is produced. Evidently the positrons have an almost mono-energetic beam with $u\approx 100\gamma_0$.

\subsubsection{Polarization}
We calculate the linear polarization degree and polarization angle as a function of time, in different wavebands, for observers located on the $x-y$ plane. Examples are shown in Figures \ref{fig:polarization_x}, \ref{fig:polarization_L45} and \ref{fig:polarization_y}. The initial equilibrium produces a high polarization $\sim 25\%$ and the polarization angle is aligned with the $x-y$ plane, as one would expect from the symmetry of the configuration. At the start of the instability, there is an overall drop in the polarization degree, which is especially obvious in the low energy band. However, for the high energy radiation, during the ``flares'' (i.e. sharp peaks in the light curves), polarization degree increases significantly, accompanied by large change in polarization angle. This is because the high energy emitting particle beams are compact and they sample a relatively strong, ordered magnetic field, and their high radiation flux outweighs the contribution from the other parts of the simulation domain. \newtext{As an example, we looked at the polarization produced by the bunch of particles we tracked in Figure \ref{fig:particle history}. We notice that these particles have gyro radius that is larger than or comparable to the scale length of the field, and the polarization angle is determined by the instantaneous orientation of the orbital plane instead of the local magnetic field, because of the presence of the electric field. The calculated polarization angle of the emission from a typical particle in the bunch is about 64$^{\circ}$, as shown by the magenta triangle in Figure \ref{fig:polarization_x}. Since this bunch dominates the total intensity, the resultant polarization angle of the whole box deviates from the nominal value $0^{\circ}$ and reaches around $60^{\circ}$.} The polarization angle of the low energy component shows only slight change during most of the evolution. When the system is fully evolved to settle into a new equilibrium (the ground state), the all-frequency integrated polarization degree again settles at $\sim 25\%$ and the polarization angle also returns to its initial value.

\begin{figure*}
  \centering
  \subfigure[]
  {
         \includegraphics[width=0.45\textwidth]{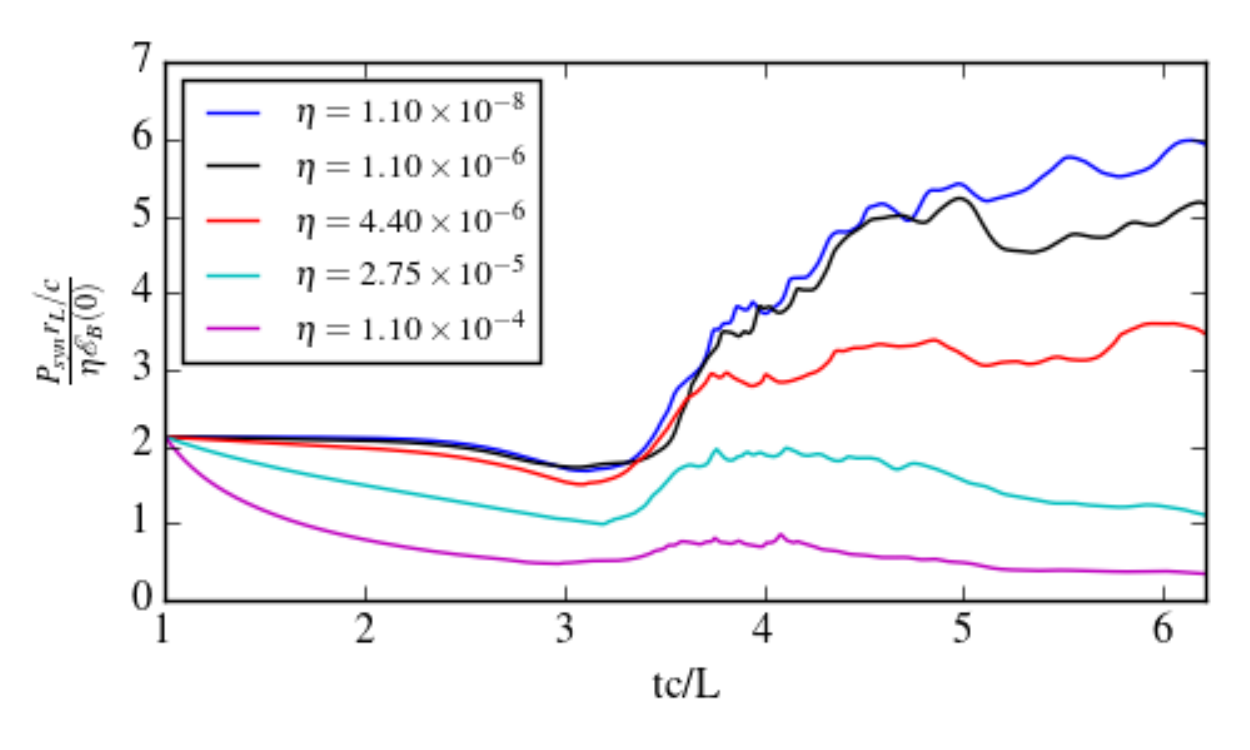}
   }
   \subfigure[]
   {
         \includegraphics[width=0.45\textwidth]{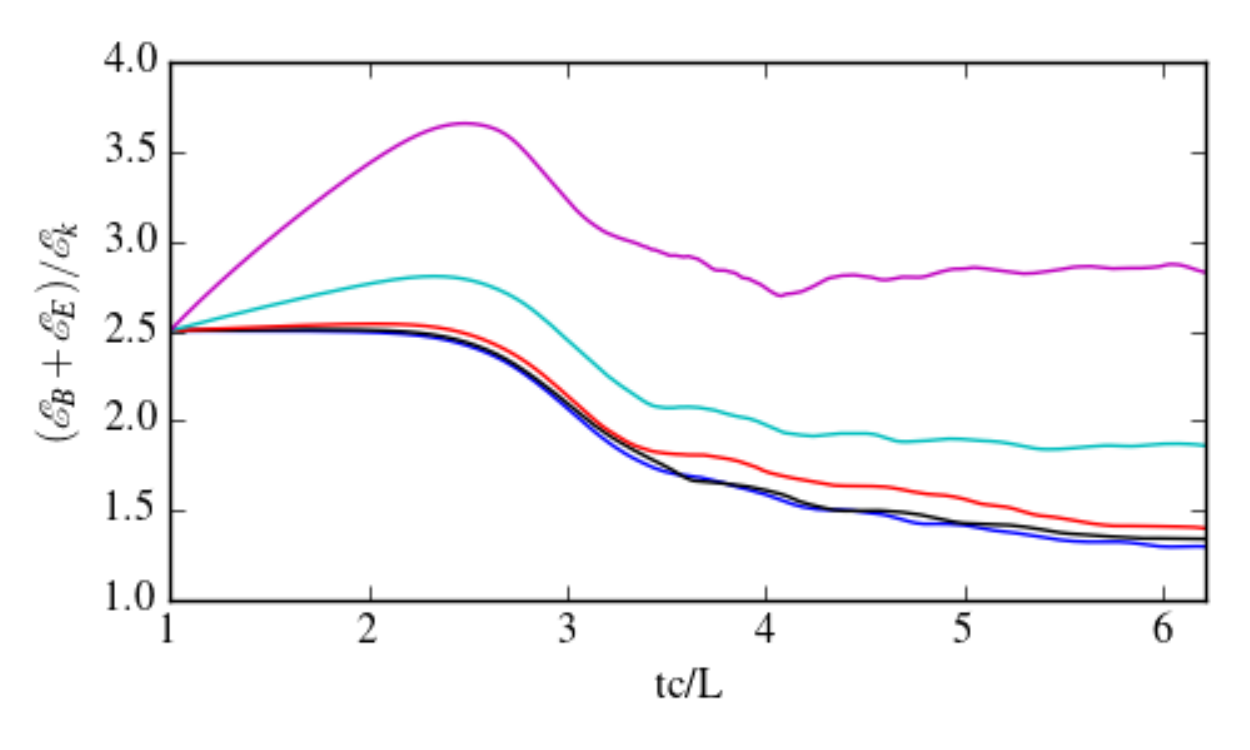}  
   }
  \caption{(a) Total radiated power as a function of time for the runs with different $\eta$. The power has been scaled by $1/\eta$. (b) The ratio between electromagnetic energy and particle kinetic energy as a function of time for the runs with different $\eta$.}\label{fig:comparison:power-sigma}
\end{figure*}

\begin{figure*}
  \centering
         \includegraphics[width=0.9\textwidth]{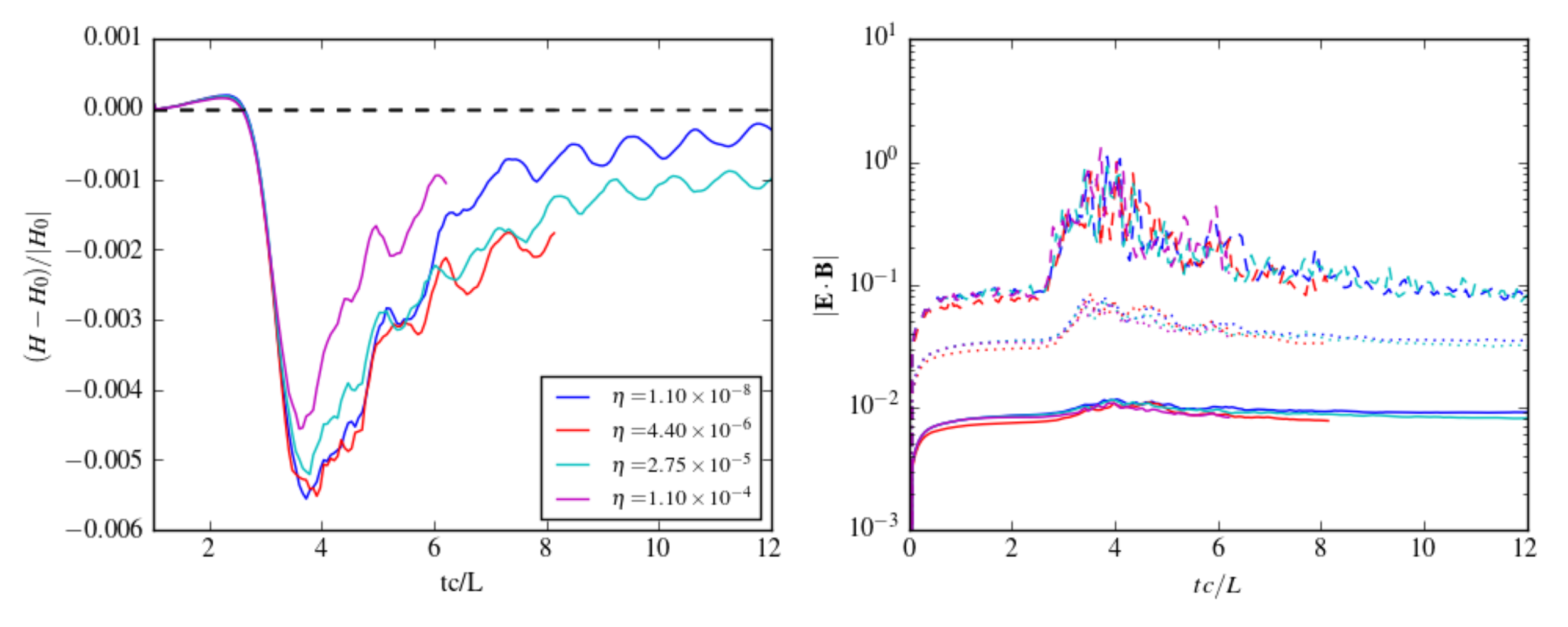}
  \caption{(a) Helicity change as a function of time for the runs with different $\eta$. (b) $|\mathbf{E}\cdot\mathbf{B}|$ values---mean (solid lines), 99 percentile (dotted lines) and maximum (dashed lines)---as a function of time for the same runs. }\label{fig:comparison:helicity}
\end{figure*}

\begin{figure*}
  \centering
  \subfigure[]
  {
        \includegraphics[width=0.46\textwidth]{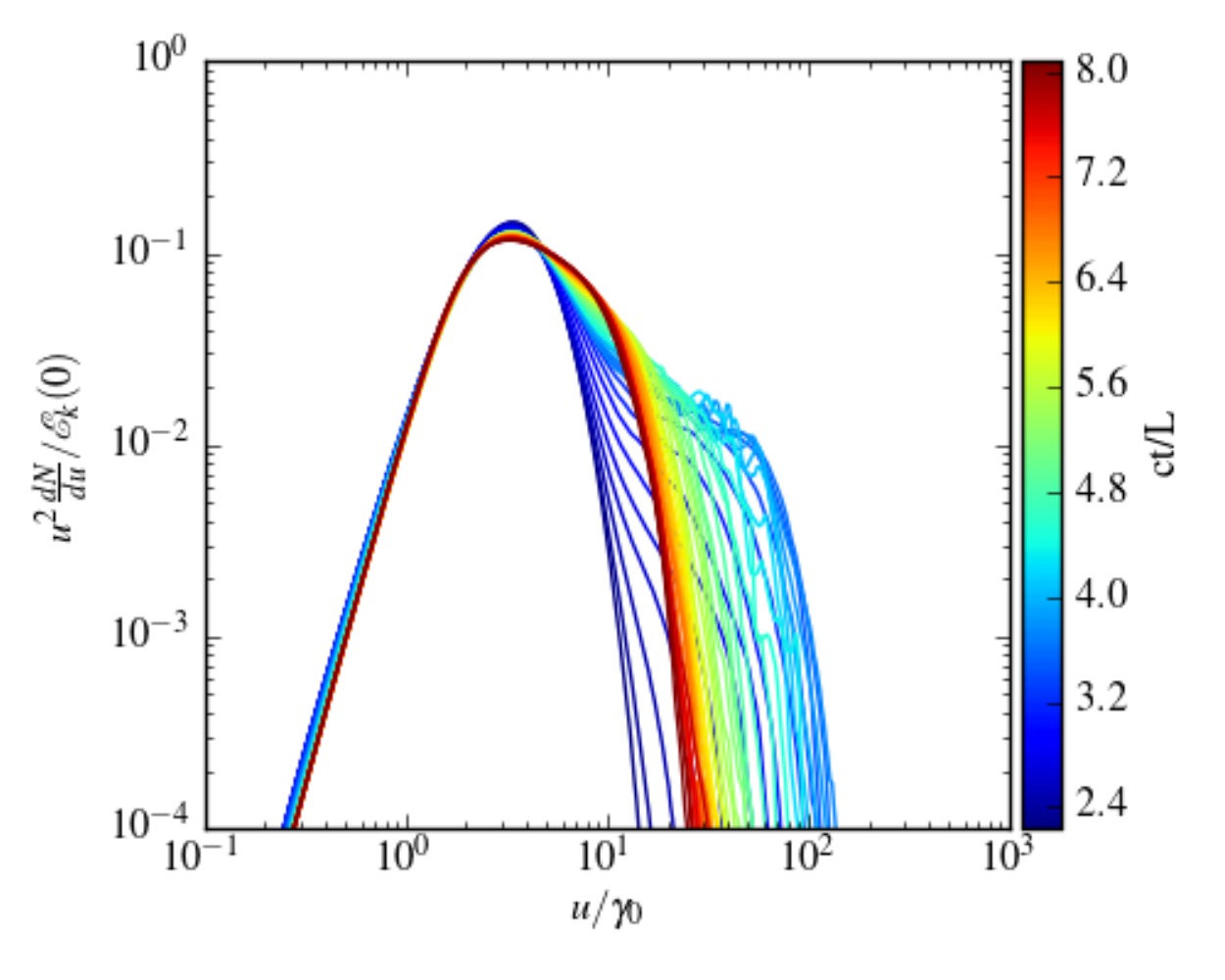}
  }
   \subfigure[]
  {
        \includegraphics[width=0.46\textwidth]{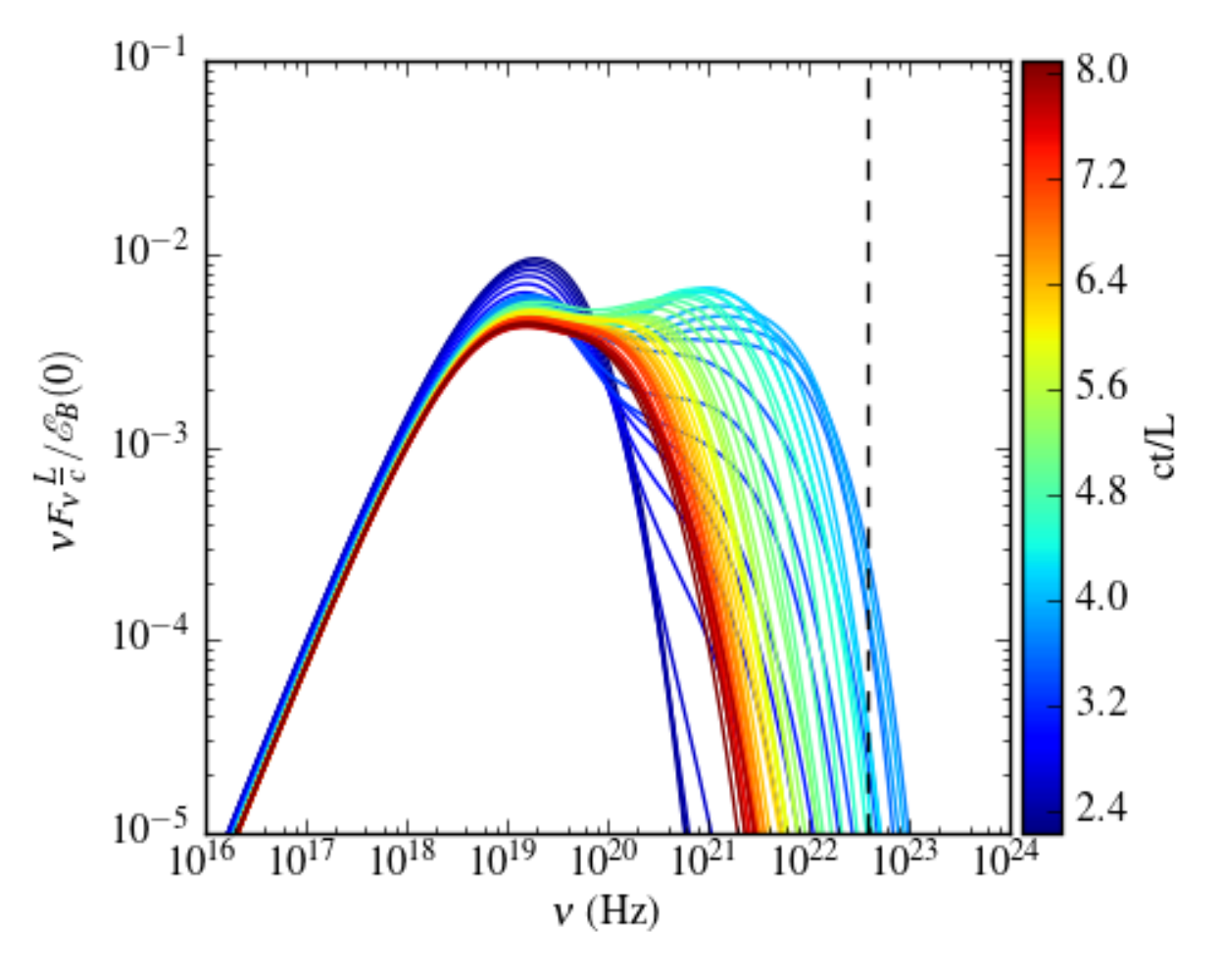}
  }

     \caption{From run 4: (a) Isotropic particle spectrum. (b) Instantaneous, isotropic radiated spectrum. The vertical dashed line corresponds to the radiation reaction limit 160 MeV.}\label{fig:rad:particle spectrum}
\end{figure*}

\begin{figure*}
  \centering
  \subfigure
    {
        \includegraphics[width=0.42\textwidth]{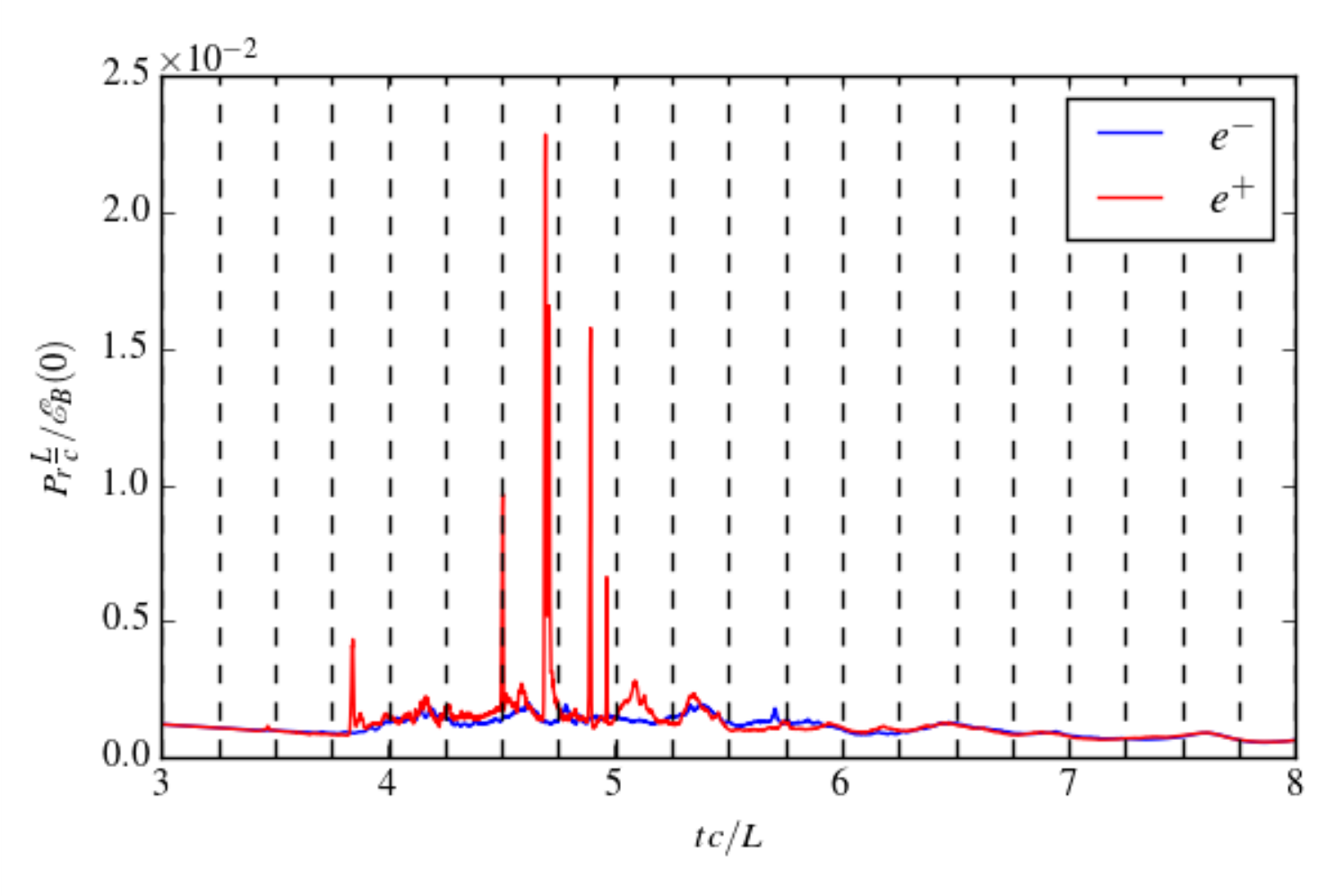}
    }
    \subfigure
   {
        \includegraphics[width=0.42\textwidth]{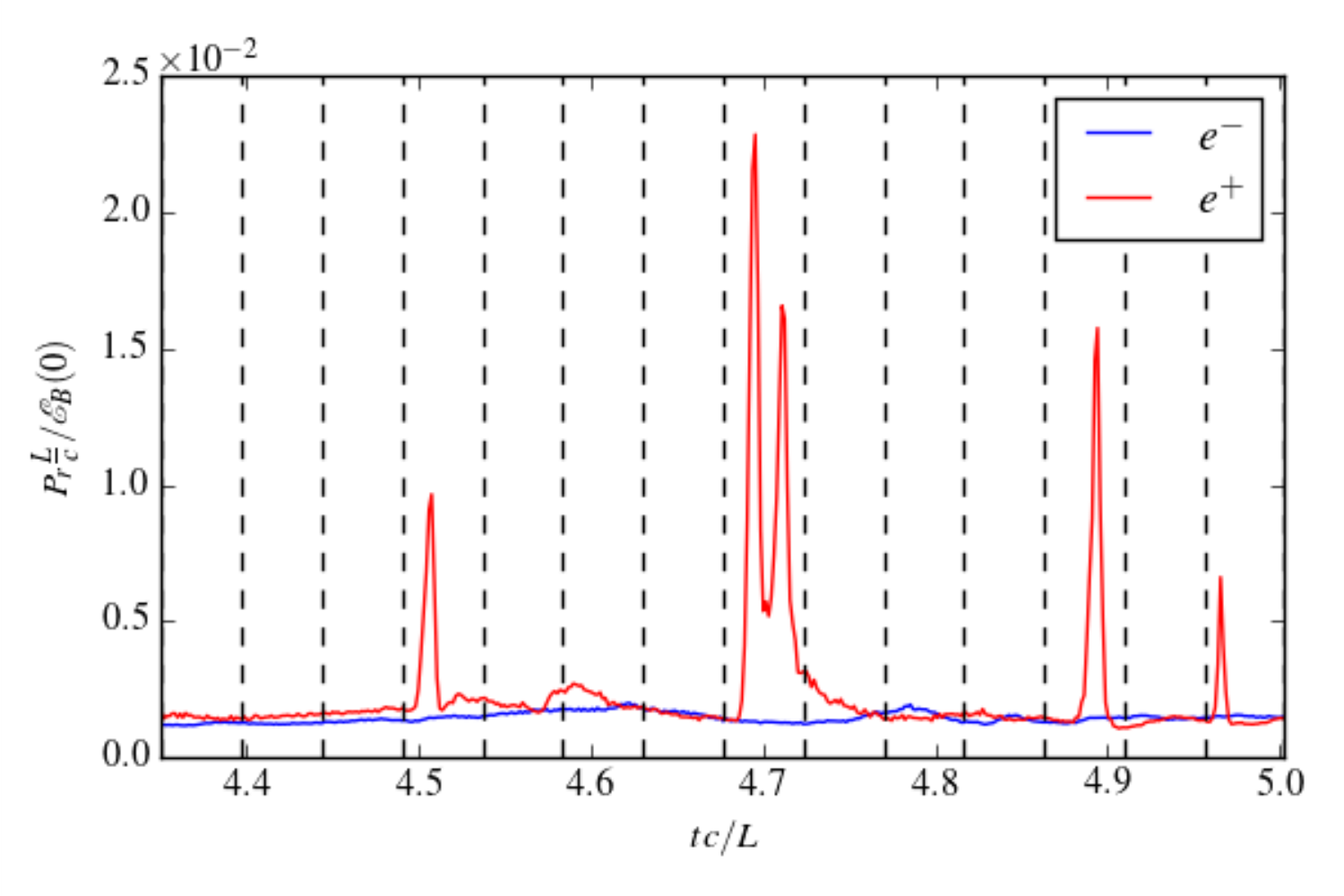}
    }\\
    \vspace{-0.5cm}
    \subfigure
   {
        \includegraphics[width=0.45\textwidth]{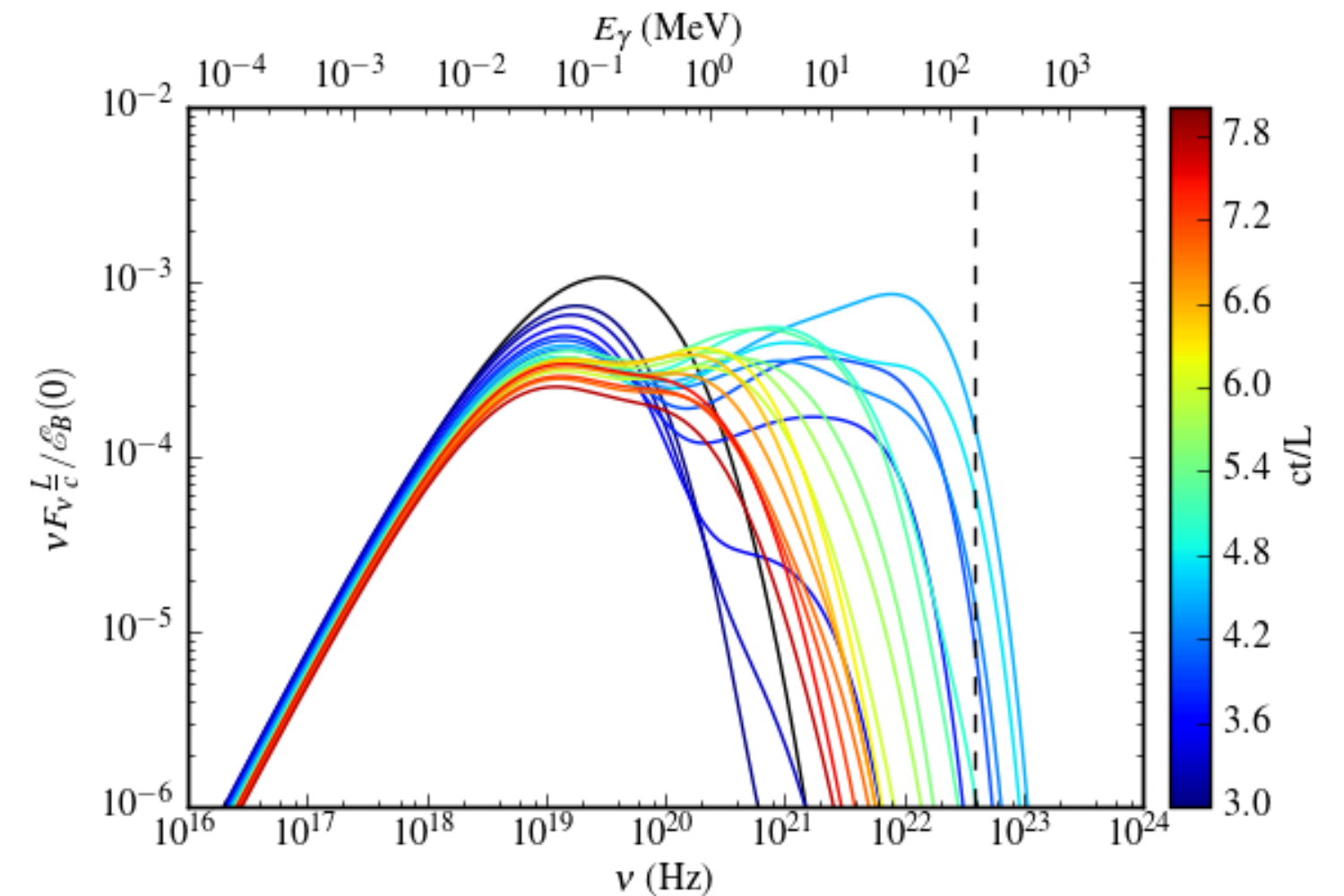}
    }
    \subfigure
   {
        \includegraphics[width=0.45\textwidth]{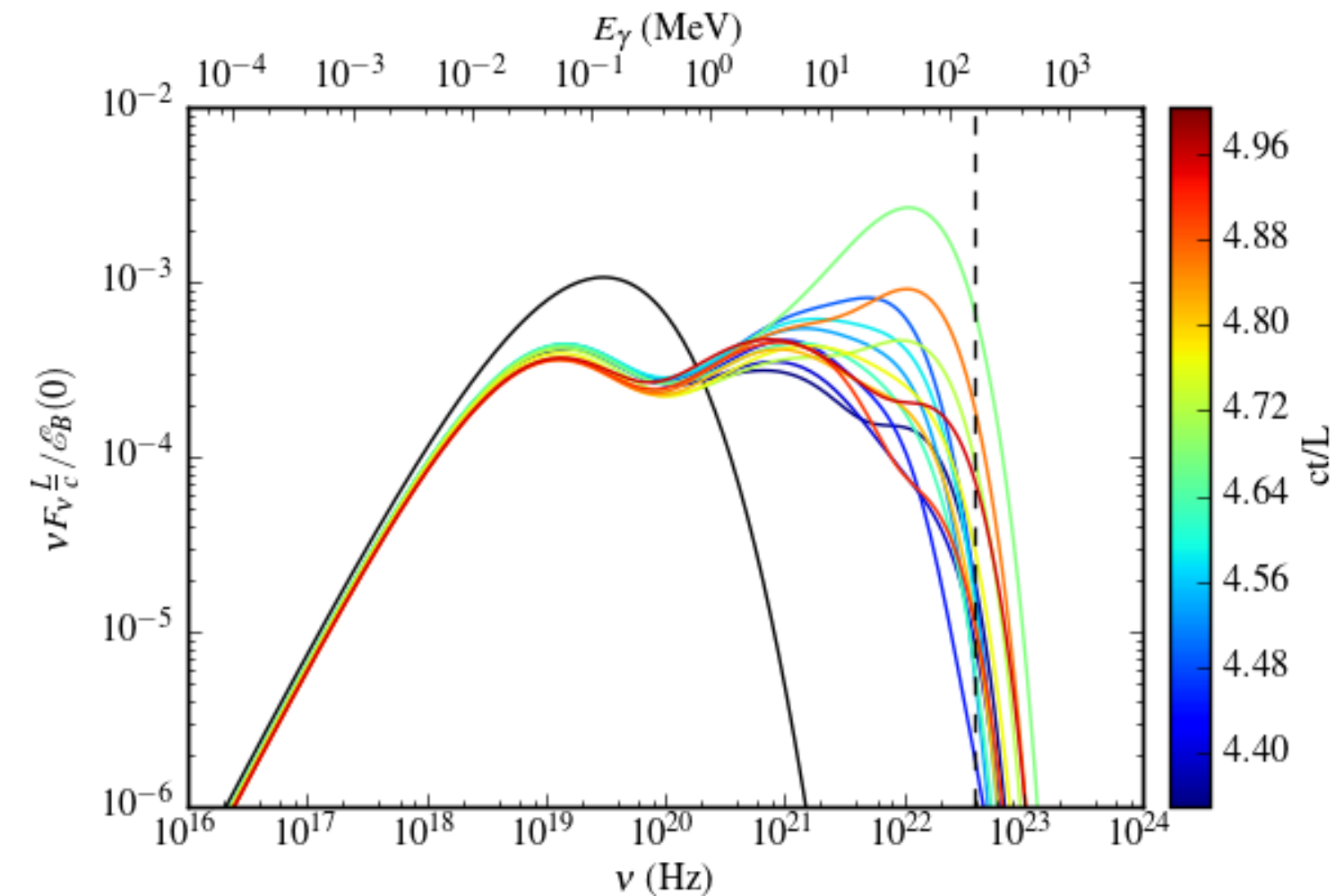}
    }
  \caption{From run 4: time dependent synchrotron spectrum as seen by an observer located at $+x$. Left panel spans across the whole simulation duration while the right panel zooms in around the highest peaks (similar to Figure \ref{fig:timespectra}). The black solid line corresponds to the quiescent spectrum at the start of the simulation. The vertical dashed line indicates the radiation reaction limit 160 MeV.}\label{fig:rad:timespectra}
\end{figure*}

\begin{figure*}
  \centering
  \subfigure
    {
        \includegraphics[width=0.42\textwidth]{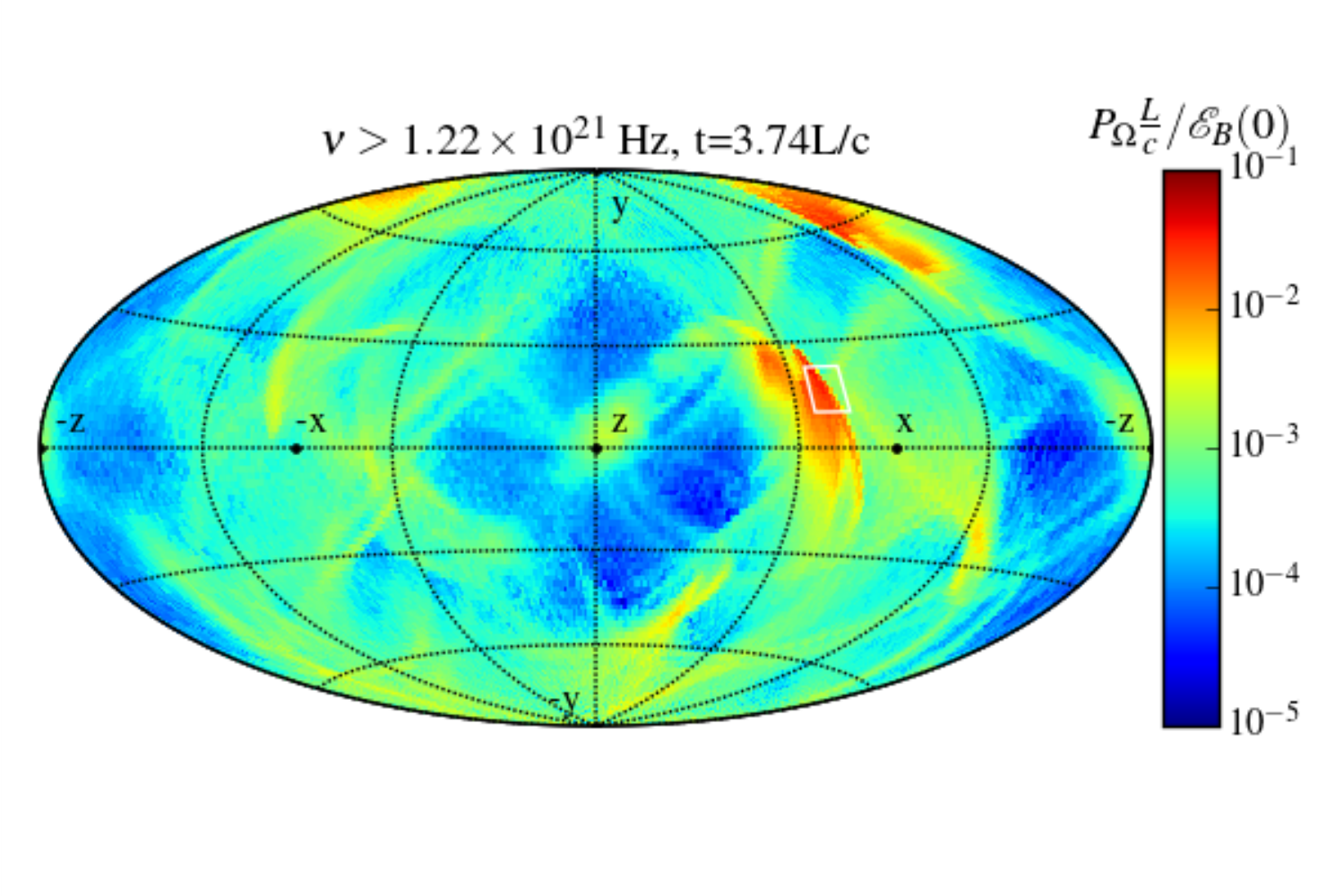}
    }
    \subfigure
   {
        \includegraphics[width=0.42\textwidth]{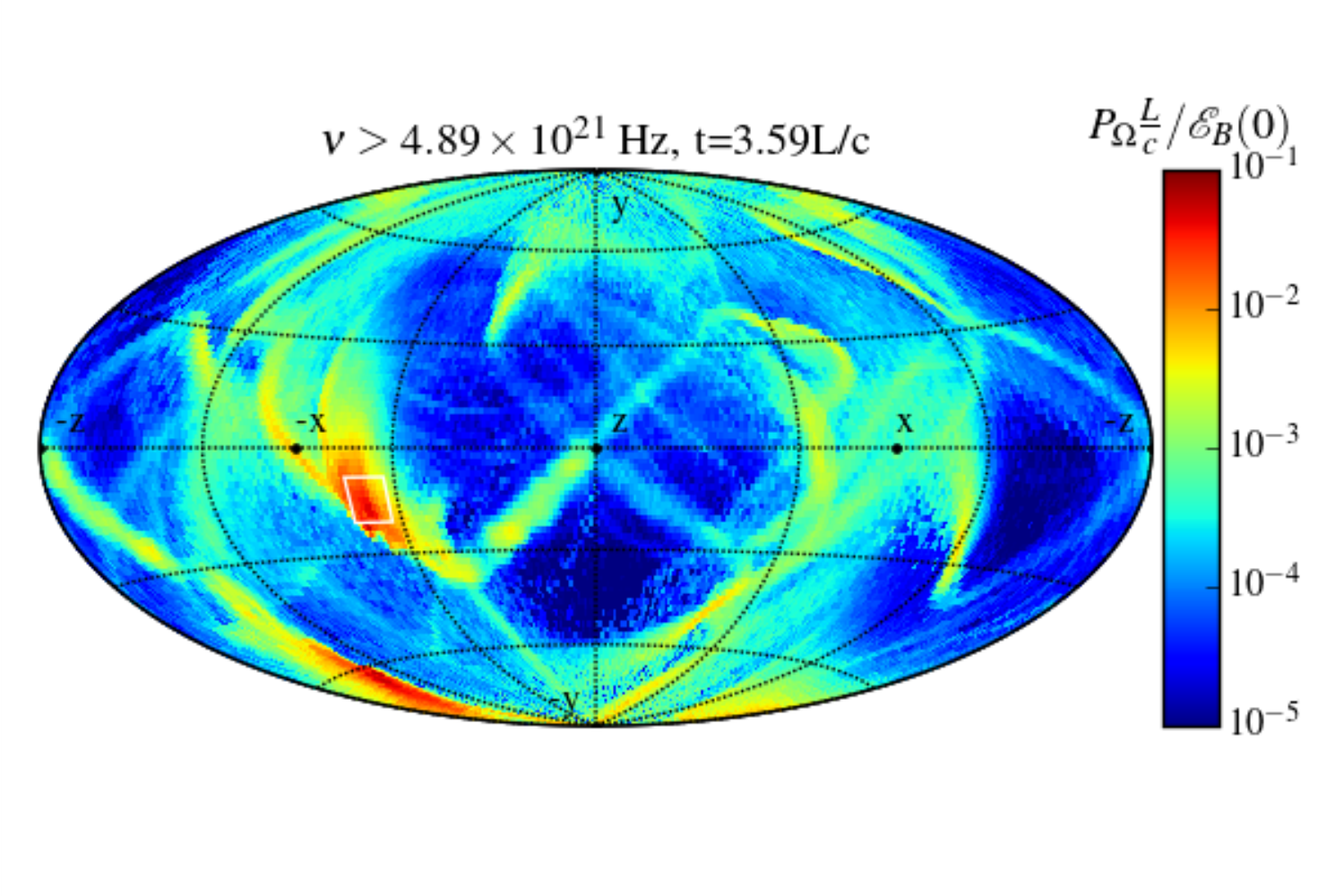}
    }\\
    \vspace{-1cm}
    \subfigure
   {
        \includegraphics[width=0.45\textwidth]{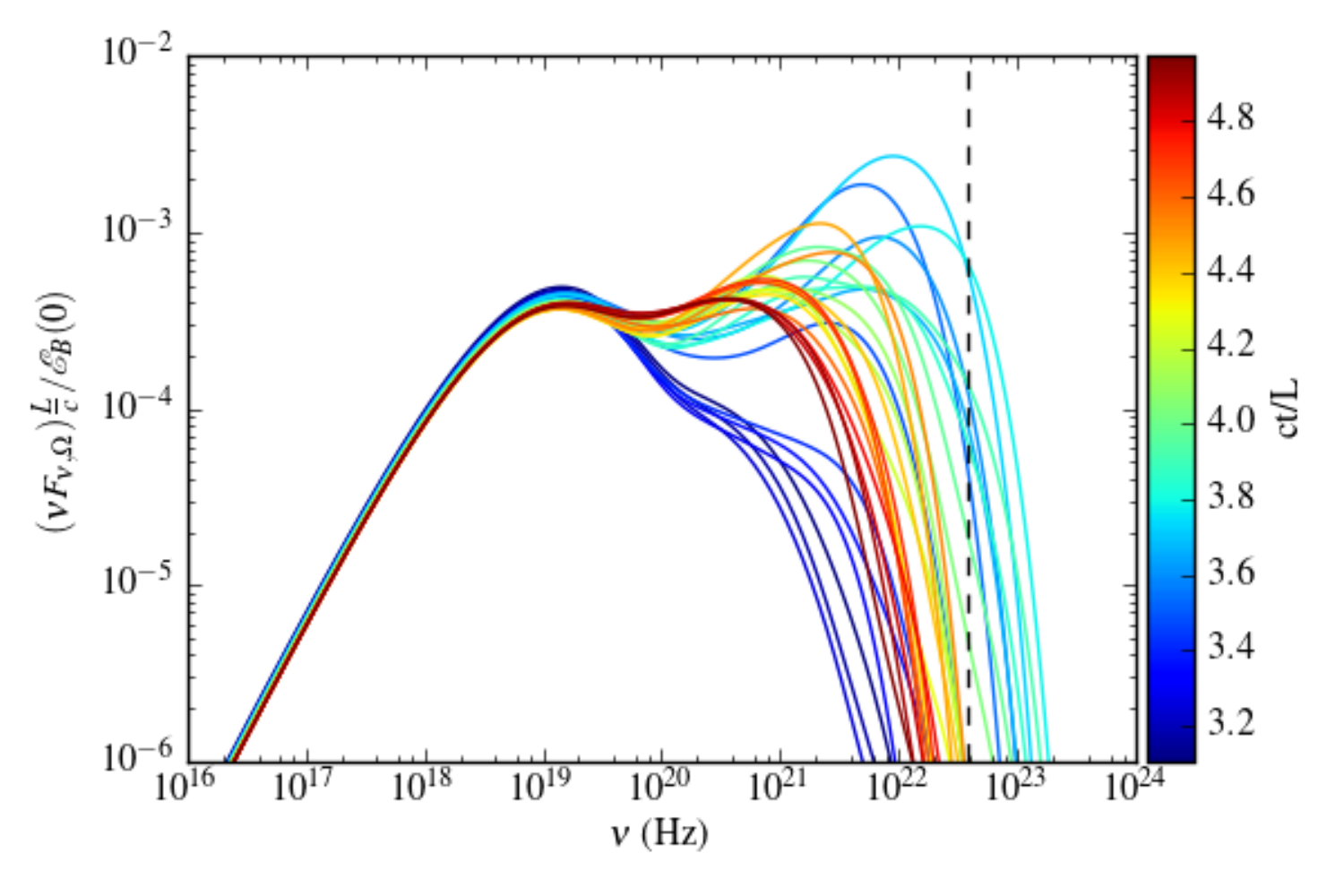}
    }
    \subfigure
   {
        \includegraphics[width=0.45\textwidth]{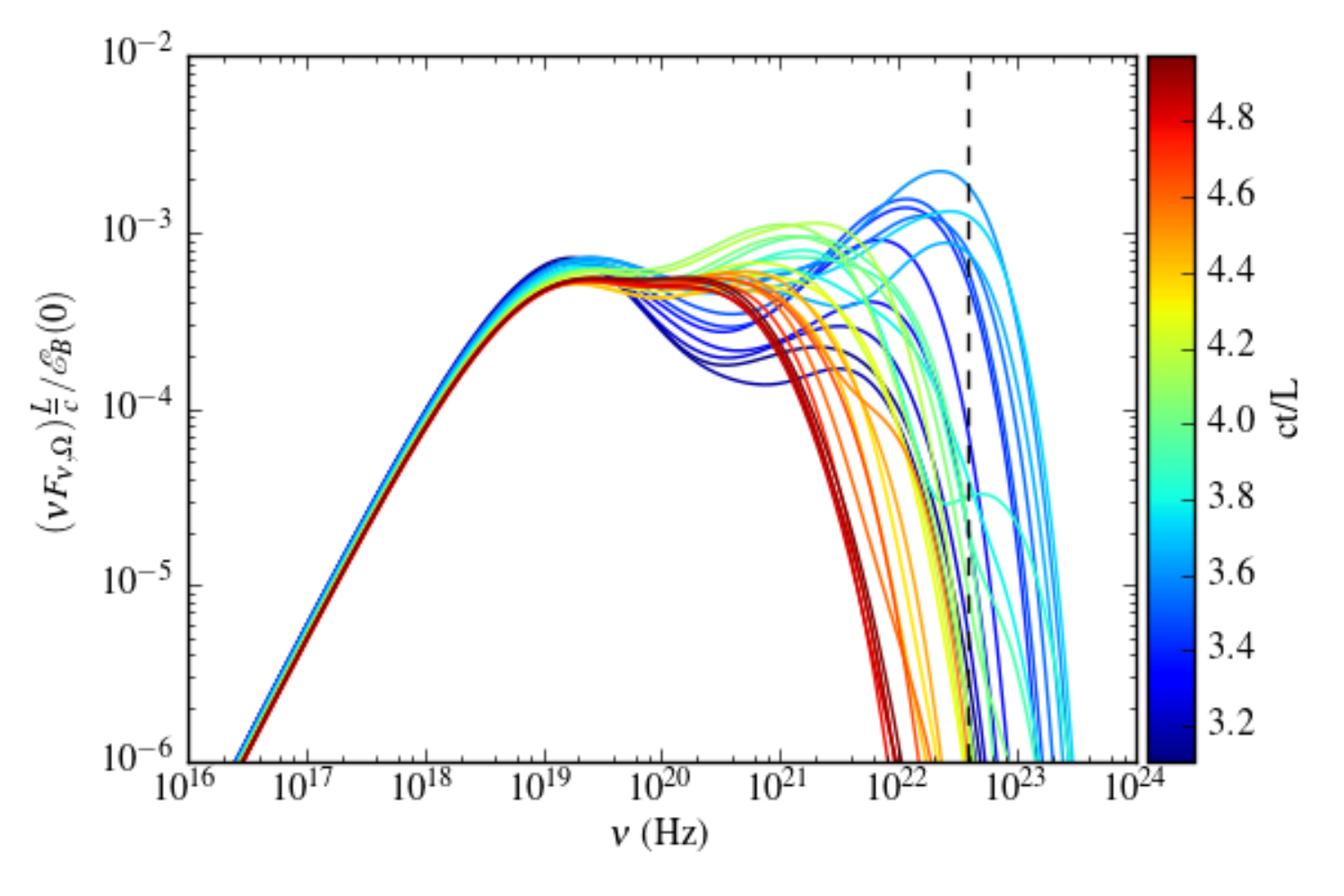}
    }
  \caption{Top: angular distribution of high energy synchrotron radiation (summed over frequency range $\omega>50\omega_0$, including contribution from both electrons and positrons) at a specific time point, for run 4 (left) and run 5 (right). Bottom: Instantaneous radiated spectrum in a $10^{\circ}\times10^{\circ}$ angular patch (shown by the white polygon in the top panel), at a series of simulation times equally spaced from $t=3.11L/c$ to $t=4.97L/c$, for run 4 (left) and run 5 (right). The power has been normalized such that the values we plot correspond to $\nu F_{\nu,\Omega}(L/c)/\mathcal{E}_{B}(0)$.}\label{fig:rad:timespectra_angular}
\end{figure*}

\begin{figure*}
  \centering
  \subfigure
    {
        \includegraphics[width=0.7\textwidth]{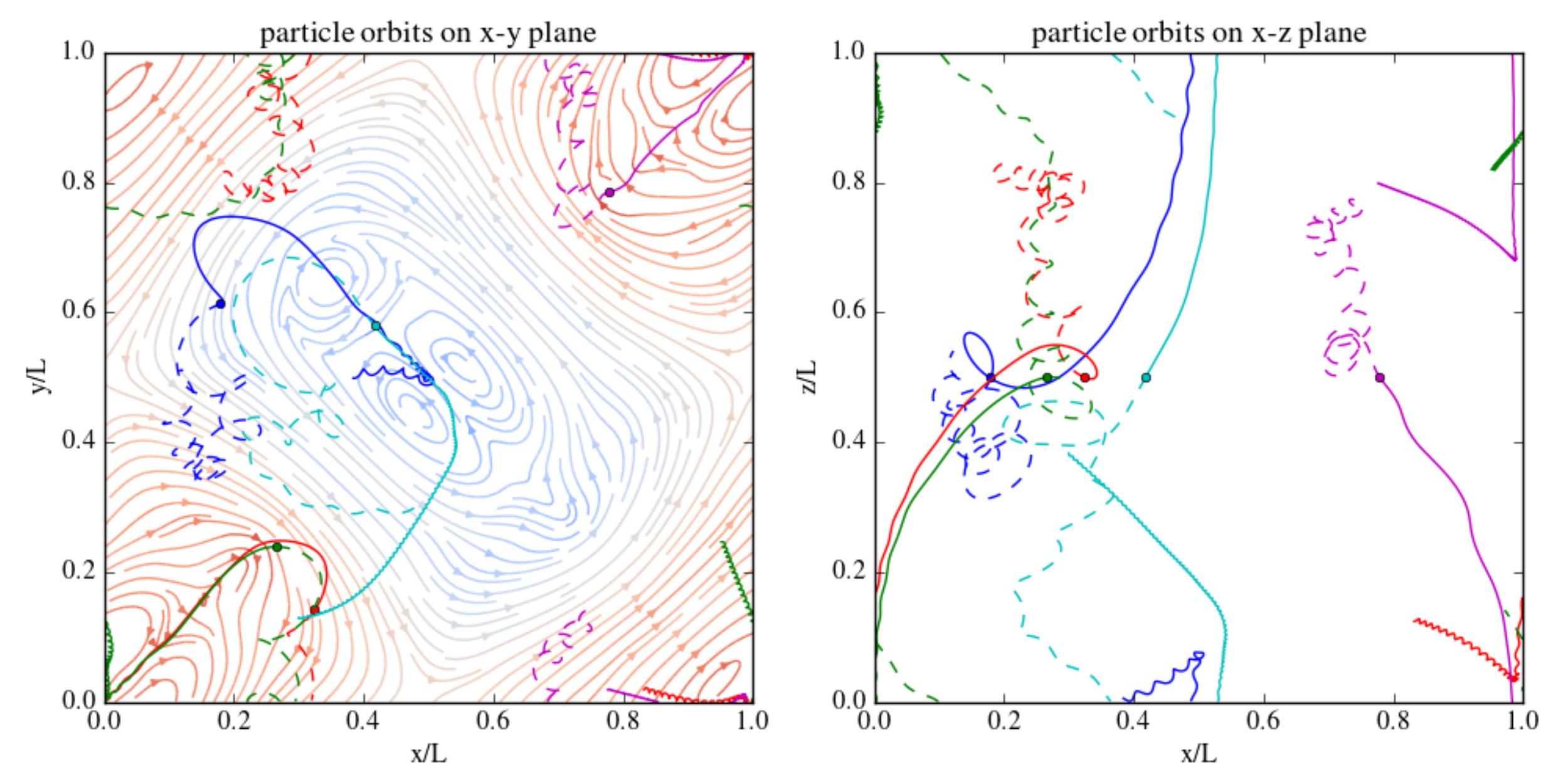}
    }\\
    \vspace{-0.5cm}
  \subfigure
    {
        \includegraphics[width=0.9\textwidth]{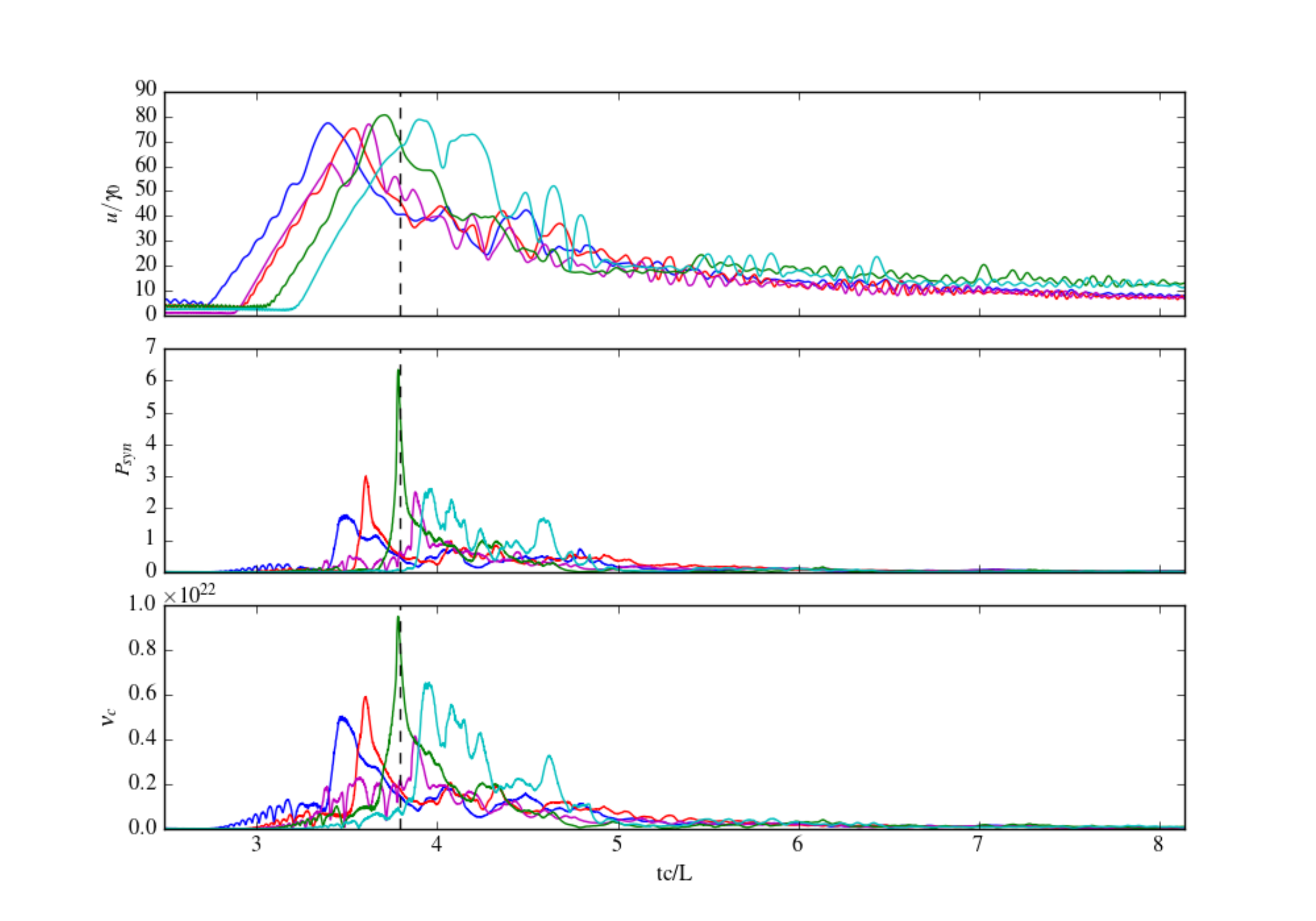}
    }
    \caption{From the strongly radiative run 4: history of selected particles that reach high energy during the simulation. Top panel shows the particle trajectory on the $x-y$ plane and $x-z$ plane, where the dots indicate particle locations at a specific time $t=3.79L/c$ (the field configuration at this time point is also shown on the $x-y$ plane); solid lines are past trajectories while dashed lines are future trajectories. The lower panels show particle energy, synchrotron power and characteristic synchrotron frequency.}\label{fig:rad:particle history}
\end{figure*}

\subsection{Dynamical consequences of radiation reaction}\label{subsec:dynamical}
In this section, we do a systematic comparison between the runs with different values of $\eta$.

\subsubsection{Total emitted energy}
A comparison of the total emitted power as a function of time for runs with different $\eta$ is shown in Figure \ref{fig:comparison:power-sigma}(a). In order to bring the synchrotron power $P_{syn}$ from different runs to the same scale, we have divided $P_{syn}$ by $\eta$ in the plot. It can be seen that when $\eta$ is sufficiently small ($\eta\lesssim10^{-6}$), the evolution of $P_{syn}$ is very similar, suggesting that in this regime radiation reaction does not affect the dynamics and the radiated energy simply scales with $\eta$. However, when $\eta$ is large enough such that accelerated particles get close to the radiation reaction limit, the scaling deviates from a linear relation with $\eta$: the back reaction of synchrotron radiation suppresses high energy particles, thus reduces the radiative output. We have tracked the energy for different components in the system, and the results are shown in Table \ref{table:energy partition}. The radiated energy is directly drawn from particle kinetic energy; in the high $\eta$ runs, the kinetic energy sees a negative growth due to the strong cooling.

In run 3 and 4, we also see sharp spikes in the received radiation light curves when the viewing directions are close to $\pm x$ or $\pm y$, with similar variation time $t_v\sim0.01L/c$ whereas the peak scales with $\eta$. From an observer's point of view, these spikes indicate radiative efficiency $\epsilon=\mathcal{E}_{spike}/\tilde{\mathcal{E}}_B\sim1$ and 10, respectively. 

\subsubsection{Evolution of the magnetization}
Figure \ref{fig:comparison:power-sigma}(b) shows the evolution of the ratio between the electromagnetic energy and particle kinetic energy $\tilde{\sigma}=(\mathcal{E}_B+\mathcal{E}_E)/\mathcal{E}_k$ as a function of time for runs with different $\eta$. When $t_{LC}/t_{\rm cool}\lesssim1\%$, the evolution of the instability heats up the particles, correspondingly we see a drop in $\tilde{\sigma}$. On the other hand, when $\eta$ starts out large such that the cooling time scale starts to be comparable with the dynamic time scale of the system, before the instability saturates the radiative cooling of particles reduces $\mathcal{E}_k$ while the field is still largely unchanged (since reducing the Lorentz factor of ultrarelativistic particles does not change the current), as a consequence $\tilde{\sigma}$ increases. The saturation of the instability again reduces $\tilde{\sigma}$, but the high level of radiative cooling keeps $\tilde{\sigma}$ from dropping to very low values.

\citet{Nalewajko:2016aa} have shown that the growth rate of the instability depends on $\bar{\sigma}_w$ (and possibly $a_0$). An effective increase in $\bar{\sigma}_w$ due to radiative cooling will affect the growth rate. We measured the growth rate $\omega_i$ during the linear evolution where the electric field can be written as $E=E_0e^{\omega_i tc/L}$; this is shown in Table \ref{table:energy partition}. The slight increase in growth rate can be attributed to increased effective magnetization as a result of radiative cooling. This is also consistent with the results of \citet{Cerutti:2013aa}, who reported that the reconnection rate increases in the strong cooling regime.

\subsubsection{Helicity}
In Figure \ref{fig:comparison:helicity}(a) we plot the relative change of helicity for runs with different $\eta$. It can be seen that for all the runs the deviation from helicity conservation is smaller than 0.6\%. Since the evolution of helicity is determined by $dH/dt=-2\int\mathbf{E}\cdot\mathbf{B}\,dV$, the negligible change in helicity means that the volume integral of $\mathbf{E}\cdot\mathbf{B}$ is small. We also checked the average and maximum values of $|\mathbf{E}\cdot\mathbf{B}|$ as a function of time, shown in Figure \ref{fig:comparison:helicity}. Although locally $|\mathbf{E}\cdot\mathbf{B}|$ can reach rather high values, the mean is small, indicating that the volume with non-negligible $\mathbf{E}\cdot\mathbf{B}$ is insignificant. And the evolution of $\mathbf{E}\cdot\mathbf{B}$ does not have any strong dependence on $\eta$. This seems to suggest that in the regime we've explored, synchrotron radiation reaction cannot support volumetric non-ideal electric field; the scale of the non-ideal region is still determined by the kinetic effects.

\subsubsection{Particle spectrum and radiation spectrum}
Radiation reaction limits the energy gain of the highest energy particles, as shown in Figure \ref{fig:rad:particle spectrum} where we plot the isotropic particle spectra at a series of simulation times for run 4. While the initial direct acceleration in the current layers produces a high energy bump in a similar fashion as run 1 (Figure \ref{fig:particle spectrum}), this new component loses its energy within roughly one light crossing time. Different portions of the light curve and time dependent synchrotron spectra, as seen by an observer located at $+x$, are shown in Figure \ref{fig:rad:timespectra}. Comparing with run 1 (Figure \ref{fig:timespectra}), we see that the peak emission frequency at the flux maximum is still efficiently boosted by $\sim3$ orders of magnitude in this case, although bumping close to the radiation reaction limit 160 MeV. This is because the direct electric field acceleration is fast enough and the radiative loss in the current layer is not significant, as we will show in the following subsection. 

We note that when the cooling is fast, observing directions on the $x-y$ plane may not be optimal for seeing the beamed highest energy emission. This is because the particle beams that are ejected from the current layers are initially pointing off the $x-y$ plane with projections on the $x-y$ plane along $\pm45^{\circ}$ (e.g., Figure \ref{fig:particleangular}); these particles then turn toward the $\pm x$, $\pm y$ directions, and the strongest radiation comes from some point in between. In Figure \ref{fig:rad:timespectra_angular}, we show a few examples of instantaneously radiated spectrum along specific directions off the $x-y$ plane, for run 4 and run 5. (This is not the spectrum received by an observer as we cannot define the receiving time unambiguously in this 2D case.) In particular, for the highest $\eta$ run we have (Figure \ref{fig:rad:timespectra_angular} right panel), the peak frequency of synchrotron radiation can reach the radiation reaction limit.

\subsubsection{Particle orbits}
In mildly radiative cases, we have shown in Figures \ref{fig:emissivity_2Dx} and \ref{fig:particle history} that the highest energy particles are initially accelerated by the parallel electric field in the current layers, where they follow Speiser orbits \citep[e.g.,][]{Uzdensky:2011aa,Cerutti:2013aa}, while at the same time being deflected toward the ends of the current layers. When they are deep in the current layer, the radiative loss is small due to reduced curvature as a result of $\mathbf{E}_{\parallel}$ acceleration; radiation only becomes significant when they get out of the current layer and start to get bent in stronger ambient magnetic field. Later on, during the large scale oscillation of the fields, some particles continue to gain energy when they bounce off an expanding magnetic domain from outside. In this phase, the acceleration is more stochastic.

When $\eta$ starts off large, the first acceleration phase, namely the action of parallel electric field, is still efficient enough to boost particles to high energies, but the second stochastic phase cannot compete with the cooling. Figure \ref{fig:rad:particle history} shows a few representative particle trajectories and their energy history. The peaks in $P_{syn}$ (e.g., near the vertical dashed line) are due to sudden increase in curvature when the particles get out of the current layer, similar to that in Figure \ref{fig:particle history}. The difference in this case is that particles are cooling significantly at the same time, so the peak drops off very quickly. After the dissolution of the current layer, all the high energy particles cool rapidly due to radiative loss and no efficient energization by stochastic processes is seen.

\section{Discussion}\label{sec:discussion}
\newtext{In our simulations, the initial configuration is an unstable force-free equilibrium that can get destroyed over a single dynamic time scale, so a natural question one might ask is whether such a structure can form in the first place. In a realistic astrophysical environment, the situation is much more complicated since there's continuous motion, energy/mass injection and/or loss, so the underlying equilibrium keeps evolving ceaselessly. We imagine that a few processes may build up plasma configurations of high magnetic free energy. The first example is that, in a pulsar wind, there's an equatorial current sheet, and random reconnection can happen at locations that are causally disconnected. This could lead to the development of highly tangled flux rope structures globally. As the flow expands, these tangled structures get frozen out as different parts stay out of causal connection. When the flow slows down eventually, they get back into causal contact and the configuration may look like some of the higher order equilibria. The second possibility is that, the pulsar wind is initially striped with a wavelength $\lambda_w$ much shorter than the termination shock radius $r_s$, e.g. $\lambda_w=10^9$ cm while $r_s=3\times10^{17}$ cm for the Crab. These stripes---small scale fluctuations---continue to go through inverse cascade, forming larger and larger flux ropes, eventually reaching a large enough scale that is relevant for the acceleration of PeV particles \citep{Zrake:2016ab}. Another possibility involves the polar jet which has relatively high magnetization and is kink unstable---suggesting that it possesses free magnetic energy. Recently \citet{Lyutikov:2016aa} also proposed that the intermediate latitude post shock flow can produce regions with high magnetization and current carrying filaments, resembling the flux tubes in force-free configurations.

Based on these considerations, we think it instructive to use the unstable force-free equilibria as a simple testbed to study the subsequent particle acceleration and radiation. With a kinetic approach, we are able to self-consistently extract the radiation signatures and study systematically the effect of radiation reaction. We find that the simple model is teaching us a lot about the generic properties of electromagnetic dissipation and radiation in magnetized, relativistic plasmas.}

The first remarkable feature is the rapid variability observed during the evolution from an initially smooth configuration. Though no current layers or other singular structures are embedded in the starting equilibrium state, its free energy is released through an ideal instability driven by the flux tubes' tendency to merge, which inevitably forces the formation of a current layer around the X-point. This is consistent with Syrovatskii's theory of current sheet formation via X-point collapse \citep{Imshennik:1967aa,Syrovatskii:1966aa,Nalewajko:2016aa}. Kinetic scales become important here: the thickness of the current layer is dictated by the plasma skin depth \citep{Nalewajko:2016aa}; the formation of plasmoids, in the situation of collisionless plasma, is also determined by kinetic processes---the tearing instability growth rate is roughly proportional to the plasma frequency $\omega_p$ \citep{Zelenyi:1979aa}, though the plasmoids can grow to macroscopic scales. The rapid variability in synchrotron radiation is produced by high energy particle beams that are ejected from the ends of the current layers, having been bunched by the tearing instability. This also leads to high energy radiation being much more variable than low energy radiation. The picture is in some sense similar to \cite{Cerutti:2013aa}. We have shown that for a single pulse, the apparent radiation efficiency can be relatively high, especially when $\eta$ is large. In particular, Figure \ref{fig:rad:particle spectrum} for run 4 ($\eta=2.75\times10^{-5}$) shows that the peak power per steradian $P_r$ seen by an observer at $x$ reaches $P_r\frac{L}{c}/\mathcal{E}_B(0)=0.01\sim0.02$ where $\mathcal{E}_B(0)$ is the total magnetic energy contained in the simulation domain at $t=0$, and the (isotropic) radiation efficiency is $\epsilon=\mathcal{E}_{spike}/\tilde{\mathcal{E}}_B\sim10$.

 This makes it attractive to associate these spikes with the Crab flares, but some caveats need to be taken into account. In the Crab flare case, we have shown that $\epsilon=\mathcal{E}_{rad}/\tilde{\mathcal{E}}_B\sim10^3 L_{36}B_{-3}^{-2}t_{10 hr}^{-2}$. This is still about two orders of magnitude larger than the maximum we got from the current simulations, and we may need to extrapolate the results using simulations with different magnetization, box sizes and $\eta$ in future studies. \newtext{In addition, if we do make a literal identification between a single spike from the simulation and a Crab flare event, this would mean that the size of the emitting PeV particle bunch is on the order of $\ell=ct_v=10^{15}t_{10 hr}\,\rm{cm}$---comparable to the Larmor radius of the PeV particles $r_L=1.7\times10^{15}\gamma_9B_{-3}^{-1}$ cm, and the simulation box would correspond to a length scale of $10^{17}$ cm---comparable to the radius of the termination shock. Such a demanding requirement is related to the intrinsic problem of insufficient energy contained in the kinetic scale beams in our current simulations. Recalling that the energy in the beams comes from the volume that has been processed by the current layers, one would hope that going to large $\sigma$ limit where both the instability growth rate and the magnetic energy density are larger, the energy deposited into the high energy particle beams would be more promising. At the same time, we should keep in mind that the scale separation in our simulation is still quite far from the astrophysical reality. In the Crab, the pressure is dominated by TeV particles, whose gyro radius is $r_L\sim10^{12}\gamma_6B_{-3}^{-1}$ cm, and the plasma skin depth $d_e\sim\sqrt{\sigma}r_L$ would be much smaller than $\ell$ (unless $\sigma\gtrsim10^6$). In contrast, our simulation has a box size $L$ that is only $800r_L=413d_e$ and the emitting region size only a few $d_e$. (A larger $L/d_e$ also means larger charge multiplicity for a fixed $\sigma$. The multiplicity $\kappa$ is defined as the ratio between the actual number of pairs and the minimum number needed to support the current. In our setup essentially $\kappa=3/a$, see Equation \ref{eq:ne}.) It is desirable to test the regime of large $L/d_e$ in the future with additional computational power \citep[c.f.][but it's a quite different setup with Harris current layer and cold, nonrelativistic background plasma]{Sironi:2016aa}. One further catch is the relatively soft spectral index $p\sim2.5$ for the nonthermal particle distribution (isotropic) we get here. The biggest Crab flare has a spectrum $F_{\nu}\propto\nu^{-0.27}$ \cite{Buehler:2012aa}, which requires a particle distribution with $p\sim1.5$, or mono-energetic if the multi-wavelength constraint is taken into account \citep{Weisskopf:2013aa}. Though the angle dependent particle distribution could have a harder spectrum (\S\ref{subsubsec:time dependent spectrum}), the actual anisotropy needs to be tested using more realistic scale separation. \citet{Nalewajko:2016aa} found that the spectral index $p$ decreases as $\sigma$ increases for a similar setup, and such a trend is also true in Harris layer reconnection \citep{Sironi:2014aa, Guo:2014aa, Werner:2016aa}, so higher $\sigma$ simulations are promising.}
 
The second important feature we notice is that as the dominant acceleration mechanism here is parallel electric field acceleration, the curvature of accelerated particles becomes small when they are in the current layer and significant synchrotron radiation only takes place after they get out of the current layer. Such a separation between the acceleration site and radiation site would allow the highest energy particles to get beyond the radiation reaction limit \citep{Uzdensky:2011aa,Cerutti:2013aa}. The premise is that the available electric potential should be enough to boost particle energy from the thermal sea to the radiation reaction limit. This is challenging to realize in PIC studies due to the limited dynamic range. In our simulations we have chosen a modest sigma and the relaxation process is able to produce highest energy particles with $\gamma\sim100\gamma_0$, so in order to test whether we can get over the radiation reaction limit, we tried to artificially increase $\eta$ for the thermal sea, or equivalently, increase the energy of the thermal particles. We find that for $\eta$ as large as $3\times10^{-5}$, we can get high energy synchrotron radiation instantaneously peaking at the radiation reaction limit. However, when $\eta$ starts out larger than $\sim3\times10^{-5}$, the thermal population would cool down back to around $\eta\sim10^{-5}$ at the saturation of the instability so it no longer makes sense to further increase $\eta$ at the beginning of the simulation. As a comparison, for the Crab, if the majority of the particles are at $1-10$ TeV energy range and the average magnetic field is $\sim$1 mG, then $\eta\sim10^{-7}-10^{-5}$. It remains to be seen whether higher $\sigma$ configurations could do a much better job. We expect this to be the case as \citet{Nalewajko:2016aa} find that both the fraction of non-thermal particles and the maximum energy of accelerated particles scale with $\sigma$. This will be tested in the near future.

As we change the cooling parameter $\eta$, we did a comparison among the various cooling regimes to see the effects of radiation reaction on the dynamics. The strong synchrotron cooling does have a significant impact on the highest energy particles, but since they are not energetically dominant, this hardly has any effect on the global dynamics. We tried to decompose the non-ideal electric field using a generalized Ohm's law (not shown here), and find that synchrotron cooling is not contributing any noticeable resistivity in the regime we've explored. This is partly because the majority of particles are not yet reaching the radiation reaction limit so that the radiation reaction force on any plasma volume remains much smaller compared to the other force terms (inertia, pressure gradient and Lorentz force), partly because the location where non-ideal electric field arises are not locations where synchrotron radiation is most significant. The situation could be quite different if the particles that are counter streaming along z in the current layer excite gyro-resonance instability, which enhances the synchrotron radiation within the current layer, or if the dominant radiation mechanism is not synchrotron but inverse Compton---in these cases radiation reaction force might contribute significantly in supporting non-ideal electric field over some extended volume \citep[e.g.,][]{Uzdensky:2016aa}.

\newtext{Besides the modest $\sigma$ and modest scale separation $L/d_e$, another limitation in our current simulations is the 2D constraint. It does not allow for any kink instability or other variations along z direction \citep[cf.][]{Cerutti:2014aa}, and the acceleration length along z is unlimited. In addition, the accelerated particles are counter streaming in the z direction---this could lead to excitation of collective modes (e.g. gyroresonance) that contribute to anomalous resistivity to support non-ideal electric field \citep[e.g.][]{Treumann:1997aa}. Also, in the strongly cooling runs, the most interesting directions for fast time variability and efficient synchrotron production might be out of the simulation plane. These will need to be tested in 3D simulations. However, 3D runs are much more computationally expensive at the moment. We plan to carry this out in the near future.}
%Here as a rough guide of practice, we give a simple scaling between the computational cost and $\sigma$, $\Lambda\equiv L/d_e$, for both 2D and 3D. We consider a similar setup as that described in \S\ref{sec:setup}, keeping $r_L/\Delta x$ and the number of particles per cell the same, then the number of grid points on each side of the box is $L/\Delta x\propto\sigma/a_0$, and $\Lambda$ scales as $\Lambda\propto\sqrt{\sigma}/a_0$, where $a_0$ is the multiplicity parameter ($|a_0|\le0.5$ has to be satisfied). Therefore, in 2D, the computational cost scales as $(\sigma/a_0)^3$ or $(\sqrt{\sigma}\Lambda)^3$, while in 3D it scales as $(\sigma/a_0)^4$ or $(\sqrt{\sigma}\Lambda)^4$.

\section{Conclusions}\label{sec:conclusion}
We have performed 2D PIC simulations of a magnetostatic equilibrium that belongs to the lowest order unstable force-free states. We work in the regime where the individual particles are ultrarelativistic, and include synchrotron radiation reaction self-consistently. A benchmark example, where $\bar{\sigma}_w=3.76$ and $L/r_L=800$, is examined in detail to obtain the radiation signatures, and a systematic study of different cooling regimes is performed to understand the dynamical effects of radiation reaction. 

We find that the evolution of the system is consistent with previous force-free, MHD \citep{East:2015aa} and PIC simulations \citep[][mildly relativistic cases]{Nalewajko:2016aa}. The ideal instability eventually leads to current sheet formation, where most of the particle acceleration and electromagnetic dissipation happens. Regions with $\mathbf{E}\cdot\mathbf{B}\ne0$ develop in the current layers, but the volume they occupy is negligibly small. As a result, helicity is pretty well conserved during the whole evolution. We also do not see regions with $E>B$ develop, because of the advection of guide field into the current layers.

The highest energy particles are first accelerated in the current layers by the parallel electric field, where they do not radiate much due to the small curvature of their trajectory, despite the presence of guide field in the current layer. Most of the radiation is produced when particles are ejected from the current layers---their trajectories start to bend significantly in the ambient magnetic field which changes direction at the end of the current layer. Such a separation between acceleration site and synchrotron radiation site could in principle facilitate acceleration beyond the synchrotron radiation reaction limit.

We find that the fastest variability in synchrotron radiation is produced when compact plasmoids that contain high energy particles are ejected from the ends of the current layer and get destroyed. These give beamed radiation as the particles released from the plasmoids start to turn in the ambient magnetic field. An observer sees high intensity radiation when the beam happens to be aligned with the line of sight. As a result, the high energy radiation is much more variable than the low energy radiation, and these flares are accompanied by an increase in the polarization degree and rapid change of polarization angle in the high energy band. The variability time scale is determined by the spatial extent of the emitting structure. In our simulations, this can be as short as $t_{v}=0.01L/c$, and the peak flux per steradian can reach $P_r\frac{L}{c}/\mathcal{E}_B(0)=0.01\sim0.02$ in the case $\eta=2.75\times10^{-5}$, giving an apparent radiation efficiency $\epsilon\sim10$. However, the total energy involved in these spikes is small. Though this setup, with the parameters that were numerically tractable here, is not yet enough to directly reproduce the feature of the Crab flares, this work suggests that runs at higher $\sigma$ are promising. We plan to explore this in future work.

\acknowledgments

We thank Beno\^{i}t Cerutti, Frederico Fiuza, Maxim Lyutikov and Lorenzo Sironi for helpful discussions. We also acknowledge the anonymous referee for many useful suggestions that helped to improve the paper. This work was supported in part by the U.S. Department of Energy contract to SLAC no. DE-AC02-76SF00515, NSF grant AST 12-12195, the Simons Foundation, the Humboldt Foundation, and the Miller Foundation (RB). YY gratefully acknowledges support from the KIPAC Gregory and Mary Chabolla fellowship and the Gabilan Fellowship awarded by Stanford University. KN was supported by the Polish National Science Centre grant 2015/18/E/ST9/00580, and by NASA through Einstein Postdoctoral Fellowship grant number PF3-140130 awarded by the Chandra X-ray Center, which is operated by the Smithsonian Astrophysical Observatory for NASA under contract NAS8-03060. Simulations were run on the Bullet Cluster at SLAC.

\bibliography{CrabFlare}

\begin{thebibliography}{47}
\expandafter\ifx\csname natexlab\endcsname\relax\def\natexlab#1{#1}\fi

\bibitem[{{Abdo} {et~al.}(2010){Abdo}, {Ackermann}, {Ajello}, {Axelsson},
  {Baldini}, {Ballet}, {Barbiellini}, {Bastieri}, {Baughman}, {Bechtol}, \&
  et~al.}]{Abdo:2010-3C279polarization}
{Abdo}, A.~A., {et~al.} 2010, \nat, 463, 919

\bibitem[{Abdo {et~al.}(2011)Abdo, Ackermann, Ajello, Allafort, Baldini,
  Ballet, Barbiellini, Bastieri, Bechtol, Bellazzini, Berenji, Blandford,
  Bloom, Bonamente, Borgland, Bouvier, Brandt, Bregeon, Brez, Brigida, Bruel,
  Buehler, Buson, Caliandro, Cameron, Cannon, Caraveo, Casandjian, {\c C}elik,
  Charles, Chekhtman, Cheung, Chiang, Ciprini, Claus, Cohen-Tanugi, Costamante,
  Cutini, D'Ammando, Dermer, de~Angelis, de~Luca, de~Palma, Digel, do~Couto~e
  Silva, Drell, Drlica-Wagner, Dubois, Dumora, Favuzzi, Fegan, Ferrara, Focke,
  Fortin, Frailis, Fukazawa, Funk, Fusco, Gargano, Gasparrini, Gehrels,
  Germani, Giglietto, Giordano, Giroletti, Glanzman, Godfrey, Grenier, Grondin,
  Grove, Guiriec, Hadasch, Hanabata, Harding, Hayashi, Hayashida, Hays, Horan,
  Itoh, J{\'o}hannesson, Johnson, Johnson, Khangulyan, Kamae, Katagiri,
  Kataoka, Kerr, Kn{\"o}dlseder, Kuss, Lande, Latronico, Lee, Lemoine-Goumard,
  Longo, Loparco, Lubrano, Madejski, Makeev, Marelli, Mazziotta, McEnery,
  Michelson, Mitthumsiri, Mizuno, Moiseev, Monte, Monzani, Morselli,
  Moskalenko, Murgia, Nakamori, Naumann-Godo, Nolan, Norris, Nuss, Ohsugi,
  Okumura, Omodei, Ormes, Ozaki, Paneque, Parent, Pelassa, Pepe, Pesce-Rollins,
  Pierbattista, Piron, Porter, Rain{\`o}, Rando, Ray, Razzano, Reimer, Reimer,
  Reposeur, Ritz, Romani, Sadrozinski, Sanchez, Parkinson, Scargle, Schalk,
  Sgr{\`o}, Siskind, Smith, Spandre, Spinelli, Strickman, Suson, Takahashi,
  Takahashi, Tanaka, Thayer, Thompson, Tibaldo, Torres, Tosti, Tramacere,
  Troja, Uchiyama, Vandenbroucke, Vasileiou, Vianello, Vitale, Wang, Wood,
  Yang, \& Ziegler}]{Abdo:2011aa-Crab}
Abdo, A.~A., {et~al.} 2011, Science, 331, 739

\bibitem[{{Aharonian} {et~al.}(2007){Aharonian}, {Akhperjanian}, {Bazer-Bachi},
  {Behera}, {Beilicke}, {Benbow}, {Berge}, {Bernl{\"o}hr}, {Boisson}, {Bolz},
  {Borrel}, {Boutelier}, {Braun}, {Brion}, {Brown}, {B{\"u}hler},
  {B{\"u}sching}, {Bulik}, {Carrigan}, {Chadwick}, {Clapson}, {Chounet},
  {Coignet}, {Cornils}, {Costamante}, {Degrange}, {Dickinson},
  {Djannati-Ata{\"\i}}, {Domainko}, {Drury}, {Dubus}, {Dyks}, {Egberts},
  {Emmanoulopoulos}, {Espigat}, {Farnier}, {Feinstein}, {Fiasson},
  {F{\"o}rster}, {Fontaine}, {Funk}, {Funk}, {F{\"u}{\ss}ling}, {Gallant},
  {Giebels}, {Glicenstein}, {Gl{\"u}ck}, {Goret}, {Hadjichristidis}, {Hauser},
  {Hauser}, {Heinzelmann}, {Henri}, {Hermann}, {Hinton}, {Hoffmann}, {Hofmann},
  {Holleran}, {Hoppe}, {Horns}, {Jacholkowska}, {de Jager}, {Kendziorra},
  {Kerschhaggl}, {Kh{\'e}lifi}, {Komin}, {Kosack}, {Lamanna}, {Latham}, {Le
  Gallou}, {Lemi{\`e}re}, {Lemoine-Goumard}, {Lenain}, {Lohse}, {Martin},
  {Martineau-Huynh}, {Marcowith}, {Masterson}, {Maurin}, {McComb}, {Moderski},
  {Moulin}, {de Naurois}, {Nedbal}, {Nolan}, {Olive}, {Orford}, {Osborne},
  {Ostrowski}, {Panter}, {Pedaletti}, {Pelletier}, {Petrucci}, {Pita},
  {P{\"u}hlhofer}, {Punch}, {Ranchon}, {Raubenheimer}, {Raue}, {Rayner},
  {Renaud}, {Ripken}, {Rob}, {Rolland}, {Rosier-Lees}, {Rowell}, {Rudak},
  {Ruppel}, {Sahakian}, {Santangelo}, {Saug{\'e}}, {Schlenker}, {Schlickeiser},
  {Schr{\"o}der}, {Schwanke}, {Schwarzburg}, {Schwemmer}, {Shalchi}, {Sol},
  {Spangler}, {Stawarz}, {Steenkamp}, {Stegmann}, {Superina}, {Tam},
  {Tavernet}, {Terrier}, {van Eldik}, {Vasileiadis}, {Venter}, {Vialle},
  {Vincent}, {Vivier}, {V{\"o}lk}, {Volpe}, {Wagner}, {Ward}, \&
  {Zdziarski}}]{Aharonian:2007aa}
{Aharonian}, F., {et~al.} 2007, \apjl, 664, L71

\bibitem[{{Albert} {et~al.}(2007){Albert}, {Aliu}, {Anderhub}, {Antoranz},
  {Armada}, {Baixeras}, {Barrio}, {Bartko}, {Bastieri}, {Becker}, {Bednarek},
  {Berger}, {Bigongiari}, {Biland}, {Bock}, {Bordas}, {Bosch-Ramon}, {Bretz},
  {Britvitch}, {Camara}, {Carmona}, {Chilingarian}, {Coarasa}, {Commichau},
  {Contreras}, {Cortina}, {Costado}, {Curtef}, {Danielyan}, {Dazzi}, {De
  Angelis}, {Delgado}, {de los Reyes}, {De Lotto}, {Domingo-Santamar{\'{\i}}a},
  {Dorner}, {Doro}, {Errando}, {Fagiolini}, {Ferenc}, {Fern{\'a}ndez}, {Firpo},
  {Flix}, {Fonseca}, {Font}, {Fuchs}, {Galante}, {Garc{\'{\i}}a-L{\'o}pez},
  {Garczarczyk}, {Gaug}, {Giller}, {Goebel}, {Hakobyan}, {Hayashida},
  {Hengstebeck}, {Herrero}, {H{\"o}hne}, {Hose}, {Hrupec}, {Hsu}, {Jacon},
  {Jogler}, {Kosyra}, {Kranich}, {Kritzer}, {Laille}, {Lindfors}, {Lombardi},
  {Longo}, {L{\'o}pez}, {L{\'o}pez}, {Lorenz}, {Majumdar}, {Maneva},
  {Mannheim}, {Mansutti}, {Mariotti}, {Mart{\'{\i}}nez}, {Mazin}, {Merck},
  {Meucci}, {Meyer}, {Miranda}, {Mirzoyan}, {Mizobuchi}, {Moralejo}, {Nieto},
  {Nilsson}, {Ninkovic}, {O{\~n}a-Wilhelmi}, {Otte}, {Oya}, {Paneque},
  {Panniello}, {Paoletti}, {Paredes}, {Pasanen}, {Pascoli}, {Pauss}, {Pegna},
  {Persic}, {Peruzzo}, {Piccioli}, {Prandini}, {Puchades}, {Raymers}, {Rhode},
  {Rib{\'o}}, {Rico}, {Rissi}, {Robert}, {R{\"u}gamer}, {Saggion}, {Saito},
  {S{\'a}nchez}, {Sartori}, {Scalzotto}, {Scapin}, {Schmitt}, {Schweizer},
  {Shayduk}, {Shinozaki}, {Shore}, {Sidro}, {Sillanp{\"a}{\"a}}, {Sobczynska},
  {Stamerra}, {Stark}, {Takalo}, {Tavecchio}, {Temnikov}, {Tescaro}, {Teshima},
  {Torres}, {Turini}, {Vankov}, {Vitale}, {Wagner}, {Wibig}, {Wittek},
  {Zandanel}, {Zanin}, \& {Zapatero}}]{Albert:2007aa}
{Albert}, J., {et~al.} 2007, \apj, 669, 862

\bibitem[{{Aleksi{\'c}} {et~al.}(2011){Aleksi{\'c}}, {Antonelli}, {Antoranz},
  {Backes}, {Barrio}, {Bastieri}, {Becerra Gonz{\'a}lez}, {Bednarek},
  {Berdyugin}, {Berger}, {Bernardini}, {Biland}, {Blanch}, {Bock}, {Boller},
  {Bonnoli}, {Borla Tridon}, {Braun}, {Bretz}, {Ca{\~n}ellas}, {Carmona},
  {Carosi}, {Colin}, {Colombo}, {Contreras}, {Cortina}, {Cossio}, {Covino},
  {Dazzi}, {De Angelis}, {De Cea del Pozo}, {De Lotto}, {Delgado Mendez},
  {Diago Ortega}, {Doert}, {Dom{\'{\i}}nguez}, {Dominis Prester}, {Dorner},
  {Doro}, {Elsaesser}, {Ferenc}, {Fonseca}, {Font}, {Fruck}, {Garc{\'{\i}}a
  L{\'o}pez}, {Garczarczyk}, {Garrido}, {Giavitto}, {Godinovi{\'c}}, {Hadasch},
  {H{\"a}fner}, {Herrero}, {Hildebrand}, {H{\"o}hne-M{\"o}nch}, {Hose},
  {Hrupec}, {Huber}, {Jogler}, {Klepser}, {Kr{\"a}henb{\"u}hl}, {Krause}, {La
  Barbera}, {Lelas}, {Leonardo}, {Lindfors}, {Lombardi}, {L{\'o}pez}, {Lorenz},
  {Makariev}, {Maneva}, {Mankuzhiyil}, {Mannheim}, {Maraschi}, {Mariotti},
  {Mart{\'{\i}}nez}, {Mazin}, {Meucci}, {Miranda}, {Mirzoyan}, {Miyamoto},
  {Mold{\'o}n}, {Moralejo}, {Nieto}, {Nilsson}, {Orito}, {Oya}, {Paneque},
  {Paoletti}, {Pardo}, {Paredes}, {Partini}, {Pasanen}, {Pauss},
  {Perez-Torres}, {Persic}, {Peruzzo}, {Pilia}, {Pochon}, {Prada}, {Prada
  Moroni}, {Prandini}, {Puljak}, {Reichardt}, {Reinthal}, {Rhode}, {Rib{\'o}},
  {Rico}, {R{\"u}gamer}, {Saggion}, {Saito}, {Saito}, {Salvati}, {Satalecka},
  {Scalzotto}, {Scapin}, {Schultz}, {Schweizer}, {Shayduk}, {Shore},
  {Sillanp{\"a}{\"a}}, {Sitarek}, {Sobczynska}, {Spanier}, {Spiro}, {Stamerra},
  {Steinke}, {Storz}, {Strah}, {Suri{\'c}}, {Takalo}, {Tavecchio}, {Temnikov},
  {Terzi{\'c}}, {Tescaro}, {Teshima}, {Thom}, {Tibolla}, {Torres}, {Treves},
  {Vankov}, {Vogler}, {Wagner}, {Weitzel}, {Zabalza}, {Zandanel}, {Zanin},
  {MAGIC Collaboration}, {Tanaka}, {Wood}, \& {Buson}}]{Aleksic:2011aa}
{Aleksi{\'c}}, J., {et~al.} 2011, \apjl, 730, L8

\bibitem[{{Aleksi{\'c}} {et~al.}(2014){Aleksi{\'c}}, {Ansoldi}, {Antonelli},
  {Antoranz}, {Babic}, {Bangale}, {Barrio}, {Gonz{\'a}lez}, {Bednarek},
  {Bernardini}, {Biasuzzi}, {Biland}, {Blanch}, {Bonnefoy}, {Bonnoli},
  {Borracci}, {Bretz}, {Carmona}, {Carosi}, {Colin}, {Colombo}, {Contreras},
  {Cortina}, {Covino}, {Da Vela}, {Dazzi}, {De Angelis}, {De Caneva}, {De
  Lotto}, {Wilhelmi}, {Mendez}, {Prester}, {Dorner}, {Doro}, {Einecke},
  {Eisenacher}, {Elsaesser}, {Fonseca}, {Font}, {Frantzen}, {Fruck}, {Galindo},
  {L{\'o}pez}, {Garczarczyk}, {Terrats}, {Gaug}, {Godinovi{\'c}}, {Mu{\~n}oz},
  {Gozzini}, {Hadasch}, {Hanabata}, {Hayashida}, {Herrera}, {Hildebrand},
  {Hose}, {Hrupec}, {Idec}, {Kadenius}, {Kellermann}, {Kodani}, {Konno},
  {Krause}, {Kubo}, {Kushida}, {La Barbera}, {Lelas}, {Lewandowska},
  {Lindfors}, {Lombardi}, {Longo}, {L{\'o}pez}, {L{\'o}pez-Coto},
  {L{\'o}pez-Oramas}, {Lorenz}, {Lozano}, {Makariev}, {Mallot}, {Maneva},
  {Mankuzhiyil}, {Mannheim}, {Maraschi}, {Marcote}, {Mariotti},
  {Mart{\'{\i}}nez}, {Mazin}, {Menzel}, {Miranda}, {Mirzoyan}, {Moralejo},
  {Munar-Adrover}, {Nakajima}, {Niedzwiecki}, {Nilsson}, {Nishijima}, {Noda},
  {Orito}, {Overkemping}, {Paiano}, {Palatiello}, {Paneque}, {Paoletti},
  {Paredes}, {Paredes-Fortuny}, {Persic}, {Poutanen}, {Moroni}, {Prandini},
  {Puljak}, {Reinthal}, {Rhode}, {Rib{\'o}}, {Rico}, {Garcia}, {R{\"u}gamer},
  {Saito}, {Saito}, {Satalecka}, {Scalzotto}, {Scapin}, {Schultz}, {Schweizer},
  {Shore}, {Sillanp{\"a}{\"a}}, {Sitarek}, {Snidaric}, {Sobczynska}, {Spanier},
  {Stamatescu}, {Stamerra}, {Steinbring}, {Storz}, {Strzys}, {Takalo},
  {Takami}, {Tavecchio}, {Temnikov}, {Terzi{\'c}}, {Tescaro}, {Teshima},
  {Thaele}, {Tibolla}, {Torres}, {Toyama}, {Treves}, {Uellenbeck}, {Vogler},
  {Zanin}, {Kadler}, {Schulz}, {Ros}, {Bach}, {Krau{\ss}}, \&
  {Wilms}}]{Aleksic:2014IC310}
---. 2014, Science, 346, 1080

\bibitem[{Arons(2012)}]{Arons:2012aa}
Arons, J. 2012, Space Science Reviews, 173, 341

\bibitem[{{Baty} {et~al.}(2013){Baty}, {Petri}, \& {Zenitani}}]{Baty:2013aa}
{Baty}, H., {Petri}, J., \& {Zenitani}, S. 2013, \mnras, 436, L20

\bibitem[{Blandford {et~al.}(2015)Blandford, East, Nalewajko, Yuan, \&
  Zrake}]{Blandford:2015aa}
Blandford, R., East, W., Nalewajko, K., Yuan, Y., \& Zrake, J. 2015,
  arXiv:1511.07515 [astro-ph], arXiv: 1511.07515

\bibitem[{Blandford {et~al.}(2014)Blandford, Simeon, \&
  Yuan}]{Blandford:2014aa}
Blandford, R., Simeon, P., \& Yuan, Y. 2014, Nuclear Physics B Proceedings
  Supplements, 256, 9

\bibitem[{{Blinov} {et~al.}(2015){Blinov}, {Pavlidou}, {Papadakis},
  {Kiehlmann}, {Panopoulou}, {Liodakis}, {King}, {Angelakis}, {Balokovi{\'c}},
  {Das}, {Feiler}, {Fuhrmann}, {Hovatta}, {Khodade}, {Kus}, {Kylafis},
  {Mahabal}, {Myserlis}, {Modi}, {Pazderska}, {Pazderski}, {Papamastorakis},
  {Pearson}, {Rajarshi}, {Ramaprakash}, {Reig}, {Readhead}, {Tassis}, \&
  {Zensus}}]{Blinov:2015aa}
{Blinov}, D., {et~al.} 2015, \mnras, 453, 1669

\bibitem[{Buehler \& Blandford(2014)}]{Buhler:2014aa}
Buehler, R., \& Blandford, R. 2014, Reports on Progress in Physics, 77, 6901

\bibitem[{Buehler {et~al.}(2012)Buehler, Scargle, Blandford, Baldini, Baring,
  Belfiore, Charles, Chiang, D'Ammando, Dermer, Funk, Grove, Harding, Hays,
  Kerr, Massaro, Mazziotta, Romani, Saz~Parkinson, Tennant, \&
  Weisskopf}]{Buehler:2012aa}
Buehler, R., {et~al.} 2012, The Astrophysical Journal, 749, 26

\bibitem[{Cerutti {et~al.}(2012)Cerutti, Werner, Uzdensky, \&
  Begelman}]{Cerutti:2012aa}
Cerutti, B., Werner, G.~R., Uzdensky, D.~A., \& Begelman, M.~C. 2012, The
  Astrophysical Journal Letters, 754, L33

\bibitem[{{Cerutti} {et~al.}(2013){Cerutti}, {Werner}, {Uzdensky}, \&
  {Begelman}}]{Cerutti:2013aa}
{Cerutti}, B., {Werner}, G.~R., {Uzdensky}, D.~A., \& {Begelman}, M.~C. 2013,
  \apj, 770, 147

\bibitem[{{Cerutti} {et~al.}(2014){Cerutti}, {Werner}, {Uzdensky}, \&
  {Begelman}}]{Cerutti:2014aa}
---. 2014, \apj, 782, 104

\bibitem[{{Courant} {et~al.}(1967){Courant}, {Friedrichs}, \&
  {Lewy}}]{Courant:1967aa}
{Courant}, R., {Friedrichs}, K., \& {Lewy}, H. 1967, IBM Journal of Research
  and Development, 11, 215

\bibitem[{{East} {et~al.}(2015){East}, {Zrake}, {Yuan}, \&
  {Blandford}}]{East:2015aa}
{East}, W.~E., {Zrake}, J., {Yuan}, Y., \& {Blandford}, R.~D. 2015, Physical
  Review Letters, 115, 095002

\bibitem[{{Giannios} {et~al.}(2009){Giannios}, {Uzdensky}, \&
  {Begelman}}]{Giannios:2009aa}
{Giannios}, D., {Uzdensky}, D.~A., \& {Begelman}, M.~C. 2009, \mnras, 395, L29

\bibitem[{{Guo} {et~al.}(2014){Guo}, {Li}, {Daughton}, \& {Liu}}]{Guo:2014aa}
{Guo}, F., {Li}, H., {Daughton}, W., \& {Liu}, Y.-H. 2014, Physical Review
  Letters, 113, 155005

\bibitem[{{Hayashida} {et~al.}(2015){Hayashida}, {Nalewajko}, {Madejski},
  {Sikora}, {Itoh}, {Ajello}, {Blandford}, {Buson}, {Chiang}, {Fukazawa},
  {Furniss}, {Urry}, {Hasan}, {Harrison}, {Alexander}, {Balokovi{\'c}},
  {Barret}, {Boggs}, {Christensen}, {Craig}, {Forster}, {Giommi},
  {Grefenstette}, {Hailey}, {Hornstrup}, {Kitaguchi}, {Koglin}, {Madsen},
  {Mao}, {Miyasaka}, {Mori}, {Perri}, {Pivovaroff}, {Puccetti}, {Rana},
  {Stern}, {Tagliaferri}, {Westergaard}, {Zhang}, {Zoglauer}, {Gurwell},
  {Uemura}, {Akitaya}, {Kawabata}, {Kawaguchi}, {Kanda}, {Moritani}, {Takaki},
  {Ui}, {Yoshida}, {Agarwal}, \& {Gupta}}]{Hayashida:2015aa}
{Hayashida}, M., {et~al.} 2015, \apj, 807, 79

\bibitem[{{Imshennik} \& {Syrovatski{\v \i}}(1967)}]{Imshennik:1967aa}
{Imshennik}, V.~S., \& {Syrovatski{\v \i}}, S.~I. 1967, Soviet Journal of
  Experimental and Theoretical Physics, 25, 656

\bibitem[{Jackson(1999)}]{Jackson:1999aa}
Jackson, J.~D. 1999, Classical electrodynamics, 3rd edn. (New York: Wiley)

\bibitem[{Kagan {et~al.}(2016)Kagan, Nakar, \& Piran}]{kagan_beaming_2016}
Kagan, D., Nakar, E., \& Piran, T. 2016, arXiv:1601.07349 [astro-ph], arXiv:
  1601.07349

\bibitem[{{Kagan} {et~al.}(2015){Kagan}, {Sironi}, {Cerutti}, \&
  {Giannios}}]{Kagan:2015aa}
{Kagan}, D., {Sironi}, L., {Cerutti}, B., \& {Giannios}, D. 2015, \ssr, 191,
  545

\bibitem[{Kiehlmann {et~al.}(2016)Kiehlmann, Savolainen, Jorstad, Sokolovsky,
  Schinzel, Marscher, Larionov, Agudo, Akitaya, Ben{\'\i}tez, Berdyugin,
  Blinov, Bochkarev, Borman, Burenkov, Casadio, Doroshenko, Efimova, Fukazawa,
  G{\'o}mez, Grishina, Hagen-Thorn, Heidt, Hiriart, Itoh, Joshi, Kawabata,
  Kimeridze, Kopatskaya, Korobtsev, Krajci, Kurtanidze, Kurtanidze, Larionova,
  Larionova, Lindfors, L{\'o}pez, McHardy, Molina, Moritani, Morozova, Nazarov,
  Nikolashvili, Nilsson, Pulatova, Reinthal, Sadun, Sasada, Savchenko, Sergeev,
  Sigua, Smith, Sorcia, Spiridonova, Takaki, Takalo, Taylor, Troitsky, Uemura,
  Ugolkova, Ui, Yoshida, Zensus, \& Zhdanova}]{kiehlmann_polarization_2016}
Kiehlmann, S., {et~al.} 2016, arXiv:1603.00249 [astro-ph], arXiv: 1603.00249

\bibitem[{Landau \& Lifshitz(1975)}]{landau_classical_1975}
Landau, L.~D., \& Lifshitz, E.~M. 1975, The classical theory of fields, 4th
  edn. (Oxford ;: Pergamon Press,)

\bibitem[{{Liu} {et~al.}(2015){Liu}, {Guo}, {Daughton}, {Li}, \&
  {Hesse}}]{Liu:2015aa}
{Liu}, Y.-H., {Guo}, F., {Daughton}, W., {Li}, H., \& {Hesse}, M. 2015,
  Physical Review Letters, 114, 095002

\bibitem[{Lyutikov {et~al.}(2016)Lyutikov, Sironi, Komissarov, \&
  Porth}]{Lyutikov:2016aa}
Lyutikov, M., Sironi, L., Komissarov, S., \& Porth, O. 2016, arXiv:1603.05731
  [astro-ph, physics:physics], arXiv: 1603.05731

\bibitem[{{Marscher} {et~al.}(2008){Marscher}, {Jorstad}, {D'Arcangelo},
  {Smith}, {Williams}, {Larionov}, {Oh}, {Olmstead}, {Aller}, {Aller},
  {McHardy}, {L{\"a}hteenm{\"a}ki}, {Tornikoski}, {Valtaoja}, {Hagen-Thorn},
  {Kopatskaya}, {Gear}, {Tosti}, {Kurtanidze}, {Nikolashvili}, {Sigua},
  {Miller}, \& {Ryle}}]{Marscher:2008aa}
{Marscher}, A.~P., {et~al.} 2008, \nat, 452, 966

\bibitem[{{Nalewajko} {et~al.}(2011){Nalewajko}, {Giannios}, {Begelman},
  {Uzdensky}, \& {Sikora}}]{Nalewajko:2011aa}
{Nalewajko}, K., {Giannios}, D., {Begelman}, M.~C., {Uzdensky}, D.~A., \&
  {Sikora}, M. 2011, \mnras, 413, 333

\bibitem[{Nalewajko {et~al.}(2016)Nalewajko, Zrake, Yuan, East, \&
  Blandford}]{Nalewajko:2016aa}
Nalewajko, K., Zrake, J., Yuan, Y., East, W.~E., \& Blandford, R.~D. 2016,
  arXiv:1603.04850 [astro-ph], arXiv: 1603.04850

\bibitem[{{Rani} {et~al.}(2013){Rani}, {Lott}, {Krichbaum}, {Fuhrmann}, \&
  {Zensus}}]{Rani:2013aa}
{Rani}, B., {Lott}, B., {Krichbaum}, T.~P., {Fuhrmann}, L., \& {Zensus}, J.~A.
  2013, \aap, 557, A71

\bibitem[{{Rudy} {et~al.}(2015){Rudy}, {Horns}, {DeLuca}, {Kolodziejczak},
  {Tennant}, {Yuan}, {Buehler}, {Arons}, {Blandford}, {Caraveo}, {Costa},
  {Funk}, {Hays}, {Lobanov}, {Max}, {Mayer}, {Mignani}, {O'Dell}, {Romani},
  {Tavani}, \& {Weisskopf}}]{Rudy:2015aa}
{Rudy}, A., {et~al.} 2015, \apj, 811, 24

\bibitem[{{Sironi} {et~al.}(2016){Sironi}, {Giannios}, \&
  {Petropoulou}}]{Sironi:2016aa}
{Sironi}, L., {Giannios}, D., \& {Petropoulou}, M. 2016, ArXiv e-prints

\bibitem[{{Sironi} \& {Spitkovsky}(2014)}]{Sironi:2014aa}
{Sironi}, L., \& {Spitkovsky}, A. 2014, \apjl, 783, L21

\bibitem[{{Syrovatskii}(1966)}]{Syrovatskii:1966aa}
{Syrovatskii}, S.~I. 1966, \sovast, 10, 270

\bibitem[{Tavani {et~al.}(2011)Tavani, Bulgarelli, Vittorini, Pellizzoni,
  Striani, Caraveo, Weisskopf, Tennant, Pucella, Trois, Costa, Evangelista,
  Pittori, Verrecchia, Del~Monte, Campana, Pilia, De~Luca, Donnarumma, Horns,
  Ferrigno, Heinke, Trifoglio, Gianotti, Vercellone, Argan, Barbiellini,
  Cattaneo, Chen, Contessi, D'Ammando, DeParis, Di~Cocco, Di~Persio, Feroci,
  Ferrari, Galli, Giuliani, Giusti, Labanti, Lapshov, Lazzarotto, Lipari,
  Longo, Fuschino, Marisaldi, Mereghetti, Morelli, Moretti, Morselli, Pacciani,
  Perotti, Piano, Picozza, Prest, Rapisarda, Rappoldi, Rubini, Sabatini,
  Soffitta, Vallazza, Zambra, Zanello, Lucarelli, Santolamazza, Giommi,
  Salotti, \& Bignami}]{Tavani:2011aa}
Tavani, M., {et~al.} 2011, Science, 331, 736

\bibitem[{{Treumann} \& {Baumjohann}(1997)}]{Treumann:1997aa}
{Treumann}, R.~A., \& {Baumjohann}, W. 1997, {Advanced space plasma physics}

\bibitem[{{Uzdensky}(2016)}]{Uzdensky:2016aa}
{Uzdensky}, D.~A. 2016, in Astrophysics and Space Science Library, Vol. 427,
  Astrophysics and Space Science Library, ed. W.~{Gonzalez} \& E.~{Parker}, 473

\bibitem[{Uzdensky {et~al.}(2011)Uzdensky, Cerutti, \&
  Begelman}]{Uzdensky:2011aa}
Uzdensky, D.~A., Cerutti, B., \& Begelman, M.~C. 2011, The Astrophysical
  Journal Letters, 737, L40

\bibitem[{Weisskopf {et~al.}(2013)Weisskopf, Tennant, Arons, Blandford,
  Buehler, Caraveo, Cheung, Costa, de~Luca, Ferrigno, Fu, Funk, Habermehl,
  Horns, Linford, Lobanov, Max, Mignani, O'Dell, Romani, Striani, Tavani,
  Taylor, Uchiyama, \& Yuan}]{Weisskopf:2013aa}
Weisskopf, M.~C., {et~al.} 2013, The Astrophysical Journal, 765, 56

\bibitem[{{Werner} {et~al.}(2016){Werner}, {Uzdensky}, {Cerutti}, {Nalewajko},
  \& {Begelman}}]{Werner:2016aa}
{Werner}, G.~R., {Uzdensky}, D.~A., {Cerutti}, B., {Nalewajko}, K., \&
  {Begelman}, M.~C. 2016, \apjl, 816, L8

\bibitem[{{Yee}(1966)}]{Yee:1966aa}
{Yee}, K. 1966, IEEE Transactions on Antennas and Propagation, 14, 302

\bibitem[{{Zelenyi} \& {Krasnoselskikh}(1979)}]{Zelenyi:1979aa}
{Zelenyi}, L.~M., \& {Krasnoselskikh}, V.~V. 1979, \sovast, 23, 460

\bibitem[{{Zrake}(2016)}]{Zrake:2016ab}
{Zrake}, J. 2016, \apj, 823, 39

\bibitem[{{Zrake} \& {East}(2016)}]{Zrake:2016aa}
{Zrake}, J., \& {East}, W.~E. 2016, \apj, 817, 89

\end{thebibliography}

\end{document}